\documentclass[letterpaper]{article}

\usepackage{adjustbox}
\usepackage[endLComment=]{algpseudocodex}
\usepackage{algorithm}
\usepackage{amsfonts}
\usepackage{amsmath}
\usepackage{authblk}
\usepackage[natbib=true, style=authoryear, giveninits=true, terseinits=true, maxcitenames=2, maxbibnames=10, uniquename=mininit, uniquename=false, uniquelist=false, dashed=false]{biblatex}
\usepackage{bm}
\usepackage{booktabs}
\usepackage{caption}
\usepackage{color}
\usepackage[T1]{fontenc}
\usepackage[margin=32mm]{geometry}
\usepackage{mathtools}
\usepackage{multirow}
\usepackage[raggedrightboxes]{ragged2e}
\usepackage{titlesec}
\usepackage[colorlinks=true, allcolors=blue]{hyperref}

\DeclareMathOperator{\diag}{diag}

\newcommand{\dd}{\mathrm{d}}
\newcommand{\defas}{:=}
\newcommand{\norm}[1]{\lVert#1\rVert}
\newcommand{\bignorm}[1]{\big\lVert#1\big\rVert}

\newcommand{\ForwardModel}{\mathcal{G}}
\newcommand{\Parametrisation}{\mathcal{P}}

\newcommand{\bData}{\bm{y}}
\newcommand{\bDataObs}{\bData_{\mathrm{obs}}}
\newcommand{\Error}{\epsilon}
\newcommand{\bError}{\bm{\Error}}
\newcommand{\Param}{\theta}
\newcommand{\bParam}{\bm{\Param}}
\newcommand{\ParamAux}{\eta}
\newcommand{\bParamAux}{\bm{\ParamAux}}

\newcommand{\ErrorDist}{\pi_{\epsilon}}

\newcommand{\LeastSq}{\Phi}

\newcommand{\ParamDim}{n}
\newcommand{\DataDim}{d}
\newcommand{\GRFDim}{q}
\newcommand{\AuxDim}{p}

\newcommand{\NumEns}{J}
\newcommand{\NumFormations}{n_{r}}
\newcommand{\NumDistributions}{n_{d}}

\newcommand{\Mean}{m}
\newcommand{\bMean}{\bm{\Mean}}
\newcommand{\Cov}{C}
\newcommand{\bCov}{\bm{\Cov}}

\newcommand{\metres}{\mathrm{m}}

\newcommand{\kg}{\mathrm{kg}}

\newcommand{\CDF}{F}

\newcommand{\Std}{\sigma}
\newcommand{\Ls}{\ell}
\newcommand{\Reg}{\nu}

\newcommand{\NormalVec}{\bm{n}}

\newcommand{\Domain}{\Omega}
\newcommand{\Boundary}{\partial\Domain}

\newcommand{\DensityLarge}{M}
\newcommand{\DensitySmall}{\rho}
\newcommand{\Enthalpy}{h}
\newcommand{\bFlux}{\bm{F}}
\newcommand{\bGravity}{\bm{g}}
\newcommand{\InternalEnergy}{u}
\newcommand{\KinematicVisc}{\nu}
\newcommand{\MassFrac}{X}
\newcommand{\Perm}{k}
\newcommand{\bPerm}{\bm{\Perm}}
\newcommand{\Porosity}{\phi}
\newcommand{\Saturation}{S}
\newcommand{\Source}{q}
\newcommand{\SpecificHeat}{c}
\newcommand{\Temperature}{T}
\newcommand{\ThermalConductivity}{K}
\newcommand{\Phase}{\chi}
\newcommand{\Pressure}{P}

\newcommand{\PermField}{\xi}
\newcommand{\PermBoundary}{\omega}

\newcommand{\Shal}{\mathcal{S}}
\newcommand{\ClayCap}{\mathcal{C}}
\newcommand{\Deep}{\mathcal{D}}
\newcommand{\Fault}{\mathcal{F}}
\newcommand{\Background}{\mathcal{B}}

\newcommand{\LevelFunc}{\varphi}

\title{\itshape\Large Ensemble Kalman Inversion for Geothermal Reservoir Modelling}

\author[1]{Alex~de~Beer\thanks{Corresponding author: \href{mailto:adeb0907@uni.sydney.edu.au}{adeb0907@uni.sydney.edu.au}.}}
\author[2]{Elvar~K.~Bjarkason}
\author[3]{Michael~Gravatt}
\author[3]{Ruanui~Nicholson}
\author[3]{John~P.~O'Sullivan}
\author[3]{Michael~J.~O'Sullivan}
\author[3]{Oliver~J.~Maclaren}
\affil[1]{School~of~Mathematics~and~Statistics,~University~of~Sydney,~New~South~Wales~\emph{2006},~Australia}
\affil[2]{Graduate~School~of~International~Resource~Sciences, Akita~University,~Akita~\emph{010-8502},~Japan}
\affil[3]{Department~of~Engineering~Science~and~Biomedical~Engineering,~University~of~Auckland, Auckland~\emph{1010},~New~Zealand}
\date{}

\titleformat{\section}{\normalsize\bfseries\uppercase}{\thesection}{1em}{}
\titleformat{\subsection}{\normalsize\bfseries}{\thesubsection}{1em}{}
\titleformat{\subsubsection}{\normalsize\itshape}{\thesubsubsection}{1em}{}

\hypersetup{
    colorlinks=True, 
    allcolors=blue
}

\bibliography{references}

\DeclareNameAlias{sortname}{family-given}
\DefineBibliographyStrings{english}{%
    andothers={\textit{et al}\adddot},
    backrefpage={page},
    backrefpages={pages},
    edition={Edition},
    mathesis={Master's thesis},
    references={\normalsize\MakeUppercase{References}},
}

\DeclareFieldFormat{pages}{#1}
\DeclareFieldFormat[article, inbook, inproceedings, thesis]{title}{#1} 
\DeclareFieldFormat[article]{volume}{\mkbibbold{#1}}
\AtEveryBibitem{\clearfield{number}}
\renewbibmacro{in:}{}

\AtBeginBibliography{%
}

\newenvironment{abs}{
    \small
    \begin{center}
        {\bfseries\MakeUppercase{\abstractname}}
    \end{center}
    \quotation
}
{\endquotation}

\begin{document}

\maketitle

\begin{abs}

Numerical models of geothermal reservoirs typically depend on hundreds or thousands of unknown parameters, which must be estimated using sparse, noisy data. However, these models capture complex physical processes, which frequently results in long run-times and simulation failures, making the process of estimating the unknown parameters a challenging task. Conventional techniques for parameter estimation and uncertainty quantification, such as Markov chain Monte Carlo (MCMC), can require tens of thousands of simulations to provide accurate results and are therefore challenging to apply in this context. In this paper, we study the ensemble Kalman inversion (EKI) algorithm as an alternative technique for approximate parameter estimation and uncertainty quantification for geothermal reservoir models. EKI possesses several characteristics that make it well-suited to a geothermal setting; it is derivative-free, parallelisable, robust to simulation failures, and in many cases requires far fewer simulations to provide an accurate characterisation of the posterior than conventional uncertainty quantification techniques such as MCMC. We illustrate the use of EKI in a reservoir modelling context using a combination of synthetic and real-world case studies. Through these case studies, we also demonstrate how EKI can be paired with flexible parametrisation techniques capable of accurately representing prior knowledge of the characteristics of a reservoir and adhering to geological constraints, and how the algorithm can be made robust to simulation failures. Our results demonstrate that EKI provides a reliable and efficient means of obtaining accurate parameter estimates for large-scale, two-phase geothermal reservoir models, with appropriate characterisation of uncertainty.

\end{abs}

\section{Introduction}

Simulating the dynamics of a geothermal reservoir involves solving a numerical model, which is typically formed by discretising the system into a large number of interacting volumes \citep{OSullivan16}. Such a model is dependent on parameters including the permeability and porosity of the rock in each volume and the location and magnitude of the hot mass upflow at the roots of the system. However, measurements of these parameters are usually sparse or unavailable entirely; instead, the data that is available consists of quantities such as downhole temperature and pressure observed at a set of wells. The problem of estimating the model parameters based on their ability to reproduce these observations is referred to as an \emph{inverse} (or calibration) problem \citep{Aster18, Kaipio06, Stuart10, Tarantola05}. This is a key component of the modelling process; improving our parameter estimates will improve the quality of forecasts produced using the model to guide decisions concerning the management of the system \citep{OSullivan16}.

Inverse problems arising in geothermal reservoir modelling are, in general, ill-posed; a solution may not exist, there may be multiple solutions, or a solution may exhibit sensitivity to small changes in the observations. Classical methods for solving inverse problems typically deal with this ill-posedness by framing the problem as the minimisation of a functional that quantifies the misfit between the model outputs and observations, supplemented with additional terms or constraints to favour solutions with desirable characteristics \citep{Aster18}; these additional terms or constraints are referred to as forms of \emph{regularisation}. An alternative approach, which we adopt in this work, is to consider the problem from the perspective of Bayesian statistics \citep{Kaipio06, Sanz23, Stuart10}. In the Bayesian approach to inverse problems, we treat both the unknown parameters and the observations as random variables. Rather than a single point estimate of the parameters, the solution to the inverse problem is an entire probability distribution, referred to as the \emph{posterior}. Once characterised, the posterior uncertainty in the parameters can be propagated through to the model predictions, allowing for a probabilistic description of the future behaviour of the system.

The complexity of a typical reservoir model means that the posterior is not available in closed form. Instead, it is generally approximated using samples. The most widely-used sampling methods for general Bayesian inverse problems include forms of Markov chain Monte Carlo \citep[MCMC;][]{Brooks11} and sequential Monte Carlo \citep[SMC;][]{Del06}. In the limit of an infinite number of samples, these methods recover the exact posterior distribution. However, they often require at least $\mathcal{O}(10^4)$ simulations of the forward model to provide an accurate characterisation \citep{Brooks11}, which can be prohibitive in a geothermal setting, where a single simulation may require hours or even days of computing time. For this reason, MCMC and SMC are seldom used to solve inverse problems arising in geothermal reservoir modelling. Notable exceptions include the studies of \citet{Cui11, Cui19, Maclaren20, Scott22}, in which physics-based surrogate models are used to accelerate the MCMC sampling process; it can, however, be challenging to develop an effective surrogate model, and even then the MCMC sampler may still require many thousands of simulations to attain statistical convergence.

The challenges associated with the use of MCMC and SMC methods have motivated the development of algorithms that characterise an approximation to the posterior, but at a significantly reduced computational cost. A feature common to many of these methods is that, under a linear forward model and Gaussian prior and error distributions, the samples they generate are distributed according to the posterior; when these conditions are violated, however, this is no longer true in general. Examples of methods that have been used in geothermal settings include local linearisation \citep[see, e.g.,][]{Omagbon21}, which involves fitting a Gaussian approximation to the posterior based on a linearisation of the forward model about the set of parameters with the greatest posterior probability \citep[the \emph{maximum-a-posteriori} estimate;][]{Kaipio06}, and randomised maximum likelihood \citep[RML; see, e.g.,][]{Tian24, Tureyen14, Zhang14}, which involves repeatedly solving stochastic optimisation problems to obtain samples distributed in regions of high posterior probability. 

An alternative class of algorithms for approximate Bayesian inference are ensemble methods, in which an interacting ensemble of \emph{particles}---sets of parameters---are combined with data, in an iterative manner, such that the distribution of the ensemble approximates the posterior \citep{Calvello22}. An advantage of ensemble methods over methods such as local linearisation and RML is that they do not require derivative information, which can be challenging to obtain; instead, derivatives are approximated using the ensemble. In addition, they are straightforward to parallelise; at each iteration, the entire ensemble, which is typically comprised of $\mathcal{O}(10^{2})$ particles, can be simulated simultaneously. The first ensemble-based algorithm was the ensemble Kalman filter \citep[EnKF;][]{Evensen94, Evensen09, Katzfuss16}. The EnKF was developed for data assimilation; that is, the process of combining a model with data to estimate the state of a dynamical system \citep{Law15, Wikle07}. It has been used extensively in areas including oceanography, petroleum engineering and weather forecasting \citep{Aanonsen09, Evensen09}, where the dimension of the unknown state may be on the order of millions but the complexity of the model limits the number of simulations that can be run to fewer than one thousand. A great deal of subsequent research, however, has focused on the development of ensemble methods for approximating the solutions to inverse problems. These methods originated in the petroleum engineering community \citep{Chen12, Chen13, Emerick13}, but have subsequently been applied in a variety of areas, including climate modelling \citep{Dunbar21, Dunbar22}, composites manufacturing \citep{Iglesias18, Matveev21, Causon24}, geophysics \citep{Muir20, Muir22, Oakley23, Tso21, Tso24}, healthcare \citep{Iglesias22}, machine learning \citep{Kovachki19}, and structural engineering \citep{De18, Iglesias24}. Both the theory and application of ensemble methods remain highly active areas of research; for further details, we refer the reader to \citet{Calvello22, Evensen22} and the references therein.

While there exist studies that apply ensemble methods within a geothermal context \citep{Bekesi20, Bjarkason21a, Bjarkason21b}, the full potential of these methods is yet to be explored in detail. Our work makes several key contributions in this direction. First, we illustrate how ensemble methods can be combined with powerful geometric, hierarchical and geologically-constrained model parametrisations capable of accurately representing prior knowledge of the characteristics of a geothermal reservoir. Second, we show how ensemble methods can be applied in a robust manner to reservoir models which encounter frequent simulation failures. Finally, we provide a thorough evaluation of the accuracy and scalability of our proposed ensemble-based framework using several case studies involving large-scale, two-phase reservoir models, including a real-world reservoir model with real data.

In this work, we focus on the ensemble Kalman inversion (EKI) algorithm \citep{Iglesias13, Iglesias21}; in a previous study involving several widely-used ensemble methods, we found that EKI provided the best approximation to a reference posterior, characterised using MCMC, for a high-dimensional subsurface flow problem \citep{deBeer24a}. We anticipate, however, that similar results to those we present in this paper could be obtained using alternative ensemble methods. We emphasise that, unlike conventional techniques such as MCMC and SMC, outside the linear-Gaussian setting the EKI algorithm will not, in general, produce samples distributed according to the posterior, even in the limit of an infinite ensemble size. There is also little theory to quantify the quality of the approximation to the posterior produced using EKI outside the linear-Gaussian setting, and there is no guarantee that if the algorithm produces accurate results when applied to a given problem, it will perform well when applied to similar problems. Nonetheless, there are a significant number of empirical results in a variety of applications (including those in the current work) that suggest this approximation is often quite accurate.

The remainder of this paper is structured as follows. In Section \ref{sec:methods}, we discuss the key components of our EKI-based approach to solving geothermal inverse problems. In Section \ref{sec:problems}, we describe the model problems we use to demonstrate the application of EKI in a geothermal context, and in Section \ref{sec:results} we describe the results of these experiments. Section \ref{sec:conclusions} summarises our conclusions and identifies avenues for future work.

\section{Methodology} \label{sec:methods}

We now provide background information on a number of important concepts used throughout the paper. Section \ref{sec:forward_prob} introduces the mathematics of geothermal reservoir modelling. Section \ref{sec:bayes} outlines the Bayesian approach to inverse problems, and Section \ref{sec:eki} introduces EKI as an efficient technique for approximating the posterior distribution. Finally, Section \ref{sec:parametrisations} outlines several key parametrisation techniques we use in combination with EKI to accurately model various geophysical characteristics.

We briefly introduce some key notation used throughout the remainder of the paper. We use boldface to denote matrices and vectors. We use $\mathbb{N}$ to denote the natural numbers, $\mathbb{R}$ to denote the real numbers, $\mathbb{R}_{+} \defas (0, \infty)$ to denote the positive real numbers, and $\bar{\mathbb{R}} \defas \mathbb{R} \cup \{-\infty, \infty\}$ to denote the extended real numbers. We write $\norm{\,\cdot\,}_{\bm{A}} \defas \norm{\bm{A}^{-1/2} \,\cdot\,}$ to denote the Euclidean norm weighted by the symmetric positive definite matrix $\bm{A}$, and we use $\{x_{i}\}_{i=1}^{n}$ as a shorthand for the set $\{x_{1}, x_{2}, \dots, x_{n}\}$.

\subsection{Forward Problem} \label{sec:forward_prob}

Mathematical models of geothermal reservoirs are governed by a set of time-dependent equations enforcing conservation of mass and conservation of energy \citep{Croucher20, OSullivan16}. Where multiple mass components are present (e.g., water and carbon dioxide), a separate mass conservation equation is solved for each. In what follows, we let $\Domain$ denote the reservoir domain, with boundary $\Boundary$ and outward-facing unit normal vector $\NormalVec$. Then, conservation of mass and energy can be expressed as
\begin{equation}
    \frac{\dd}{\dd t}\int_{\Domain}\DensityLarge^{\kappa}\,\dd\bm{x} = -\int_{\Boundary}\bFlux^{\kappa}\cdot\NormalVec\,\dd\bm{\sigma} + \int_{\Domain}\Source^{\kappa}\,\dd\bm{x}, \,\, \kappa = 1, 2, \dots, N+1,
\end{equation}
where components $\kappa = 1, 2, \dots, N$ are mass components, and component $N+1$ is the energy component. For each mass component ($\kappa \leq N$), $\DensityLarge^{\kappa}$ denotes mass density (kg\,m$^{-3}$), $\bFlux^{\kappa}$ denotes mass flux (kg\,m$^{-2}$\,s$^{-1}$), and $\Source^{\kappa}$ denotes mass sources or sinks (kg\,m$^{-3}$\,s$^{-1}$). For the energy component ($\kappa = N + 1$), $\DensityLarge^{\kappa}$ denotes energy density (J\,m$^{-3}$), $\bFlux^{\kappa}$ denotes energy flux (J\,m$^{-2}$\,s$^{-1}$), and $\Source^{\kappa}$ denotes energy sources or sinks (J\,m$^{-3}$\,s$^{-1}$). The mass and energy densities are given by
\begin{equation}
    \DensityLarge^{\kappa} = \begin{cases}
        \Porosity\sum_{\Phase}\DensitySmall_{\Phase}\Saturation_{\Phase}\MassFrac^{\kappa}_{\Phase}, \,\,& \kappa \leq N, \\
        (1 - \Porosity)\DensitySmall_{r}\SpecificHeat_{r} \Temperature + \Porosity\sum_{\Phase}\DensitySmall_{\Phase}\InternalEnergy_{\Phase}\Saturation_{\Phase}, \,\,& \kappa = N + 1.
    \end{cases}
\end{equation}
Here, $\Porosity$ denotes porosity (dimensionless), $\Saturation_{\Phase}$ denotes the saturation of phase $\Phase$ (dimensionless), $\MassFrac_{\Phase}^{\kappa}$ denotes the mass fraction of phase $\Phase$ of component $\kappa$ (dimensionless), $\DensitySmall_{\Phase}$ denotes the density of phase $\Phase$ (kg\,m$^{-3}$), $\InternalEnergy_{\Phase}$ denotes the specific internal energy of phase $\Phase$ (J\,kg$^{-1}$), $\DensitySmall_{r}$ denotes the rock grain density (kg\,m$^{-3}$) $\SpecificHeat_{r}$ denotes the specific heat capacity of rock (J\,kg$^{-1}$\,K$^{-1}$), and $\Temperature$ denotes temperature (K). The mass and energy fluxes are given by
\begin{equation}
    \bFlux^{\kappa} = \begin{cases}
        \sum_{\Phase}\bFlux_{\Phase}^{\kappa}, \quad& \kappa \leq N, \\
        -\ThermalConductivity\nabla\Temperature + \sum_{m=1}^{N}\sum_{\Phase}\Enthalpy_{\Phase}\bFlux_{\Phase}^{m}, \quad& \kappa = N + 1, 
    \end{cases}
\end{equation}
where the fluxes of each phase of each component are described by a multiphase version of Darcy's law:
\begin{equation}
    \bFlux_{\Phase}^{\kappa} = -\frac{\bPerm\Perm_{r\Phase}}{\KinematicVisc_{\Phase}}\MassFrac^{\kappa}_{\Phase}(\nabla \Pressure - \rho_{\Phase}\bGravity).
\end{equation}
In the above, $\Enthalpy_{\Phase}$ denotes the specific enthalpy of phase $\Phase$ (J\,kg$^{-1}$), $\ThermalConductivity$ denotes thermal conductivity (J\,s$^{-1}$\,m$^{-1}$\,K$^{-1}$), $\bPerm$ denotes the permeability tensor (m$^{2}$), $\Perm_{r\Phase}$ denotes the relative permeability of phase $\Phase$ (dimensionless), $\KinematicVisc_{\Phase}$ denotes the kinematic viscosity of phase $\Phase$ (m$^{2}$\,s$^{-1}$), $\Pressure$ denotes pressure (kg\,m$^{-2}$), and $\bGravity$ denotes gravitational acceleration (m\,s$^{-2}$). 

\subsection{Bayesian Inverse Problems} \label{sec:bayes}

Posing an inverse problem within the Bayesian framework requires us to specify the nature of the relationship between the unknown parameters, $\bParam \in \mathbb{R}^{\ParamDim}$, and the observations, $\bData \in \mathbb{R}^{\DataDim}$. In a geothermal setting, the unknown parameters are typically comprised of quantities such as the components of the permeability tensor in each cell of the (discretised) reservoir model, and the magnitude of the hot mass upflow entering through each cell at the base of the model domain. The observations, by contrast, are typically comprised of quantities such as the reservoir temperature, pressure and enthalpy, recorded at a set of wells. Throughout this work, we relate the parameters and observations using the so-called \emph{additive error model} \citep{Kaipio06}, given by
\begin{equation}
    \bData = \ForwardModel(\bParam) + \bError,
\end{equation}
where $\ForwardModel : \mathbb{R}^{\ParamDim} \rightarrow \mathbb{R}^{\DataDim}$ denotes the forward model, and $\bError \in \mathbb{R}^{\DataDim}$ denotes unknown errors resulting from factors such as noise in the observations or discrepancies between the model and the true data-generating mechanism. In a geothermal setting, the forward model can typically be decomposed as 
\begin{equation}
    \ForwardModel = \mathcal{B} \circ \mathcal{F},
\end{equation}
where the mapping $\mathcal{F}$ involves simulating the dynamics of the reservoir, as outlined in Section \ref{sec:forward_prob}, using a given set of parameters, while the mapping $\mathcal{B}$ extracts the observed quantities from the full set of simulation outputs.

In a Bayesian setting, we model the parameters, observations and errors as random variables. Our aim is to characterise the \emph{posterior} density; that is, the conditional density of $\bParam$ given a particular realisation of the data, $\bDataObs$, denoted by $\pi(\bParam|\bDataObs)$. To do this, we proceed by first specifying a \emph{prior} density, $\bParam \sim \pi(\bParam)$, for the parameters, and a density, $\bError \sim \ErrorDist(\bError)$, for the errors. We assume that the parameters and errors are independent; in this case, it is straightforward to show that the conditional density, or \emph{likelihood}, of $\bDataObs$ given $\bParam$ inherits the density of the errors \citep{Kaipio06}; that is, 
\begin{equation}
    \pi(\bDataObs|\bParam) = \ErrorDist(\bDataObs - \ForwardModel(\bParam)).
\end{equation}
By Bayes' theorem, the posterior can then be expressed as
\begin{equation}
    \pi(\bParam|\bDataObs) = \frac{1}{Z}\pi(\bDataObs|\bParam)\pi(\bParam) = \frac{1}{Z}\ErrorDist(\bDataObs - \ForwardModel(\bParam))\pi(\bParam), \label{eq:bayes}
\end{equation}
where the normalising constant, $Z$, is defined as
\begin{equation}
    Z \defas \int_{\mathbb{R}^{\ParamDim}} \pi(\bDataObs|\bParam)\pi(\bParam) \, \dd\bParam = \pi(\bDataObs).
\end{equation}

Throughout this paper, we model the distribution of the errors as a centred Gaussian with known covariance; that is, $\bError \sim \mathcal{N}(\bm{0}, \bCov_{\Error})$. The likelihood can therefore be expressed as
\begin{equation}
    \ErrorDist(\bDataObs - \ForwardModel(\bParam)) \propto \exp\left(-\LeastSq(\bParam; \bDataObs)\right), \label{eq:post_gauss}
\end{equation}
where the \emph{data misfit functional} $\LeastSq(\bParam; \bDataObs)$, which quantifies the fit of the model to the data, is defined as
\begin{equation}
    \LeastSq(\bParam; \bDataObs) \defas \frac{1}{2}\norm{\bDataObs - \ForwardModel(\bParam)}_{\bCov_{\Error}}^{2}. \label{eq:least_sq}
\end{equation}



\subsection{Ensemble Kalman Inversion} \label{sec:eki}

Ensemble Kalman inversion (EKI) was originally introduced by \cite{Iglesias13} as a derivative-free optimisation technique aimed at minimising the least-squares functional \eqref{eq:least_sq} within the subspace spanned by the initial ensemble. Subsequent studies, however, have developed variants of EKI aimed at approximating the posterior \citep{Iglesias18, Iglesias21}. In this setting, the algorithm can be viewed as approximating a sequence of densities, $\{\pi_{i}(\cdot)\}_{i=1}^{\NumDistributions}$, that transition smoothly from prior to posterior. In particular, the $i$th density approximated by the algorithm takes the form
\begin{equation}
    \pi_{i}(\bParam) \propto \pi(\bParam)^{1-t^{(i)}}\pi(\bParam|\bDataObs)^{t^{(i)}}, \label{eq:densities}
\end{equation}
where $0 = t^{(1)} < t^{(2)} < \cdots < t^{(\NumDistributions)} = 1$. We note that these variants of EKI possess the same update dynamic (discussed below) as the ensemble smoother with multiple data assimilation (ES-MDA), introduced by \citet{Emerick13} in the context of petroleum engineering. They differ, however, in terms of the choice of the intermediate densities that are approximated by the ensemble.

To initialise the EKI algorithm, we sample an ensemble of $\NumEns \in \mathbb{N}$ particles, $\{\bParam_{j}^{(1)}\}_{j=1}^{\NumEns}$, from the prior. At each subsequent iteration, we apply the forward model to each particle of the ensemble, to obtain the set of simulation outputs $\{\ForwardModel(\bParam_{j}^{(i)})\}_{j=1}^{\NumEns}$. We note that this can be done in parallel. Then, we estimate the (cross-)covariances $\bCov_{\Param\ForwardModel}^{(i)}$ and $\bCov_{\ForwardModel\ForwardModel}^{(i)}$ using the ensemble:
\begin{align}
    \bCov_{\Param\ForwardModel}^{(i)} &= \frac{1}{\NumEns-1} \sum_{j=1}^{\NumEns} (\bParam_{j}^{(i)} - \bMean_{\Param}^{(i)})(\ForwardModel(\bParam_{j}^{(i)}) - \bMean_{\ForwardModel}^{(i)})^{\top}, \label{eq:cov_tg} \\
    \bCov_{\ForwardModel\ForwardModel}^{(i)} &= \frac{1}{\NumEns - 1} \sum_{j=1}^{\NumEns} (\ForwardModel(\bParam_{j}^{(i)}) - \bMean_{\ForwardModel}^{(i)})(\ForwardModel(\bParam_{j}^{(i)}) - \bMean_{\ForwardModel}^{(i)})^{\top}, \label{eq:cov_gg}
\end{align}
where the means $\bMean_{\Param}^{(i)}$ and $\bMean_{\ForwardModel}^{(i)}$ are given by
\begin{equation}
    \bMean_{\Param}^{(i)} = \frac{1}{\NumEns}\sum_{j=1}^{\NumEns} \bParam_{j}^{(i)}, \quad \bMean_{\ForwardModel}^{(i)} = \frac{1}{\NumEns}\sum_{j=1}^{\NumEns} \ForwardModel(\bParam_{j}^{(i)}).
\end{equation}
Finally, we update each particle according to
\begin{equation}
    \bParam^{(i+1)}_{j} = \bParam^{(i)}_{j} + \bCov_{\Param\ForwardModel}^{(i)}(\bCov_{\ForwardModel\ForwardModel}^{(i)} + \alpha^{(i)}\bCov_{\Error})^{-1}(\bData - \ForwardModel(\bParam_{j}^{(i)}) + \bError^{(i)}_{j}), \label{eq:eki_update}
\end{equation}
where $\{\bError_{j}^{(i)}\}_{j=1}^{\NumEns} \sim \mathcal{N}(\bm{0}, \alpha^{(i)}\bCov_{\Error})$ denotes a set of artificial error terms, and the parameter $\alpha^{(i)}$ (often termed the regularisation parameter) is given by
\begin{equation}
    \alpha^{(i)} = \frac{1}{t^{(i+1)}-t^{(i)}}. \label{eq:timestep}
\end{equation}

An important choice that must be made when implementing EKI is the selection of an appropriate value for the regularisation parameter, $\alpha^{(i)}$ (which determines $t^{(i+1)}$), at each iteration. Loosely speaking, increasing the regularisation parameter will increase the number of intermediate densities approximated by the EKI algorithm as it transitions from prior to posterior, providing the particles with an increased opportunity to explore the parameter space and giving them a greater chance of moving towards regions of high posterior density. This is, however, associated with an increased computational cost; each additional intermediate density that is approximated requires an additional $\NumEns$ runs of the forward model. A variety of techniques for choosing $\alpha^{(i)}$ have been proposed, both in the context of ES-MDA \citep[see, e.g.,][]{Emerick19, Le16, Rafiee17} and EKI \citep{Iglesias18, Iglesias21}. In this paper, we use the method proposed by \cite{Iglesias21}, termed the data misfit controller (DMC), in which the regularisation parameter is selected adaptively to control the symmetrised Kullback-Leibler divergence between the densities approximated at each iteration of the algorithm. More specifically, the DMC selection of $\alpha^{(i)}$ is given by
\begin{equation}
    \alpha^{(i)} = \min\left\{\max\left\{\frac{\DataDim}{2\Mean_{\LeastSq}}, \sqrt{\frac{\DataDim}{2\sigma_{\LeastSq}^{2}}}\right\}, 1-t^{(i)}\right\}^{-1}, \label{eq:eki_reg}
\end{equation}
where $\DataDim$ denotes the dimension of the observations, and $\Mean_{\LeastSq}$ and $\sigma^{2}_{\LeastSq}$ are defined as
\begin{align}
    \Mean_{\LeastSq} &= \frac{1}{\NumEns}\sum_{j=1}^{\NumEns} \LeastSq(\bParam^{(i)}_{j}; \bData),\\ 
    \sigma^{2}_{\LeastSq} &= \frac{1}{\NumEns-1}\sum_{j=1}^{\NumEns}(\LeastSq(\bParam^{(i)}_{j}; \bData) - m_{\LeastSq})^{2}.
\end{align}
A drawback of selecting $\alpha^{(i)}$ using \eqref{eq:eki_reg} is that we are not able to specify the number of iterations required by the EKI algorithm \emph{a-priori}. In the numerical experiments discussed in Section \ref{sec:results}, however, we find that between $4$ and $7$ iterations are required.

\subsubsection{Convergence and Convergence Diagnostics}

The implementation of EKI we use in this work is considered converged once the value of $t^{(i+1)}$, related to the value of the regularisation parameter $\alpha^{(i)}$ through Equation \eqref{eq:timestep}, is equal to $1$. This is motivated by the theoretical properties of EKI in the linear-Gaussian setting; in the limit of an infinite number of particles, under a linear forward model and Gaussian prior and error distributions, this convergence criterion ensures that the particles of the final ensemble are distributed exactly according to the posterior. For general Bayesian inverse problems this is no longer true; nonetheless, this criterion is still, in general, adhered to.

Given the lack of theoretical results regarding the ability of EKI to characterise the posterior distribution for general Bayesian inverse problems, it is important to check that the approximation to to the posterior computed using the algorithm appears reasonable. In the case studies presented in Section \ref{sec:problems}, we make extensive use of posterior predictive checks \citep[see, e.g.,][]{Gelman13, Gelman20} to confirm that the data is consistent with the simulated outputs generated using the final ensemble. We note also that quantitative measures of the fit to the data have been suggested for the same purpose in some studies \citep[see, e.g.,][]{Iglesias24, Tso24}. However, it is important to emphasise that any of these types of checks should be interpreted with caution. It is possible for the particles of the final EKI ensemble to fit the data well, but provide a poor representation of the posterior. Conversely, if the particles of the final ensemble fit the data poorly, it is possible that this is an issue with the modelling of the problem under consideration rather than a shortcoming of the EKI algorithm.

\subsubsection{Localisation and Inflation}

While the standard EKI algorithm we have described is often used successfully to solve real-world inverse problems, it can suffer from issues that arise when the size of the ensemble is significantly smaller than the dimension of the parameter space; that is, $\NumEns \ll \ParamDim$ \citep{Evensen22, Lacerda19}. In particular, the ensemble estimates of the covariances $\bCov_{\Param\ForwardModel}$ and $\bCov_{\ForwardModel\ForwardModel}$ (defined in Eqs \ref{eq:cov_tg} and \ref{eq:cov_gg}) can be affected by spurious correlations that arise when a small ensemble is used, resulting in updates to model parameters being influenced by observations with which they have no relationship, as well as the underestimation of posterior variances and even, in extreme cases, the collapse of the ensemble onto a single point in parameter space \citep{Lacerda19}. In addition, EKI possesses the invariant subspace property \citep{Iglesias13}; that is, at each iteration, the updated ensemble is contained within the subspace spanned by the initial ensemble. If the size of the ensemble is small in comparison to the dimension of the parameter space, there may exist regions of parameter space with significant posterior probability that cannot be reached by the ensemble.

A variety of ad-hoc techniques have been developed to address these issues. Two of the most widely used of these are \emph{localisation} and \emph{inflation} \citep{Evensen22}. Localisation techniques aim to reduce the effect of spurious correlations when the ensemble is updated at each iteration of EKI by modifying the ensemble estimates of the covariance matrices $\bCov_{\Param\ForwardModel}$ and $\bCov_{\ForwardModel\ForwardModel}$, or the so-called \emph{Kalman gain} \citep{Evensen22}. These modifications typically have the additional effect of breaking the subspace property. By contrast, inflation techniques artificially increase the variance of the ensemble at each iteration. 

In the Supplementary Material, we provide an illustration of the differences in the approximation to the posterior obtained, for each of the synthetic case studies considered in Section \ref{sec:problems}, when EKI is used without localisation or inflation and when it used with well-known localisation and inflation techniques applied. Namely, we apply the bootstrap-based localisation method proposed by \citet{Zhang10}, and the adaptive inflation method proposed by \citet{Evensen09}. In all cases, the results produced using localisation and inflation, and those produced without either technique applied, are similar. However, as expected, both localisation and inflation act to increase the variance of the final EKI ensemble and result in a greater number of the parameters of each reservoir model being contained within the final ensemble. Our results suggest that, if one's aim is to capture the true values of the model parameters and predictive quantities of interest within the final ensemble, the use of these techniques is advisable. We note, however, that these results tell us little in terms of whether the use of localisation and inflation result in an improved approximation to the posterior; it is possible that, for the particular problems we study, a significant proportion of the true values of the parameters are located in the tails of the posterior. To answer this definitively, we would need to characterise the posterior using an exact sampling method such as MCMC \citep[see, e.g.,][]{Emerick13b, Iglesias13b}, which in a geothermal setting is infeasible due to the computational cost. For this reason, the results we report on in the main body of the paper are obtained without the use of localisation or inflation. Nonetheless, the preliminary results in the Supplementary Material suggest that these techniques are potentially useful when applying the EKI algorithm in a geothermal setting, and should be further investigated.

\subsubsection{Simulation Failures} \label{sec:resampling}

A consideration of particular importance in a geothermal setting is how failed simulations should be dealt with when applying EKI; the failure rate of a geothermal reservoir model varies significantly depending on the characteristics of the model and the choice of prior parametrisation, but it is not uncommon for it to exceed 50 percent \citep[see, e.g.,][]{Dekkers23, Omagbon21}. Here, we use the approach of \citet{Lopez22} and maintain the size of the ensemble throughout the inversion by resampling failed particles from a Gaussian distribution with moments estimated using the successful particles. At each iteration, we update the successful particles, whose indices we denote using $\mathcal{J}_{s}$, using Equation \eqref{eq:eki_update}. Subsequently, we compute their empirical mean and covariance:
\begin{align}
    \hat{\bMean}_{\Param}^{(i)} &\defas \frac{1}{|\mathcal{J}_{s}|}\sum_{j \in \mathcal{J}_{s}} \bParam_{j}^{(i)}, \label{eq:succ_mean} \\
    \hat{\bCov}_{\Param\Param}^{(i)} &\defas \frac{1}{|\mathcal{J}_{s}| - 1} \sum_{j \in \mathcal{J}_{s}} (\bParam_{j}^{(i)} - \hat{\bMean}_{\Param}^{(i)})(\bParam_{j}^{(i)} - \hat{\bMean}_{\Param}^{(i)})^{\top}. \label{eq:succ_cov}
\end{align}
We then resample the failed particles from the Gaussian distribution given by $\mathcal{N}(\hat{\bMean}_{\Param}^{(i)}, \hat{\bCov}_{\Param\Param}^{(i)} + \delta \bCov_{0})$, where $\bCov_{0}$ denotes the prior covariance, and $\delta > 0$ is a small constant. 

The addition of a multiple of the prior covariance to the empirical covariance of the ensemble acts to increase its rank, which is otherwise bounded from above by the number of successful particles, and means that particles sampled from the resulting Gaussian distribution can be distributed in regions outside the span of the successful particles. If the empirical covariance were used without this modification, each resampled particle would be confined to the subspace spanned by the successful particles, which would reduce in size with each failed simulation. 

The selection of the parameter $\delta$ is an important part of the resampling process. If $\delta$ is chosen to be too small, the resampled particles will lie close to the subspace spanned by the successful particles, and it may remain challenging for the particles to explore regions of the parameter space outside of this subspace. On the other hand, if $\delta$ is chosen to be too large, the variance of the resampled particles will be larger than that of the successful particles, which will likely result in an increase in subsequent simulation failures, and the potential overestimation of posterior variances. In our numerical experiments, we use a value of $\delta=10^{-4}$. We note, however, that this is by no means the best possible choice of $\delta$. In future work, it would be valuable to investigate the effects of modifying the value of $\delta$ on the number of subsequent simulation failures and the quality of the solution to the inverse problem, and to develop guidelines for the selection of this parameter.

Algorithm \ref{alg:eki} provides a summary of the complete EKI-DMC procedure, including the resampling step.

\begin{algorithm}[hbt!]
    
    \caption{EKI-DMC with Resampling} \label{alg:eki}
    
    \begin{algorithmic}[1]
        
        \State Set $i \gets 1$, $t^{(1)} \gets 0$

        \LComment{Sample initial ensemble from prior}

        \For{$j \in \{1, 2, \dots, \NumEns\}$}
            \State Sample $\bParam^{(1)}_{j} \sim \pi(\bParam)$
        \EndFor
        
        \While{$t^{(i)} < 1$}
            
            \State Set $\mathcal{J}_{s} \gets \emptyset$, $\mathcal{J}_{f} \gets \emptyset$

            \LComment{Simulate ensemble (parallelisable)}
            \For{$j \in \{1, 2, \dots, \NumEns\}$}
                
                \State Simulate $ \ForwardModel(\bParam_{j}^{(i)})$
                
                \If{\Call{FailedSimulation}{$\ForwardModel(\bParam_{j}^{(i)})$}}
                    \State{Set $\mathcal{J}_{f} \gets \mathcal{J}_{f} \cup \{j\}$}
                \Else
                    \State{Set $\mathcal{J}_{s} \gets \mathcal{J}_{s} \cup \{j\}$}
                \EndIf
            \EndFor

            \LComment{Update successful particles}
            
            \State Compute $\alpha^{(i)}$ (Eq.~\ref{eq:eki_reg})

            \State Compute $\bCov_{\Param\ForwardModel}^{(i)}$ and $\bCov_{\ForwardModel\ForwardModel}^{(i)}$ (Eqs \ref{eq:cov_tg}, \ref{eq:cov_gg})

            \For{$j \in \mathcal{J}_{s}$}
                \State Compute $\bParam_{j}^{(i+1)}$ (Eq.~\ref{eq:eki_update})
            \EndFor

            \LComment{Resample failed particles}

            \State Compute $\hat{\bMean}_{\Param}^{(i+1)}$ and $\hat{\bCov}_{\Param\Param}^{(i+1)}$ (Eqs \ref{eq:succ_mean}, \ref{eq:succ_cov})

            \For{$j \in \mathcal{J}_{f}$}
                \State Sample $\bParam^{(i+1)}_{j} \sim \mathcal{N}(\hat{\bMean}_{\Param}^{(i+1)}, \hat{\bCov}_{\Param\Param}^{(i+1)} + \delta \bCov_{0})$
            \EndFor

            \State Set $t^{(i+1)} \gets t^{(i)} + \frac{1}{\alpha^{(i)}}$
            \State Set $i \gets i + 1$
    
        \EndWhile
        
        \State \Output $\{\bParam^{(i)}_{1}, \bParam^{(i)}_{2}, \dots, \bParam^{(i)}_{\NumEns}\}$
        
    \end{algorithmic}
\end{algorithm}

\subsection{Prior Parametrisations} \label{sec:parametrisations}

A key part of solving an inverse problem using the Bayesian framework is the specification of a suitable prior distribution for the unknown parameters. This can be challenging, as it typically involves translating qualitative information provided by experts on the likely values of these parameters into the form of a probability density \citep{Kaipio06}. 

Some challenges arise if we aim to use EKI to infer the values of the geological parameters of interest directly. Firstly, there may be some \emph{hyperparameters}---that is, parameters of the prior itself (such as the standard deviation and lengthscale associated with a Gaussian random field; see Section \ref{sec:grfs})---that are poorly known and should also be incorporated into the prior parametrisation. Additionally, ensemble methods do not naturally adhere to constraints and discontinuities in the parametrisation when updating the particles of the ensemble at each iteration.

To circumvent these difficulties, we first define a set of unconstrained auxiliary parameters, $\bParamAux \in \mathbb{R}^{\AuxDim}$ (which include values for the hyperparameters), and a mapping, $\Parametrisation : \mathbb{R}^{\AuxDim} \rightarrow \mathbb{R}^{\ParamDim}$, which acts to transform realisations of these axillary parameters into corresponding realisations of the geological parameters, $\bParam \in \mathbb{R}^{\ParamDim}$. We then use EKI to infer the values of the axillary parameters, using the augmented forward model defined by
\begin{equation}
    \ForwardModel = \mathcal{B} \circ \mathcal{F} \circ \Parametrisation,
\end{equation}
before transforming the particles of the final ensemble using $\Parametrisation$ to obtain the EKI approximation to the posterior of the geological parameters. 

In the remainder of this section, we elaborate on some of the key transformations we use when defining the mapping $\Parametrisation$, that allow us to represent a diverse range of prior beliefs regarding the characteristics of given geothermal system. A particularly flexible class of models we use to describe a range of spatially-varying phenomena are Gaussian random fields (GRFs), which we review in Section \ref{sec:grfs}. Section \ref{sec:level_sets} describes how we use can pair a GRF with the level set method to parametrise layered rock formations with variable interfaces, and Section \ref{sec:constraints} discusses how we can use variable transformations to impose general geophysical constraints as part of the prior.



\subsubsection{Gaussian Random Fields} \label{sec:grfs}

A Gaussian random field (GRF) on $\mathbb{R}^{\GRFDim}$ is a collection of random variables, indexed by $\mathbb{R}^{\GRFDim}$, with the property that the joint distribution of any finite subset is Gaussian. A one-dimensional GRF is often referred to as a Gaussian process \citep{Rasmussen06}. In the same way a multivariate Gaussian distribution is completely characterised by its mean, $\bm{m}$, and covariance matrix, $\bm{C}$, a GRF is completely characterised by its mean function, $m(\bm{x})$, and covariance function, $\mathcal{C}(\bm{x}, \bm{x})$.

The covariance function encodes important properties of the GRF, including the degree of smoothness and the correlation length of realisations of the field \citep{Rasmussen06}, and should be selected carefully such that the resulting GRF provides an accurate representation of one's prior knowledge. A flexible choice of covariance function that has been used to model a diverse range of phenomena is the (anisotropic) Whittle-Mat{\'e}rn covariance function \citep{Lindgren11, Roininen14}, given by
\begin{equation}
	\mathcal{C}_{\mathrm{WM}}(\bm{x}, \bm{x}') = \Std^{2}\frac{\bignorm{\bm{x}-\bm{x}'}^{\nu}_{\bm{L}}}{2^{\Reg-1}\Gamma(\Reg)}K_{\Reg}\left(\bignorm{\bm{x}-\bm{x}'}_{\bm{L}}\right), \label{eq:wm_cov}
\end{equation}
where $K_{\Reg}(\cdot)$ denotes the modified Bessel function of the second kind of order $\Reg$, $\Gamma(\cdot)$ denotes the Gamma function, $\Std$ and $\Reg$ are standard deviation and regularity parameters respectively, and $\bm{L} \in \mathbb{R}^{\GRFDim \times \GRFDim}$ is a diagonal matrix with elements equal to the squares of the characteristic lengthscales along each coordinate axis. In two dimensions, $\bm{L} \defas \diag(\Ls_{1}^{2}, \Ls_{2}^{2})$, and in three dimensions, $\bm{L} \defas \diag(\Ls_{1}^{2}, \Ls_{2}^{2}, \Ls_{3}^{2})$. We note that it is possible to align these lengthscales in directions other than the coordinate axes through an appropriate transformation of $\bm{L}$ \citep[see, e.g.,][]{Roininen14}; however, we do not experiment with this here.

In practice, there is often significant uncertainty regarding appropriate values for the standard deviations and lengthscales of these fields, and the incorrect specification of these hyperparameters can result in poor reconstructions of the truth \citep{Chada18, Chen19}. For this reason, we typically place hyperpriors on these parameters; we elaborate on this process below.

A variety of methods for sampling from a Whittle-Mat{\'e}rn field exist, including computing a (truncated) Karhunen-Lo{\`e}ve expansion of the field, or the Cholesky decomposition of the covariance matrix \citep{Chen19}. An alternative approach, which we use throughout this work, is to sample from the field by solving a stochastic PDE \citep{Lindgren11, Roininen14}. The benefit of this approach is that it provides a computationally efficient way to generate realisations of the field as the hyperparameters and, therefore, the covariance function of the field change; rather than computing a factorisation of the updated covariance matrix, sampling from the field amounts to forming and solving a sparse linear system, which can be carried out very quickly. In particular, it can be shown \citep{Lindgren11} that the solution, $u(\bm{x})$, to the stochastic PDE given by
\begin{equation}
    (I - \nabla \cdot \bm{L} \nabla)^{\frac{\Reg+\GRFDim/2}{2}}(u(\bm{x}) - \Mean(\bm{x})) = \alpha^{\frac{1}{2}}\prod_{i=1}^{\GRFDim}\Ls_{i}^{\frac{1}{2}} \xi(\bm{x}), \label{eq:wm_spde_full}
\end{equation}
is a draw from a Whittle-Mat{\'e}rn field with mean function $m(\bm{x})$ and covariance function given by \eqref{eq:wm_cov}. In \eqref{eq:wm_spde_full}, $\xi(\bm{x})$ denotes Gaussian white noise, $I$ denotes the identity operator, $\nabla$ denotes the gradient operator, and the coefficient $\alpha$ is defined as
\begin{equation}
    \alpha \defas \Std^{2}\frac{2^{\GRFDim}\pi^{\GRFDim/2}\Gamma(\Reg+\GRFDim/2)}{\Gamma(\Reg)}.
\end{equation}
Generally, we set the regularity parameter to $\nu = 2 - \GRFDim/2$, which allows for a straightforward discretisation of \eqref{eq:wm_spde_full} using finite differences or finite elements \citep{Lindgren11, Roininen14}. Solving \eqref{eq:wm_spde_full} on a bounded domain requires the specification of suitable boundary conditions, which affect the covariance structure of the field close to the boundary. We follow the approach of \citet{Roininen14}; that is, we specify a set of Robin boundary conditions, 
\begin{equation}
    u(\bm{x}) + \lambda \frac{\partial u(\bm{x})}{\partial \bm{n}} = 0, \quad \bm{x} \in \Boundary,
\end{equation}
and tune the coefficient $\lambda$ to minimise the effect of the boundary conditions on the covariance structure. We note, however, that there exist other approaches to this issue, including generating realisations of the field on a domain that is larger than the physical domain of interest \citep{Roininen14}. In summary, generating a realisation from the (discretised) GRF amounts to sampling a set of white noise and a set of values for the standard deviation and lengthscales of the field, then solving Equation \eqref{eq:wm_spde_full}.

\subsubsection{The Level Set Method} \label{sec:level_sets}

A geothermal reservoir is typically composed of distinct layers of rock with sharp interfaces. There is often significant uncertainty, however, in the properties of each layer, as well as the location of the interfaces between the layers. The level set method \citep{Iglesias16} provides a way of parametrising unknowns of this nature \citep{Nicholson20}, and has been used successfully in combination with ensemble methods in a range of geophysical applications \citep{Muir20, Muir22, Tso21, Tso24}. Distinct formations with variable interfaces are defined using the contours of a continuous auxiliary function, $\LevelFunc(\bm{x})$, referred to as the \emph{level set function}. 

When using the level set method, we select a number of possible rocktypes, $\NumFormations$, with associated physical parameters $\{\kappa_{i}\}_{i=1}^{\NumFormations} \in \mathbb{R}$, and choose constants $\{c_{i}\}_{i=0}^{\NumFormations} \in \bar{\mathbb{R}}$ such that $-\infty = c_{0} < c_{1} < \dots < c_{\NumFormations} = \infty$. Then, the region of the domain occupied by rocktype $\Domain_{i}$ is given by
\begin{equation}
    \Domain_{i} = \{\bm{x} \in \Domain \,|\, c_{i-1} \leq \varphi(\bm{x}) < c_{i}\}, \label{eq:subdomains}
\end{equation}
and the physical property at location $\bm{x} \in \Domain$ is given by
\begin{equation}
    \kappa(\bm{x}) = \sum_{i=1}^{\NumFormations}\kappa_{i}\mathbb{I}_{\Domain_{i}}(\bm{x}), \label{eq:level_map}
\end{equation}
where $\mathbb{I}_{\Domain_{i}}(\bm{x})$ denotes the indicator function of set $\Domain_{i}$. We note that it is possible to treat the values of the physical parameters, $\{\kappa_{i}\}_{i=1}^{\NumFormations}$, as uncertain, or even replace them with random fields \citep{Iglesias22, Matveev21}. However, we do not experiment with these ideas in this work.

\begin{figure}
    \centering
    \includegraphics[width=0.5\linewidth]{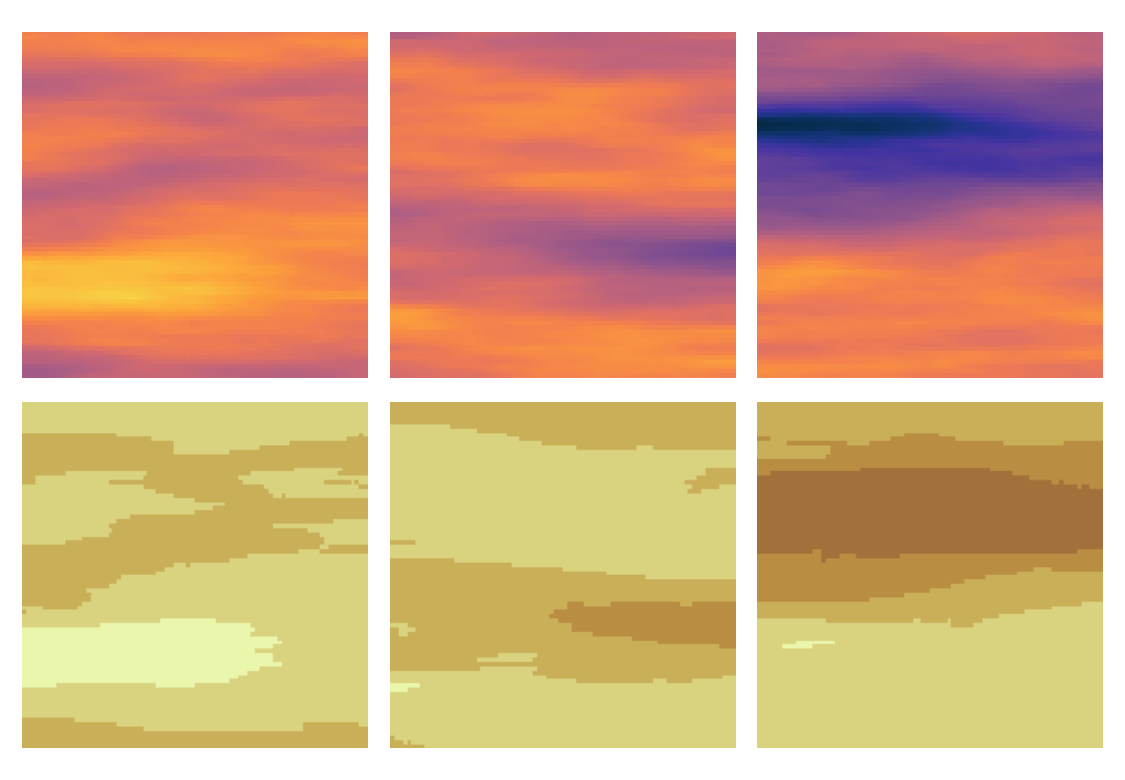}
    \caption{Draws of a level set function, $\LevelFunc(\bm{x})$ \emph{(top row)}, and the corresponding geometric fields obtained by applying a mapping of the form \eqref{eq:level_map} \emph{(bottom row)}. The level set function is a GRF, defined on $(0, 1000)^{2}$, with a Whittle-Mat{\'e}rn covariance function with $\Reg=3/2$, $\Std=1.0$, $\Ls_{1}=1250$ and $\Ls_{2}=200$. The thresholding constants used as part of the level set mapping are $\{-\infty, -3/2, -1/2, 1/2, 3/2, \infty\}$.}
    \label{fig:level_sets}
\end{figure}

The choice of level set function is crucial; this (in combination with the choice of coefficients $\{c_{i}\}_{i=0}^{\NumFormations}$) determines the characteristics of the formations obtained after applying the level set mapping \eqref{eq:level_map}. A common choice of level set function is a Gaussian random field \citep[see, e.g.,][]{Iglesias22, Muir20, Tso21, Tso24}. Figure \ref{fig:level_sets} shows examples of draws of a GRF with a Whittle-Mat{\'e}rn covariance function, and the corresponding physical fields obtained after applying a mapping of the form \eqref{eq:level_map}. Despite the thresholding of the level set function to produce the corresponding physical field, the roles of the hyperparameters remain similar; increasing the lengthscale increases the size of the resulting formations, while increasing the standard deviation increases the probability of observing rocktypes corresponding to more extreme values of the function. We note that, if a GRF is used as the level set function, it is possible to condition it to hard data (e.g., from borehole logs) on the rock properties at a given set of locations using Gaussian process regression, or \emph{kriging} \citep[see, e.g.,][]{Iglesias24}.

We remark that the continuity of the level set function imposes a fixed ordering to the rocktypes (that is, rocktype $i$ can only border rocktypes $i-1$ and $i+1$), and prevents the existence of ``junctions'', at which three or more rocktypes intersect \citep{Iglesias16}. If we do not possess the prior knowledge to justify these restrictions, the use of a vector-valued level set function \citep[see, e.g.,][]{Tso24} allows us to generate fields with arbitrary arrangements of the regions. This comes, however, at the cost of increasing the dimensionality of the inverse problem.

\subsubsection{Constraints and Geological Realism} \label{sec:constraints}

There are a variety of constraints we typically wish to enforce when solving geothermal inverse problems. For instance, we often require that parameters, such as the reservoir permeability structure or the magnitude of the mass upflows at the base of the system, are positive or bounded within an appropriate range \citep{Bjarkason21b, Nicholson20}. Additionally, geological constraints may dictate the relative values of model parameters \citep{deBeer23a}; for instance, the permeability along the strike of a fault should be at least as great as that of the surrounding formation, while the permeability across the strike of a fault should be at most as great as that of the surrounding formation.

To work with a parameter with an arbitrary target distribution, we first define an unconstrained auxillary parameter, $\ParamAux$, distributed according to the unit normal distribution. To transform samples of $\ParamAux$ to samples from the target distribution, we apply a mapping of the form
\begin{equation}
    \ParamAux \mapsto (\CDF_{t}^{-1}\circ\CDF_{n})(\ParamAux) \label{eq:inv_cdf}
\end{equation}
where $\CDF_{t}$ denotes the cumulative distribution function (CDF) of the target distribution, and $\CDF_{n}$ denotes the CDF of the unit normal distribution. A common target distribution we consider is the uniform distribution, $\mathcal{U}(\Param_{0}, \Param_{1})$; in this case, Equation \eqref{eq:inv_cdf} becomes
\begin{equation}
    \ParamAux \mapsto \Param_{0} + (\Param_{1} - \Param_{0})\CDF_{n}(\ParamAux).
\end{equation}
In some cases, the target distribution of a given parameter may be defined conditionally on the value of another parameter; we illustrate this idea when parametrising the permeability structure of the reservoir model introduced in Section \ref{sec:setup_vol}.

\section{Model Problems} \label{sec:problems}

We now introduce three model problems we use to demonstrate the application of the methodology outlined in Section \ref{sec:methods}. In particular, we consider a synthetic vertical slice model, a synthetic three dimensional model, and a real-world volcanic reservoir model. The synthetic problems allow us to demonstrate the use of the parametrisation techniques outlined in Section \ref{sec:parametrisations}, and to evaluate the ability of the EKI algorithm to recover the true values of the model parameters. The final problem allows us to illustrate some of the challenges involved in modelling large-scale, real-world geothermal systems. We note that the synthetic problems are adapted from those presented in the Master's thesis \citep{deBeer24a}. Additionally, preliminary results for the first synthetic problem were presented at the 2024 New Zealand Geothermal Workshop \citep{deBeer23b}.

We simulate each model using the parallel, open-source simulator Waiwera \citep{Croucher20}, which uses a finite volume discretisation of the governing equations introduced in Section \ref{sec:forward_prob}. Timestepping is carried out using the backward Euler method.

In all of our experiments, at each EKI iteration we simulate the entire ensemble (which consists of 100-200 particles) in parallel on a high-performance computing platform provided by New Zealand eScience Infrastructure (NeSI). This accelerates the EKI algorithm significantly.

\subsection{Vertical Slice Model}

The first test case we consider is a vertical ``slice'' model, with domain $(0\,\metres, 1500\,\metres) \times (0\,\metres, 60\,\metres) \times (-1500\,\metres, 0\,\metres)$. When simulating the reservoir dynamics using Waiwera, we consider a single mass component---water---and an energy component.

\begin{figure}
    \centering
    \includegraphics[width=0.5\linewidth]{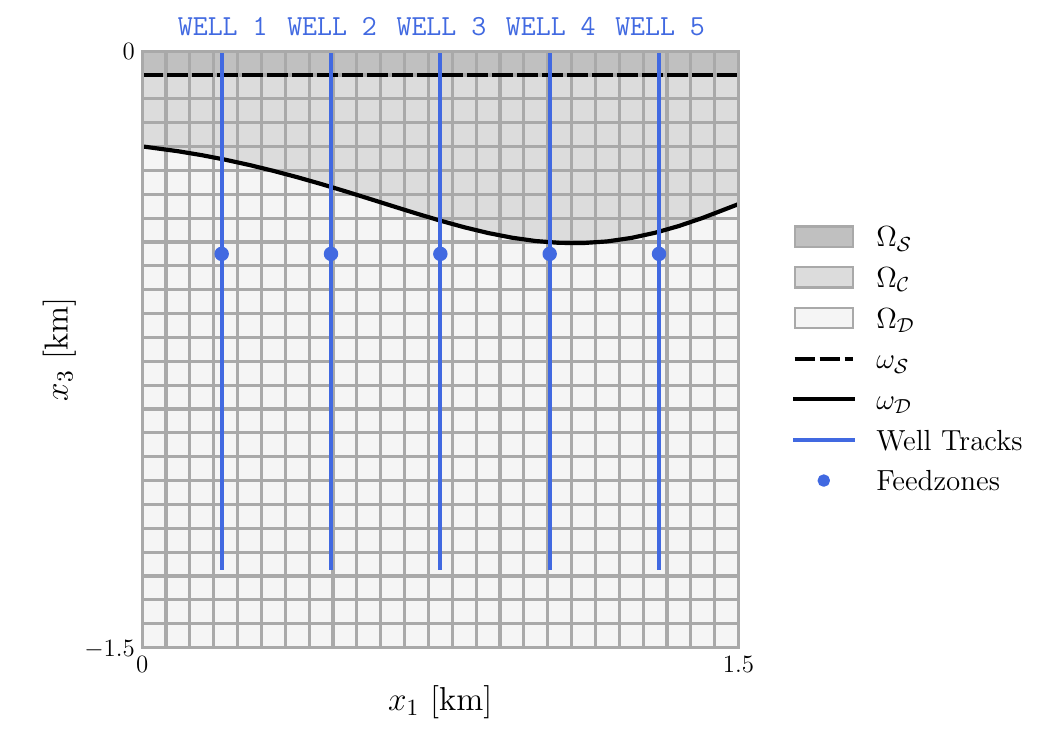}
    \caption{The mesh of the vertical slice model. Also indicated are the well tracks and feedzones, as well as a possible partition of the domain into regions $\Domain_{\Shal}$, $\Domain_{\ClayCap}$, and $\Domain_{\Deep}$.}
    \label{fig:mesh_slice}
\end{figure}

The model contains five production wells, which are shown in Figure~\ref{fig:mesh_slice}. Each well contains a single feedzone at a depth of $500\,\metres$. We consider a combined natural state and production history simulation \citep{OSullivan16}; that is, we simulate the dynamics of the system until steady state conditions are reached, then use the resulting state of the system as the initial condition for the subsequent production simulation. During the production period, which lasts for two years, each well extracts fluid at a rate of $2\,\mathrm{kg}\,\mathrm{s}^{-1}$. 

\begin{figure}
    \centering
    \includegraphics[width=0.3\linewidth]{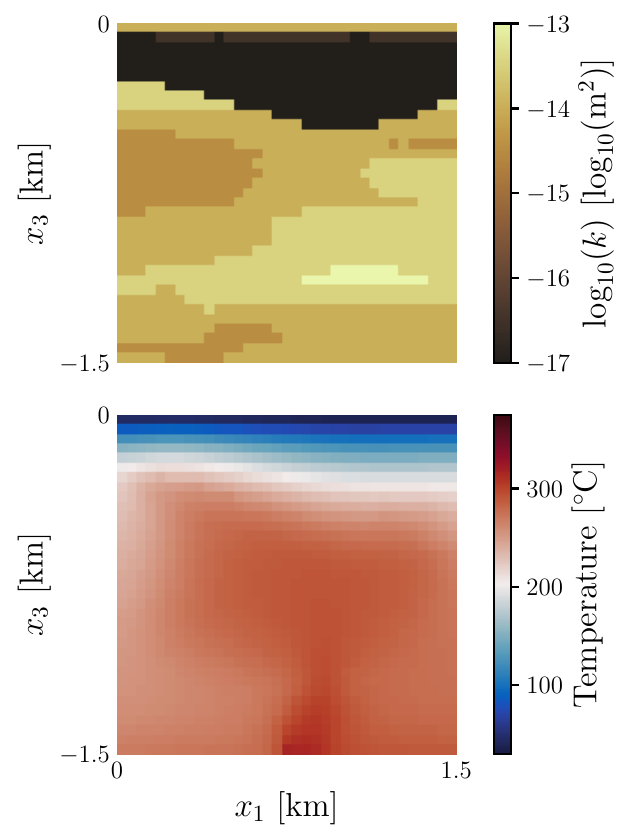}
    \caption{The true permeability structure (\emph{top}) and natural state convective plume (\emph{bottom}) of the vertical slice model. The mass upflow through the bottom boundary of the true system has a rate of $0.149\,\mathrm{kg}\,\mathrm{s}^{-1}$.}
    \label{fig:truth_slice}
\end{figure}

Figure \ref{fig:truth_slice} shows the permeability structure and natural state convective plume of the true system, generated using a draw from the prior (outlined in Section \ref{sec:prior_slice}). To avoid the ``inverse crime'' of using the same model to generate the data and to carry out the inversion \citep{Kaipio06, Kaipio07}, the true system is discretised on a $35\times1\times35$ mesh, while the inversion is carried out using a $25\times1\times25$ mesh. 

\subsubsection{Data}

\begin{figure*}
    \centering
    \includegraphics[width=0.75\textwidth]{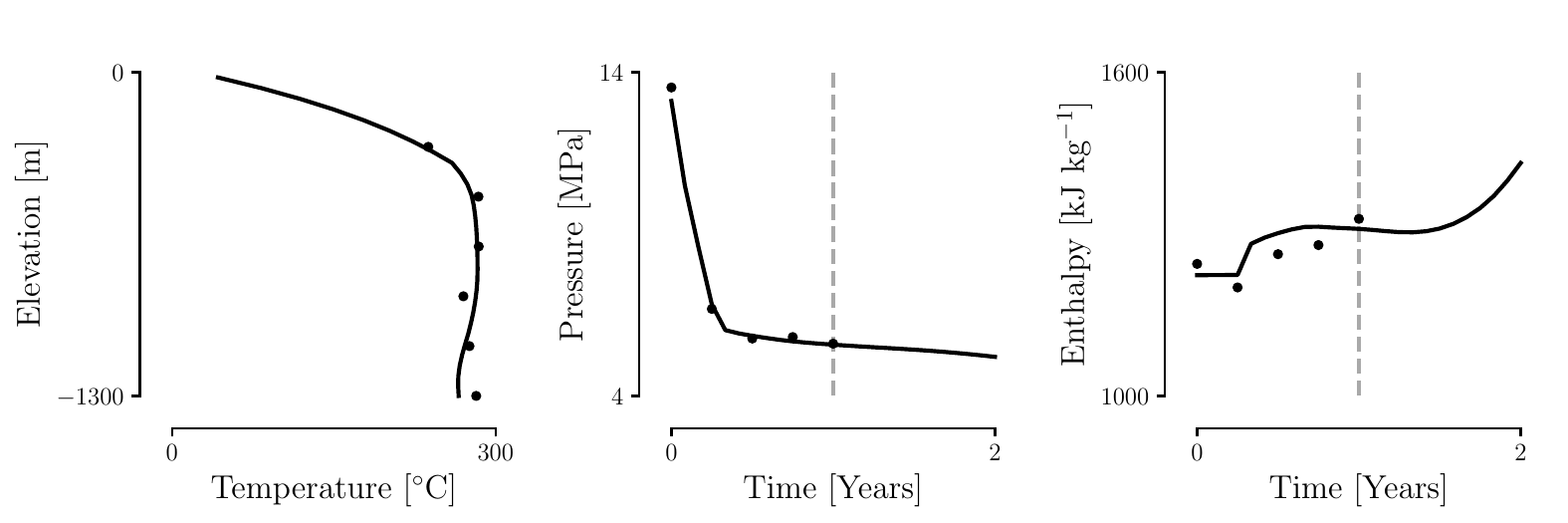}
    \caption{Natural state temperature data (\emph{left}), feedzone pressure data (\emph{centre}), and feedzone enthalpy data (\emph{right}) at well 2 of the vertical slice model. The solid line in each plot denotes the true model state, and the dots denote the noisy observations. The dashed grey line in the pressure and enthalpy plots denotes the end of the data collection period.}
    \label{fig:data_slice}
\end{figure*}

The data we use to carry out the inversion is composed of the temperature recorded at six equispaced points down each well prior to the start of production, and the pressure and enthalpy of the fluid extracted at each well at three-month intervals over the first year of production. This gives a total of $80$ measurements. We add independent Gaussian noise to each observation with a standard deviation of $2$ percent of the maximum modelled value of the corresponding data type. Figure \ref{fig:data_slice} shows the observations recorded at well 2.

\subsubsection{Prior Parametrisation} \label{sec:prior_slice}

We consider the problem of estimating the permeability structure of the system and the mass rate of the upflow at the base of the reservoir. All other reservoir properties are assumed to be known. The rock of the reservoir is assumed to have a porosity of $0.1$, a density of $2500\,\kg\,\metres^{-3}$, a thermal conductivity of $2.5 \,\mathrm{W}\,\metres^{-1}\,\mathrm{K}^{-1}$, and a specific heat of $1000\,\mathrm{J}\,\kg^{-1}\,\mathrm{K}^{-1}$. The top boundary of the model is set to a constant pressure of $1\,\mathrm{bar}$ and a temperature of $20^{\circ}\mathrm{C}$, representing an atmospheric boundary condition. We impose a background heat flux (representing conductive heat flow) of $200\,\mathrm{mW}\,\metres^{-2}$ through the bottom boundary, except for the cell at the centre of the boundary, which is given a mass flux (of unknown rate) of fluid with an enthalpy of $1500\,\mathrm{kJ}\,\kg^{-1}$. The side boundaries are closed.

To parametrise the permeability of the model, we first partition the model domain, $\Domain$, into three subdomains with variable interfaces: a shallow high-permeability region ($\Domain_{\mathcal{S}}$), a low-permeability clay cap ($\Domain_{\mathcal{C}}$), and a deep high-permeability region ($\Domain_{\mathcal{D}}$). The permeability at an arbitrary $\bm{x} = (x_{1}, x_{2}, x_{3}) \in \Domain$ is given by
\begin{equation}
    \Perm(\bm{x}) = \begin{cases}
        \PermField_{\Shal}(\bm{x}), & \quad \PermBoundary_{\Shal}(x_{1}) < x_{3}, \\
        \PermField_{\ClayCap}(\bm{x}), & \quad \PermBoundary_{\Deep}(x_{1}) < x_{3} \leq \PermBoundary_{\Shal}(x_{1}), \\
        \PermField_{\Deep}(\bm{x}), & \quad x_{3} \leq \PermBoundary_{\Deep}(x_{1}).
    \end{cases} \label{eq:partition_slice}
\end{equation}
In \eqref{eq:partition_slice}, the functions $\PermField_{i} : \Domain \rightarrow \mathbb{R}_{+}$, $i \in \{\Shal, \ClayCap, \Deep\}$, define the permeability in each region, and the functions $\PermBoundary_{\Shal} : (0, 1500) \rightarrow \mathbb{R}$ and $\PermBoundary_{\Deep} : (0, 1500) \rightarrow \mathbb{R}$ define the interfaces between $\Domain_{\Shal}$ and $\Domain_{\ClayCap}$, and $\Domain_{\ClayCap}$ and $\Domain_{\Deep}$ (the top and bottom surfaces of the clay cap) respectively. Figure \ref{fig:mesh_slice} shows a possible partition of the model domain.

We set $\PermBoundary_{\Shal}(x_{1}) = -60\,\metres$, which reflects a prior assumption that the location of the top surface of the clay cap is known. However, we treat the location of the bottom surface of the clay cap as unknown, giving 
$\PermBoundary_{\Deep}$ a Gaussian process prior with a mean of $-350\,\metres$, and a squared-exponential covariance function \citep[see, e.g.,][]{Rasmussen06} with a standard deviation, $\Std$, of $80\,\metres$, and a lengthscale, $\Ls$, of $500\,\metres$.

For each permeability field, we select a number of possible rock types, each with associated permeabilities. We then choose a set of constants to threshold the level set function at to produce each rock type; for instance, the value of permeability field $\PermField_{\ClayCap}$ at an arbitrary $\bm{x} \in \Domain$ is given by
\begin{equation}
    \PermField_{\ClayCap}(\bm{x}) = \begin{cases}
        10^{-17.0}\,\metres^{2}, & \quad \phantom{+}\LevelFunc_{\ClayCap}(\bm{x}) < -0.5, \\
        10^{-16.5}\,\metres^{2}, & \quad -0.5 \leq \LevelFunc_{\ClayCap}(\bm{x}) < 0.5, \\
        10^{-16.0}\,\metres^{2}, & \quad \phantom{+}0.5 \leq \LevelFunc_{\ClayCap}(\bm{x}).
    \end{cases} \label{eq:levels_deep}
\end{equation}
The permeabilities of fields $\PermField_{\Shal}$ and $\PermField_{\Deep}$ are defined similarly, such that they vary between $10^{-15}\,\metres^{2}$ and $10^{-13}\,\metres^{2}$. For completeness, we list the definitions of these in Appendix \ref{sec:level_set_slice}.

\begin{figure}
    \centering
    \includegraphics[width=0.5\linewidth]{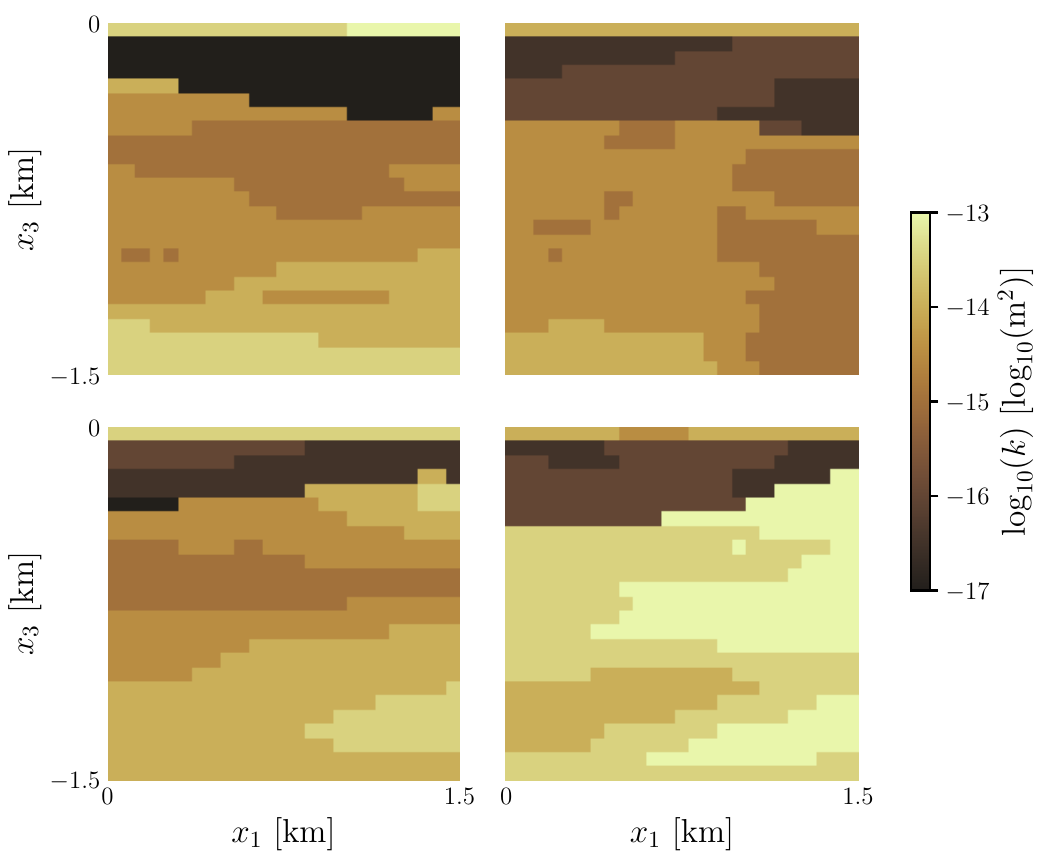}
    \caption{The permeability structures of particles drawn from the prior of the vertical slice model.}
    \label{fig:particles_pri_2d}
\end{figure}

We use a centred Whittle-Mat{\'e}rn field with an anisotropic lengthscale as the underlying level set function for each permeability field (note that because the discretisation of the model domain is refined only in the $x_{1}$ and $x_{3}$ dimensions, each field is two-dimensional). The lengthscale in the horizontal ($x_{1}$) direction of all three fields is uniformly distributed between $1000\,\metres$ and $2000\,\metres$, and the lengthscale in the vertical ($x_{3}$) direction is uniformly distributed between $200\,\metres$ and $500\,\metres$. In regions $\Domain_{\Shal}$ and $\Domain_{\ClayCap}$, the standard deviation is uniformly distributed between $0.5$ and $1.0$, and in region $\Domain_{\Deep}$ it is uniformly distributed between $0.75$ and $1.25$. Figure \ref{fig:particles_pri_2d} shows a set of permeability structures corresponding to particles drawn from the prior, after applying the level set transformation to each field and partitioning the domain into regions $\Domain_{\Shal}$, $\Domain_{\ClayCap}$ and $\Domain_{\Deep}$.

We impose a uniform prior on the rate of the mass upflow (which we denote using $Q^{1}_{\mathrm{up}}$) with bounds of $0.1\,\mathrm{kg}\,\mathrm{s}^{-1}$ and $0.2\,\mathrm{kg}\,\mathrm{s}^{-1}$.

Table \ref{tab:prior_2d} provides a summary of the prior parametrisation.

\begin{table*}
    \centering
    \caption{The prior parametrisation for the vertical slice model.}
    \scriptsize
    \begin{adjustbox}{center}
    \begin{tabular}{ p{30mm} p{50mm} p{25mm} p{58mm} }
        \toprule
        \textbf{Model Component} & \textbf{Parameter Description} & \textbf{Symbol} & \textbf{Prior} \\
        \midrule 
        Geometry & Elevation of boundary between clay cap and deep region ($\metres$) & $\PermBoundary_{\Deep}$ & $\mathcal{N}(-350, \mathcal{C}_{\mathrm{SE}}[\Std=80, \Ls=500])$ \\
        \midrule
        Permeability & Level set function in each subdomain & $\LevelFunc_{i}, i \in \{\Shal, \ClayCap, \Deep\}$ & $\mathcal{N}(0, \mathcal{C}_{\mathrm{WM}}[\Std=\Std_{\LevelFunc_{i}}, \bm{L}=\bm{L}_{\LevelFunc_{i}}])$ \\
        & Standard deviation of level set function in subdomains $\Shal, \ClayCap$ & $\Std_{\LevelFunc_{i}}, i \in \{\Shal, \ClayCap\}$ & $\mathcal{U}(0.5, 1.0)$ \\
        & Standard deviation of level set function in subdomain $\Deep$ & $\Std_{\LevelFunc_{\Deep}}$ & $\mathcal{U}(0.75, 1.25)$ \\
        & Horizontal lengthscale of level set function in each subdomain ($\metres$) & $\Ls_{1,\LevelFunc_{i}}, i \in \{\Shal, \ClayCap, \Deep\}$ & $\mathcal{U}(1000, 2000)$ \\
        & Vertical lengthscale of level set function in each subdomain ($\metres$) & $\Ls_{3,\LevelFunc_{i}}, i \in \{\Shal, \ClayCap, \Deep\}$ & $\mathcal{U}(200, 500)$ \\
        \midrule
        Mass upflow & Magnitude of mass upflow through central cell at base of domain ($\kg\,\mathrm{s}^{-1}$) & $Q^{1}_{\mathrm{up}}$ & $\mathcal{U}(0.1, 0.2)$ \\
        \bottomrule
    \end{tabular}
    \end{adjustbox}
    \label{tab:prior_2d}
\end{table*}

\subsection{Synthetic Three-Dimensional Model}

The second case study we consider is a three-dimensional model with a vertical fault running through the centre of the reservoir, which acts as a pathway for hot mass upflow from the roots of the system. As in the vertical slice case study, when simulating the reservoir dynamics, we consider a single mass component---water---and an energy component.

\begin{figure*}
    \centering
    \includegraphics[width=0.70\textwidth]{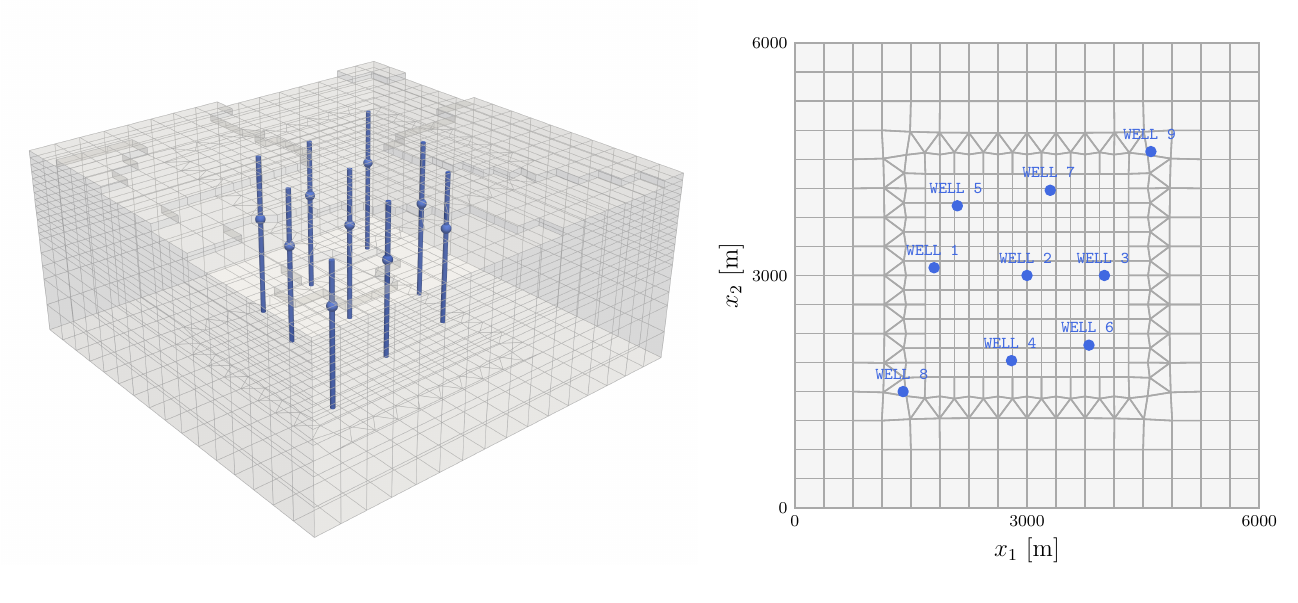}
    \caption{The mesh of the synthetic three-dimensional model. The plot on the right shows a top-down view of the plot on the left. In both plots, the wells are indicated using blue.}
    \label{fig:mesh_fault}
\end{figure*}

\begin{figure*}
    \centering
    \includegraphics[width=0.75\textwidth]{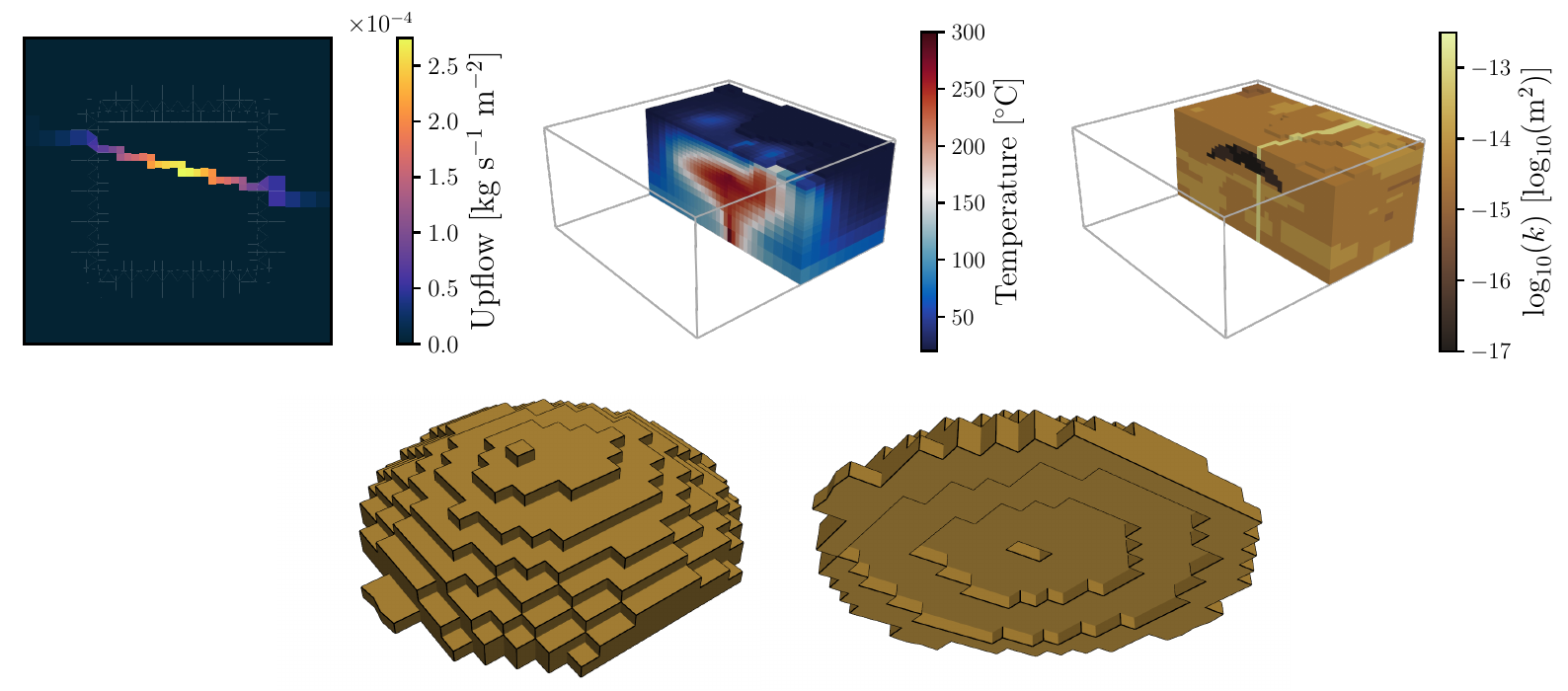}
    \caption{The true upflow (\emph{top left}), natural state convective plume (\emph{top centre}), permeability structure (\emph{top right}), and the top and bottom surfaces of the clay cap (\emph{bottom}), of the synthetic three-dimensional model.}
    \label{fig:truth_fault}
\end{figure*}

The model domain, shown in Figure \ref{fig:mesh_fault}, spans $6000\,\metres$ in the horizontal ($x_{1}$ and $x_{2}$) directions, extends to a depth of $3000\,\metres$ in the vertical ($x_{3}$) direction, and has a variable topography. The system contains nine production wells, the locations of which are also indicated in Figure \ref{fig:mesh_fault}. Each has a single feedzone at a depth of $1200\,\metres$. Figure \ref{fig:truth_fault} shows the mass upflow, permeability structure and clay cap of the true system, generated using a draw from the prior. Also shown is the true natural state convective plume.

\subsubsection{Data}

\begin{figure*}
    \centering
    \includegraphics[width=0.75\textwidth]{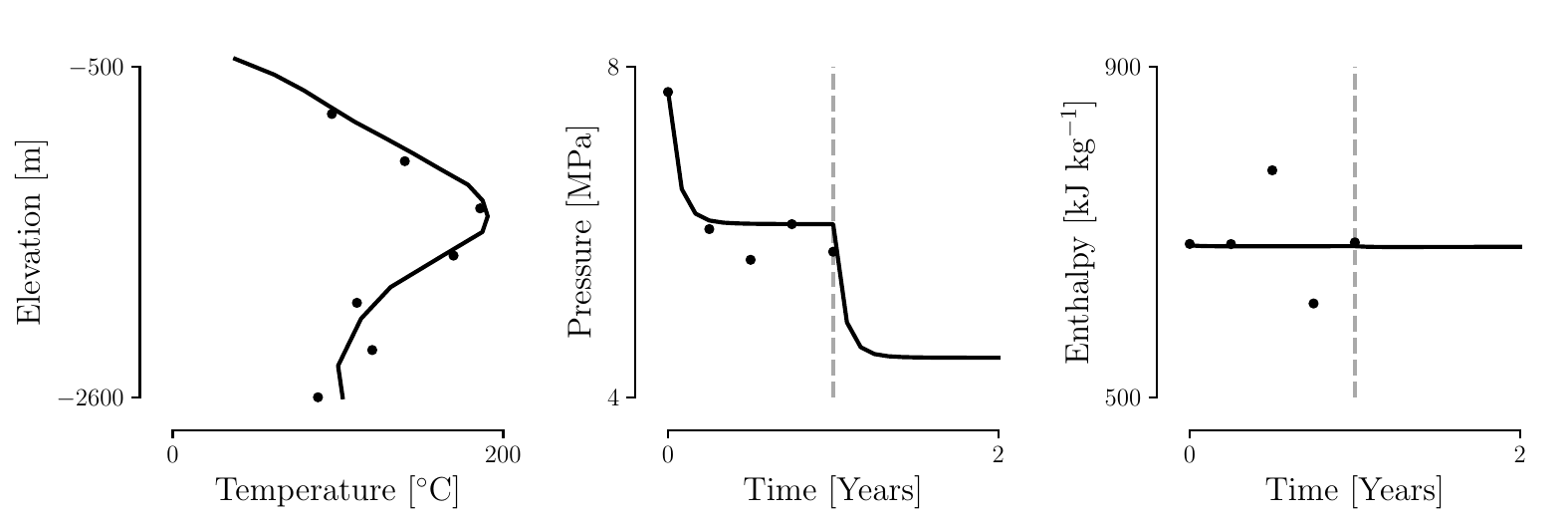}
    \caption{Natural state temperatures (\emph{left}), feedzone pressures (\emph{centre}) and feedzone enthalpies (\emph{right}) at well 4 of the synthetic three-dimensional model. The solid line in each plot denotes the true model state, and the dots denote the noisy observations. The dashed grey line in the pressure and enthalpy plots indicates the end of the data collection period.}
    \label{fig:data_fault}
\end{figure*}

As in the vertical slice problem, we consider a combined natural state and production history simulation. The production period lasts for two years. During the first year, each well extracts fluid at a rate of $0.25\,\mathrm{kg}\,\mathrm{s}^{-1}$. During the second year, this is increased to $0.5\,\mathrm{kg}\,\mathrm{s}^{-1}$. The data we generate is of the same form as in the vertical slice problem. However, we increase the standard deviation of the Gaussian noise corrupting each measurement to $5$ percent of the maximum modelled value of the corresponding data type. Figure \ref{fig:data_fault} shows the data recorded at well 4. To avoid an inverse crime, the data is generated using a mesh containing $13,383$ cells, while the inversion is carried out using a coarsened mesh containing $8,788$ cells.

\subsubsection{Prior Parametrisation}

We consider the problem of estimating the permeability structure of the system, as well as the magnitude and location of the mass flux at the base of the system. All other rock properties and boundary conditions are assumed known, and take the same values as in the vertical slice case.

\begin{figure*}
    \centering
    \includegraphics[width=0.8\linewidth]{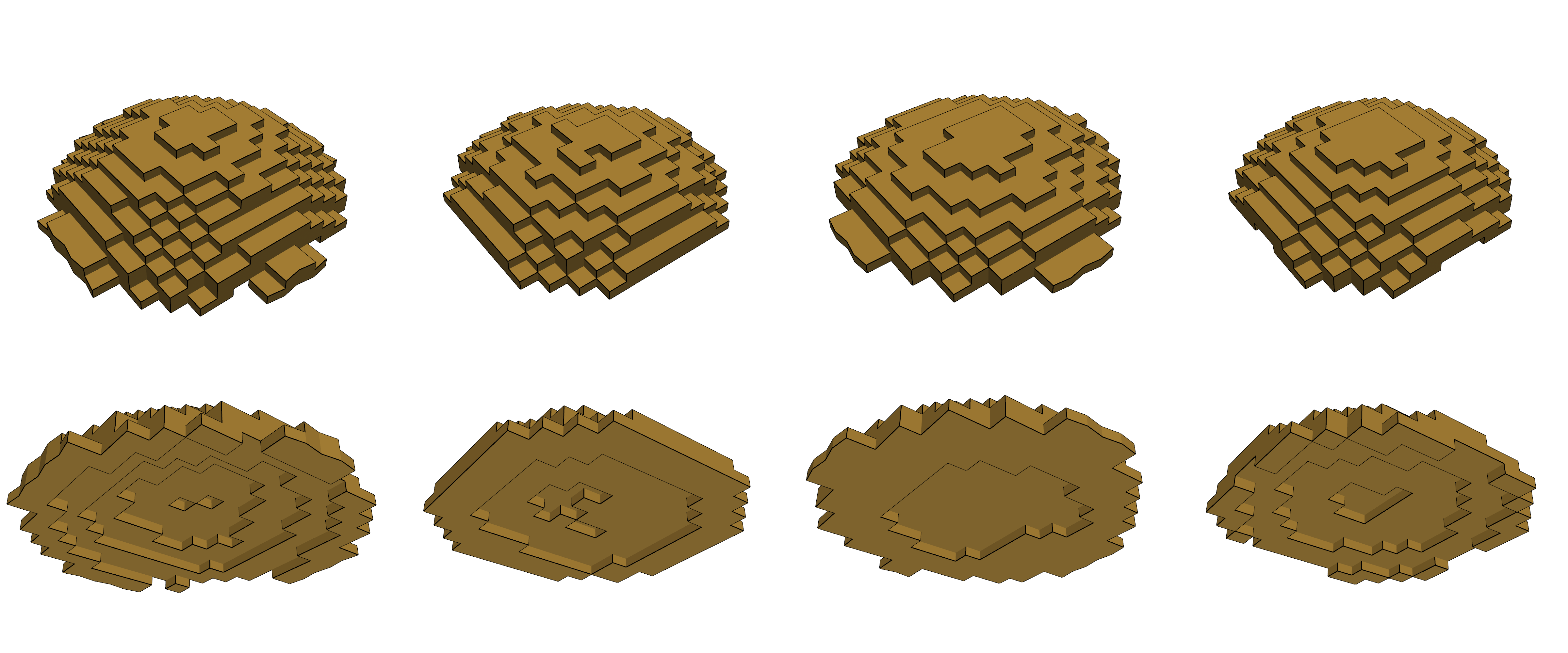}
    \caption[Clay cap geometries drawn from the prior of the synthetic three-dimensional model.]{The top surfaces (\emph{top row}) and bottom surfaces (\emph{bottom row}) of clay cap geometries generated using draws from the prior.}
    \label{fig:caps_fault_pri}
\end{figure*}

To parametrise the permeability of the model, we partition the domain into three subdomains: a low-permeability clay cap ($\Domain_{\ClayCap}$), a high-permeability fault ($\Domain_{\Fault}$) and a background region of moderate permeability ($\Domain_{\Background})$. We model $\Domain_{\ClayCap}$ as the deformation of a star-shaped set, which we denote using $\mathcal{S}$, with central point $\bm{s} = (s_{1}, s_{2}, s_{3}) \in \Domain$. We set $s_{1}=s_{2}=3000\,\metres$, reflecting a prior belief that the clay cap is approximately centred within the horizontal bounds of the model domain. However, we introduce variability into the depth of the clay cap by modelling $s_{3}$ as uniformly distributed; that is, $s_{3} \sim \mathcal{U}(-900\,\metres, -775\,\metres)$. Because $\mathcal{S}$ is star-shaped, the line segment connecting $\bm{s}$ and any point on the boundary of $\mathcal{S}$ is contained within $\mathcal{S}$; therefore, we can define $\mathcal{S}$ by working in spherical coordinates and specifying the distance, $r(\theta, \varphi)$, from $\bm{s}$ to the boundary of $\mathcal{S}$ in the direction of each angle $(\theta, \varphi) \in [0, 2\pi)^{2}$. To introduce uncertainty into the shape of the boundary of $\mathcal{S}$, we represent $r(\theta, \varphi)$ as the sum of two quantities: the distance from $\bm{s}$, in direction $(\theta, \varphi)$, to the ellipsoid (also centred at $\bm{s}$) with principal axes in the $x_{1}$ and $x_{2}$ directions of length $2w$ and a principal axis in the $x_{3}$ direction of length $2h$, and a truncated Fourier series composed of 20 basis functions with random coefficients $\{f_{i}\}_{i=1}^{20}$. Parameter $w$ of the ellipsoid is modelled as being uniformly distributed on $[1400\,\metres, 1600\,\metres]$, and parameter $h$ is modelled as being uniformly distributed on $[200\,\metres, 300\,\metres]$. The coefficients of the Fourier series are modelled as independent and normally distributed; $f_{i} \sim \mathcal{N}(0\,\metres, 25\,\metres^{2})$. Finally, we define $\Domain_{\mathcal{C}}$ as
\begin{equation}
    \Domain_{\ClayCap} \defas \Bigg\{\bm{x} \in \Domain \,\,\Big|\,\, \left(x_{1}, x_{2}, x_{3}+d\frac{x_{1}^{2}+x_{2}^{2}}{w^{2}}\right) \in \mathcal{S} \Bigg\}.
\end{equation}
Increasing parameter $d$ has the effect of shifting the outer edges of $\Domain_{\ClayCap}$ downward. We model $d$ as being uniformly distributed on $[300\,\metres,\, 600\,\metres]$. When working with the discretisation of the model domain, we treat all cells of the model mesh with centres located within $\Domain_{\ClayCap}$ as part of the clay cap. Figure \ref{fig:caps_fault_pri} shows a set of (discretised) clay cap geometries generated using draws from the prior.

We model $\Domain_{\Fault}$ as the intersection between a vertical plane and the model domain, excluding points located in the clay cap; that is,
\begin{equation}
    \Domain_{\Fault} \defas \Bigg\{\bm{x} \in \Domain \,\,\Big|\,\, x_{2} = b_{w} + \frac{b_{e}-b_{w}}{6000}x_{1}, \bm{x} \notin \Domain_{\ClayCap}\Bigg\}. \label{eq:dom_fault}
\end{equation}
The background region is then defined as $\Domain_{\Background} \defas \Domain \setminus (\Domain_{\ClayCap} \cup \Domain_{\Fault})$. In \eqref{eq:dom_fault}, the parameters $b_{e}$ and $b_{w}$ define the locations (in the $x_{2}$ direction) at which the fault intersects the eastern and western boundaries of $\Domain$. We model these as being uniformly distributed on $[1500\,\metres,\, 4500\,\metres]^{2}$. When working with the discretisation of the model domain, we treat all cells of the model mesh that intersect with $\Domain_{\Fault}$ as part of the fault (with the exception of those that are part of the clay cap). The region of the bottom boundary of the model domain that is part of the fault is associated with a mass flux. We parametrise the magnitude of this flux, denoted by $\Source_{\mathrm{up}}^{1}$, using a Whittle-Mat{\'e}rn field with a mean and standard deviation that reduce as the horizontal distance to the centre of the mesh increases, reflecting a prior belief that the majority of the upflow in the fault is likely to be close to the centre of the domain, beneath the clay cap. In particular, the mean, denoted by $\Mean_{\mathrm{up}}^{\Source}(\bm{x})$, is given by $\Mean_{\mathrm{up}}^{\Source}(\bm{x}) = 2.5 \times 10^{-4} \Delta(\bm{x})$,
and the standard deviation, denoted by $\Std^{\mathrm{up}}_{\Source}(\bm{x})$, is given by $\Std_{\mathrm{up}}^{\Source}(\bm{x}) \sim \mathcal{U}(1.0\times10^{-5}\Delta(\bm{x}), 5.0\times10^{-5}\Delta(\bm{x}))$, where $\Delta(\bm{x})$ is defined as
\begin{equation}
    \Delta(\bm{x}) = \exp\left(-\frac{(x_{1}-3000)^2 + (x_{2}-3000)^2}{1600}\right).
\end{equation}
We use an (isotropic) lengthscale of $500\,\metres$.

\begin{figure}
    \centering
    \includegraphics[width=0.5\linewidth]{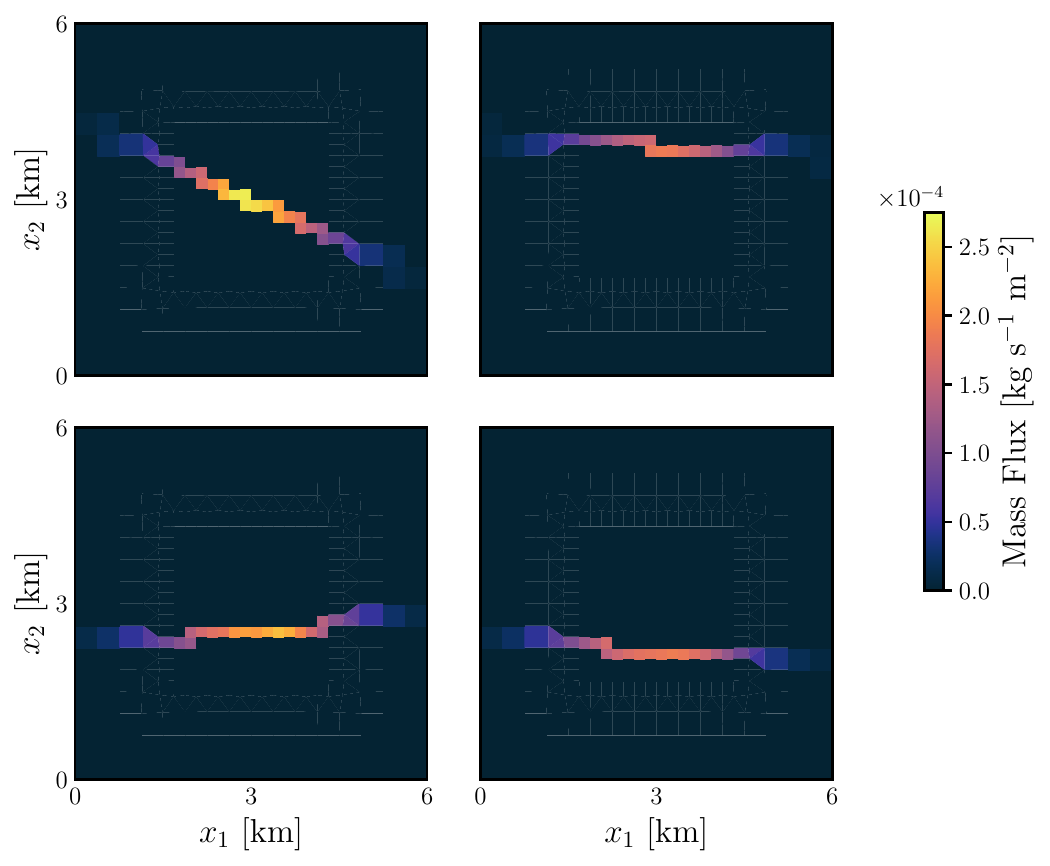}
    \caption{The location and magnitude of the mass flux associated with particles drawn from the prior of the synthetic three-dimensional model.}
    \label{fig:upflows_pri_fault}
\end{figure}

Figure \ref{fig:upflows_pri_fault} shows the geometry and mass flux of faults corresponding to particles drawn from the prior. We note that this type of parametrisation is far from the only way to represent the upflow at the base of a reservoir model; for alternative parametrisation techniques capable of representing different types of prior knowledge, see \citet{Bjarkason21b, Nicholson20}.

\begin{figure}
    \centering
    \includegraphics[width=0.5\linewidth]{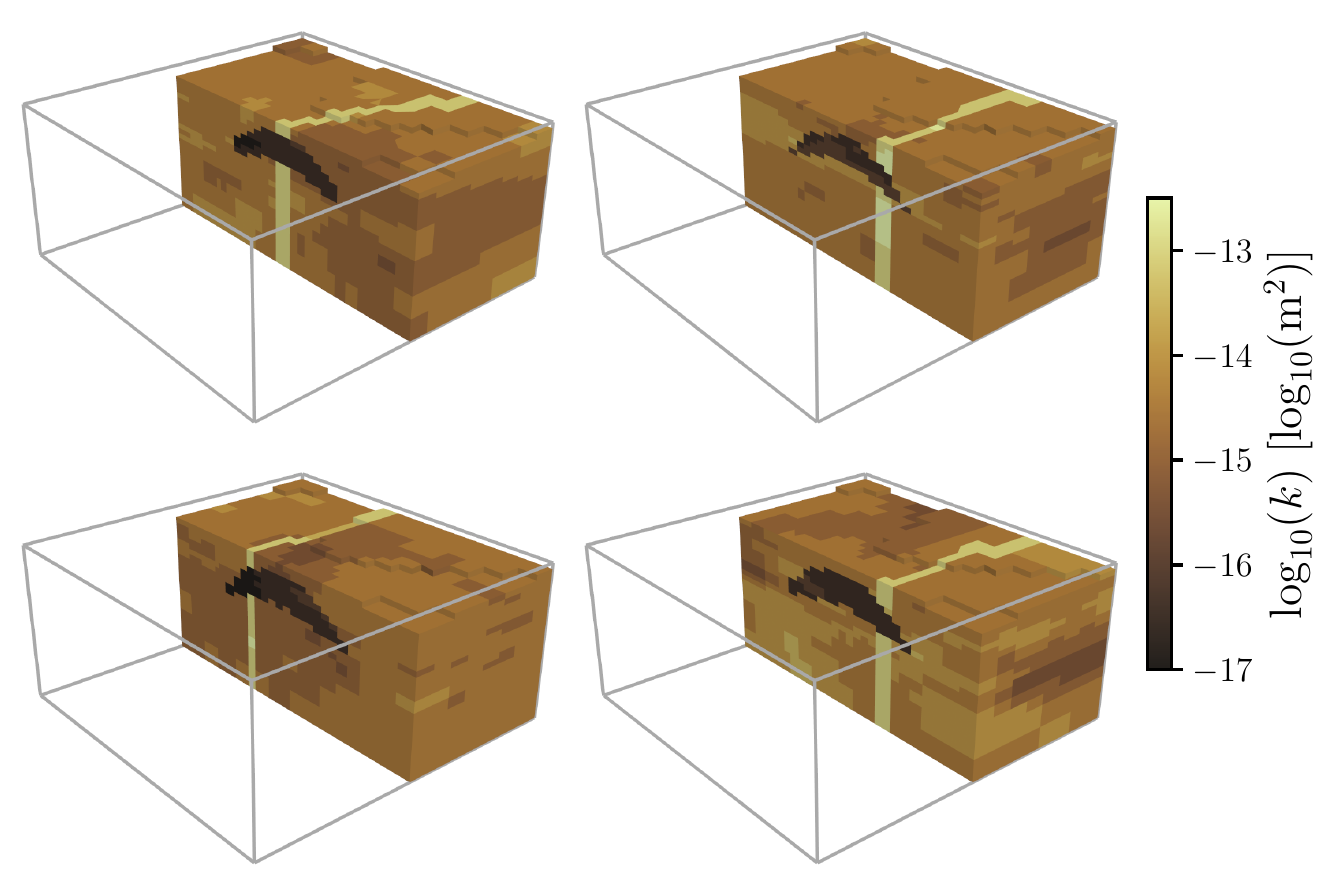}
    \caption{The permeability structures of particles drawn from the prior of the synthetic three-dimensional model.}
    \label{fig:particles_pri_fault}
\end{figure}

To model the permeability in each subdomain, we use the level set method in the same manner as for the vertical slice problem. We select the formations in each subdomain such that the permeability in $\Domain_{\mathcal{C}}$ varies between $10^{-17}\,\metres^{2}$ and $10^{-16}\,\metres^{2}$, the permeability in $\Domain_{\Fault}$ varies between $10^{-13.5}\,\metres^{2}$ and $10^{-12.5}\,\metres^{2}$, and the permeability in $\Domain_{\Background}$ varies between $10^{-15.5}\,\metres^{2}$ and $10^{-13.5}\,\metres^{2}$; for completeness, Appendix \ref{sec:level_set_fault} contains each level set mapping. Figure \ref{fig:particles_pri_fault} shows the permeability structures associated with a set of particles drawn from the prior.

Table \ref{tab:prior_3d} provides a summary of the prior parametrisation.

\begin{table*}
    \centering
    \caption{The prior parametrisation for the synthetic three-dimensional model.}
    \scriptsize
    \begin{adjustbox}{center}
    \begin{tabular}{ p{30mm} p{50mm} p{25mm} p{58mm} }
        \toprule
        \textbf{Model Component} & \textbf{Parameter Description} & \textbf{Symbol} & \textbf{Prior} \\
        \midrule 
        Clay cap geometry ($\Domain_{\ClayCap}$) & Vertical depth of centre of clay cap ($\metres$) & $s_{3}$ & $\mathcal{U}(-900, -775)$ \\
        & Width ($x_{1}$ and $x_{2}$ directions) parameter of set $\mathcal{S}$ ($\metres$) & $w$ & $\mathcal{U}(1400, 1600)$ \\
        & Height ($x_{3}$ direction) parameter of set $\mathcal{S}$ ($\metres$) & $h$ & $\mathcal{U}(200, 300)$ \\
        & Fourier coefficients used to define boundary of set $\mathcal{S}$ ($\metres$) & $f_{i}$, $i\in\{1, \dots, 20\}$ & $\mathcal{N}(0, 25)$ \\
        & Curvature parameter ($\metres$) & $d$ & $\mathcal{U}(300, 600)$ \\
        \midrule
        Fault geometry ($\Domain_{\Fault}$) & Point at which fault intersects eastern boundary ($\metres$) & $b_{e}$ & $\mathcal{U}(1500, 4500)$ \\
        & Point at which fault intersects western boundary ($\metres$) & $b_{w}$ & $\mathcal{U}(1500, 4500)$ \\
        \midrule
        Permeability & Level set function in each subdomain & $\LevelFunc_{i}, i \in \{\ClayCap, \Fault, \Background\}$ & $\mathcal{N}(0, \mathcal{C}_{\mathrm{WM}}[\Std=\Std_{\LevelFunc_{i}}, \bm{L}=\bm{L}_{\LevelFunc_{i}}])$ \\
        & Standard deviation of level set function in subdomains $\ClayCap, \Fault$ & $\Std_{\LevelFunc_{i}}, i \in \{\ClayCap, \Fault\}$ & $\mathcal{U}(0.5, 1.0)$ \\
        & Standard deviation of level set function in subdomain $\Background$ & $\Std_{\LevelFunc_{\Background}}$ & $\mathcal{U}(0.75, 1.50)$ \\
        & Horizontal ($x_{1}$ and $x_{2}$) lengthscales of level set function in each subdomain ($\metres$) & $\Ls_{j,\LevelFunc_{i}}$, $i \in \{\Shal, \ClayCap, \Deep\}$, $j \in \{1, 2\}$ & $\mathcal{U}(1000, 2000)$ \\
        & Vertical lengthscale of level set function in each subdomain ($\metres$) & $\Ls_{3,\LevelFunc_{i}}, i \in \{\Shal, \ClayCap, \Deep\}$ & $\mathcal{U}(200, 500)$ \\
        \midrule
        Mass upflow & Magnitude of mass upflow through bottom boundary of domain contained in $\Domain_{\Fault}$ ($\kg\,\mathrm{m}^{-2}\,\mathrm{s}^{-1}$) & $\Source_{\mathrm{up}}^{1}$ & $\mathcal{N}(\Mean_{\mathrm{up}}^{\Source}, \mathcal{C}_{\mathrm{WM}}[\Std=\Std_{\mathrm{up}}^{\Source}, \bm{L}=\diag(500, 500)])$ \\
        & Standard deviation of mass upflow through bottom boundary of domain contained in $\Domain_{\Fault}$ ($\kg\,\mathrm{m}^{-2}\,\mathrm{s}^{-1}$) & $\Std_{\mathrm{up}}^{\Source}$ & $\mathcal{U}(1.0\times10^{-5}\Delta(\bm{x}), 5.0\times10^{-5}\Delta(\bm{x}))$ \\
        \bottomrule
    \end{tabular}
    \end{adjustbox}
    \label{tab:prior_3d}
\end{table*}

\subsection{Real Volcanic Geothermal System} \label{sec:setup_vol}

Finally, we apply EKI to a model of a large-scale, real-world volcanic geothermal system.\footnote{Information about this system is commercially sensitive; for this reason, we do not specify its name or location. The units of the data have also been redacted.} Prior modelling of the system has been carried out using the standardised geothermal modelling framework outlined in \cite{OSullivan23}. First, a conceptual model incorporating key characteristics of the system, including the geological structure, heat source, and reservoir boundaries, was developed. These features were then mapped onto the grid of the numerical model, which we now describe.

\subsubsection{Numerical Model}

\begin{figure*}
    \centering
    \includegraphics[width=0.8\textwidth]{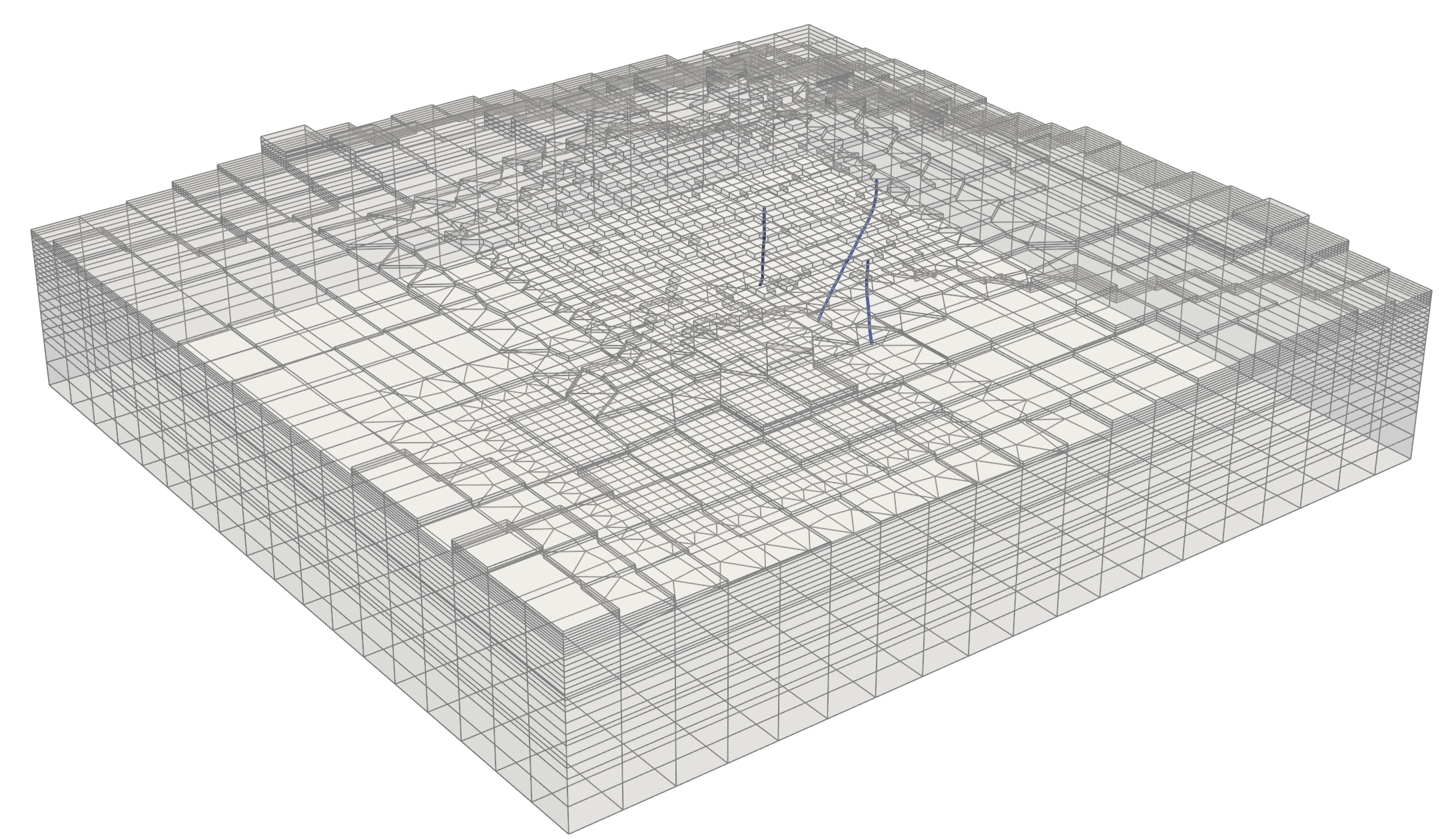}
    \caption{The mesh of the volcanic geothermal model. The blue tubes denote the three well tracks.}
    \label{fig:mesh_vol}
\end{figure*}

\begin{figure}
    \centering
    \includegraphics[width=0.5\linewidth]{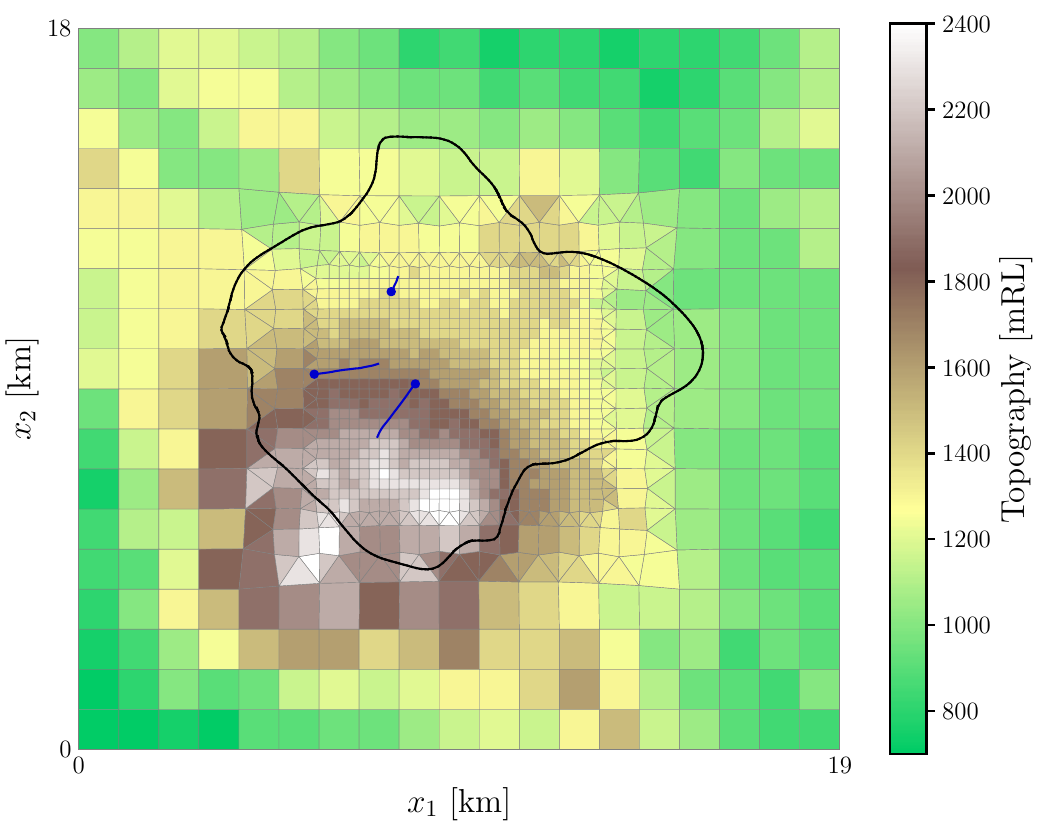}
    \caption{The topography of the volcanic geothermal model, including the three well tracks (blue) and the lateral extent of the clay cap (black).}
    \label{fig:elevs_vol}
\end{figure}

The grid of the numerical model is plotted in Figure \ref{fig:mesh_vol}. It covers an area of almost $400\,\mathrm{km}^{2}$, which is sufficient to contain the entire convective geothermal system, and contains approximately $40,000$ cells. The grid extends to the ground surface, capturing the vadose zone. We therefore consider two mass components---water and air---as well as an energy component when simulating the reservoir dynamics using Waiwera. Figure \ref{fig:elevs_vol} shows the model topography.

\begin{figure*}
    \centering
    \includegraphics[width=0.75\textwidth]{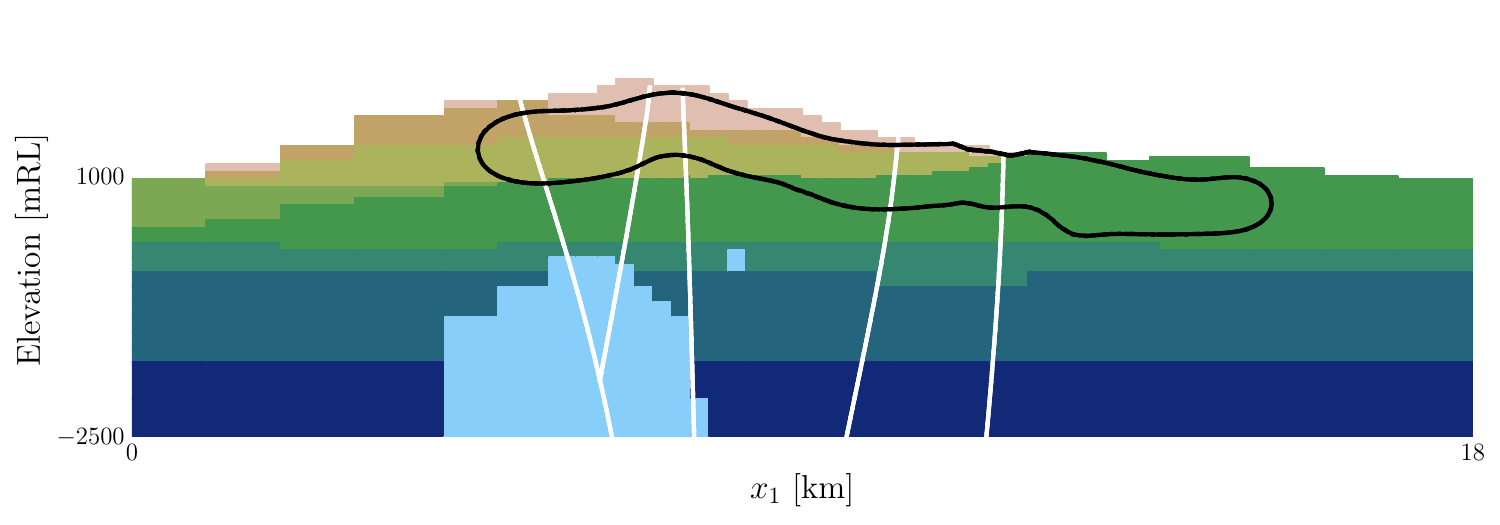}
    \caption{A vertical slice through the centre of the volcanic geothermal model. Each colour represents a different rock formation (the intrusion at the base of the model is denoted using light blue). The black line indicates the boundary of the clay cap, and the white lines denote faults.}
    \label{fig:rocktypes_vol}
\end{figure*}

Figure \ref{fig:rocktypes_vol} shows a vertical slice through the centre of the model, providing an indication of the geological structure. Notably, there is a large intrusion at the base of the system which acts as a source of hot mass upflow. Figures \ref{fig:elevs_vol} and \ref{fig:rocktypes_vol} also depict the location of the clay cap of the system, which has been inferred from resistivity data.

The top boundary of the model is set to a constant temperature and pressure, representing an atmospheric boundary condition. Additionally, rainfall is modelled by imposing a constant mass flux through the boundary with an appropriate enthalpy. The shallow sections of the side boundaries are associated with pressure-dependent flow conditions, allowing fluid to flow between the model domain and the surrounding regions as the water table fluctuates. The deeper sections of the side boundaries are closed. 

\begin{figure}
    \centering
    \includegraphics[width=0.5\linewidth]{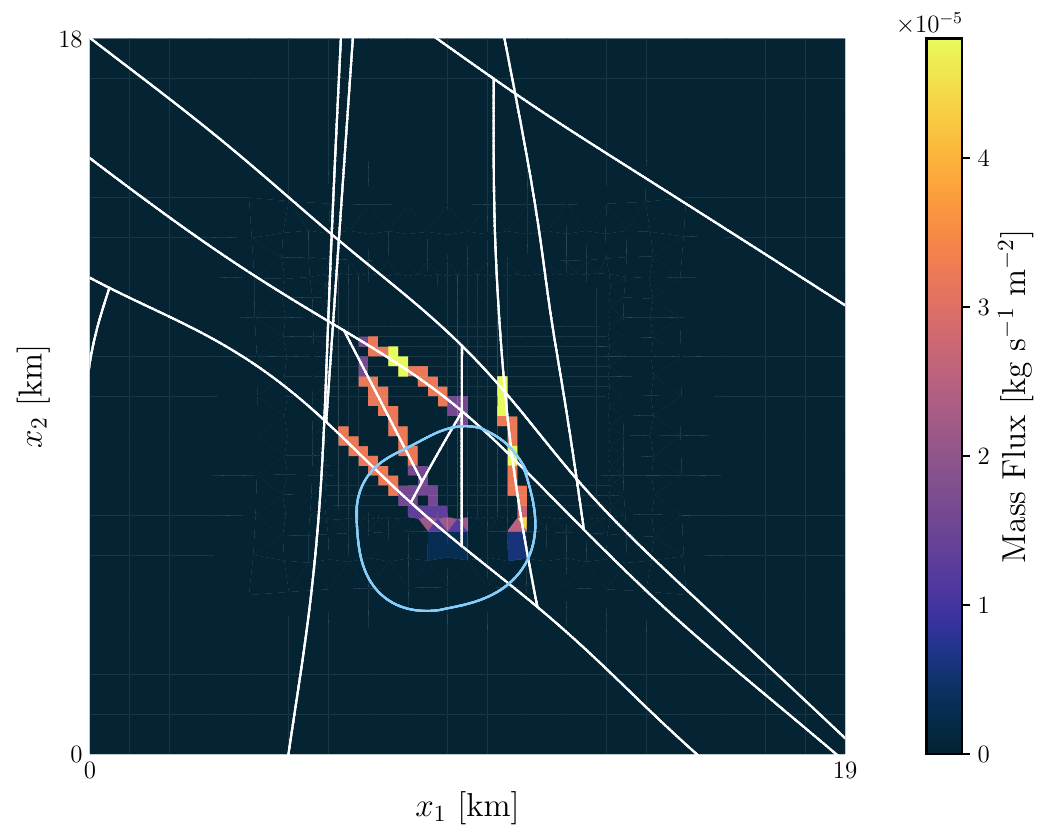}
    \caption{The mass flux at the base of the volcanic geothermal model. The white lines indicate faults, and the light blue line indicates the boundary of the intrusion.}
    \label{fig:upflow_vol}
\end{figure}

The bottom boundary of the model domain is associated with a constant heat flux of $80\,\mathrm{mW}\,\metres^{-2}$, representing conductive heat flow. Additionally, the sections corresponding to some of the key faults, as well as sections of the intrusion, are associated with a constant, high-enthalpy mass flux, simulating hot fluid rising from deeper within the system. The mass flux in each region is indicated in Figure \ref{fig:upflow_vol}.

\subsubsection{Data}

Data on the natural state of the system has been collected at three exploration wells, indicated in Figures \ref{fig:mesh_vol} and \ref{fig:elevs_vol}. However, no production has been carried out yet. For this reason, we consider only natural state simulations in this case study.

\begin{figure*}
    \centering
    \includegraphics[width=0.75\textwidth]{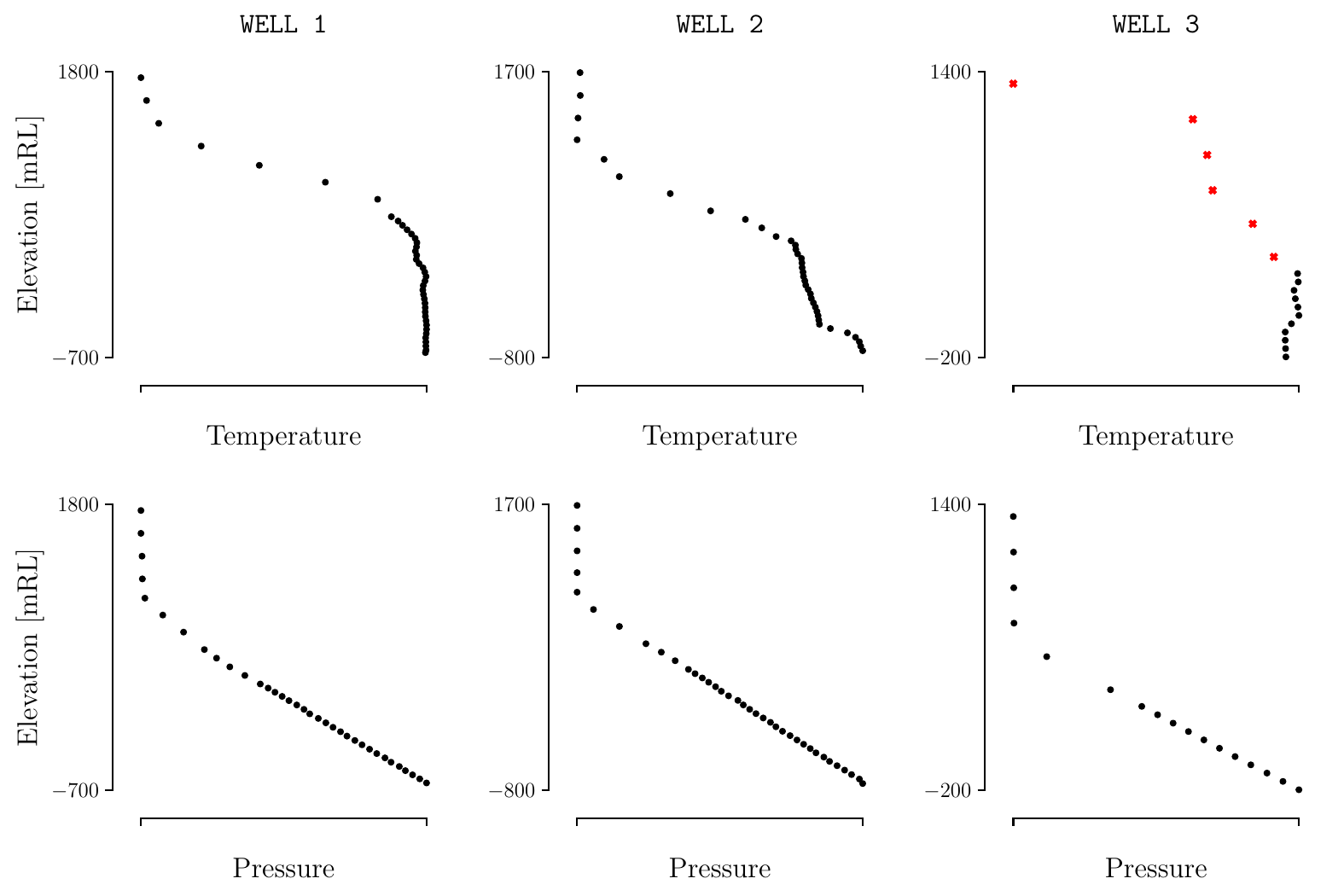}
    \caption{Natural state temperatures (\emph{top row}) and pressures (\emph{bottom row}) at each of the observation wells of the volcanic geothermal system. The red crosses denote data that was not used when approximating the posterior of the model parameters using EKI. Note that the units of the measurements have been redacted for confidentiality reasons.}
    \label{fig:data_vol}
\end{figure*}

Figure \ref{fig:data_vol} shows the data collected at each exploration well. Downhole temperature and pressure profiles such as these can be challenging to interpret due to internal flow effects within the well which mean the measured values are not representative of the actual state of the reservoir \citep{Grant83, OSullivan16}. Notably, the measured temperatures at the top of well 3 (indicated using red crosses) are significantly higher than would be expected; these are likely to be the product of steam flow within the wellbore. For the numerical model to be able to replicate these temperature profiles, we would need to couple Waiwera with a wellbore simulator, which is outside the scope of the current study. Instead, we simply discard these measurements.

When applying EKI to this system, we model the errors in the measurements as independent and normally distributed. We assume that the errors in the temperature measurements have a standard deviation of $15^{\circ}\mathrm{C}$, and the errors in the pressure measurements have a standard deviation of $10\,\mathrm{bar}$.

\subsubsection{Prior Parametrisation}

We aim to use EKI to estimate the (anisotropic) permeability structure of the rock formations, faults, and clay cap of the system. To remain consistent with the modelling framework of \citet{OSullivan23}, we treat the boundaries of each of these features as known; it would, however, be worthwhile to study the effect of introducing uncertainty into these boundaries using the level set method. The remainder of the rock properties are treated as known; we assume that all rock in the reservoir has a thermal conductivity of $2.5 \mathrm{W}\,\metres^{-1}\,\mathrm{K}^{-1}$ and a specific heat of $1000\,\mathrm{J}\,\kg^{-1}\,\mathrm{K}^{-1}$.

As is standard, we treat the permeability tensor as diagonal; that is, $\bPerm \defas \diag(\Perm_{1}, \Perm_{2}, \Perm_{3})$, where $\Perm_{1}$ and $\Perm_{2}$ denote the permeabilities in the horizontal directions and $\Perm_{3}$ denotes the vertical permeability. We characterise the prior distribution of the permeability in each formation using the geological rules outlined by \citet{deBeer23a}, which we briefly describe here. We first partition each formation into a number of ``rocktypes'': a ``base'' rocktype, and a rocktype that represents the clay cap. Within each base and clay cap rocktype, we add additional rocktypes to represent singular faults and the intersections of multiple faults. For simplicity, we treat the permeabilities of different formations as independent. Within each formation, we also model all clay cap and fault rocktype permeabilities as conditionally independent given the permeabilities of the base rocktype.

For a given formation, we model the log-permeability of the base rocktype, denoted here using $\bPerm^{b}$, using a truncated Gaussian distribution \citep[see, e.g.,][]{Burkhardt14}; that is,
\begin{equation} 
    \begin{bmatrix}
        \log_{10}(\Perm^{b}_{1}) \\
        \log_{10}(\Perm^{b}_{2}) \\
        \log_{10}(\Perm^{b}_{3})
    \end{bmatrix}
    \sim \mathcal{TN}(\bMean^{b}, \bCov^{b}, \bm{a}^{b}, \bm{b}^{b}).
\end{equation}
The mean, $\bMean^{b}$, is the same in all directions, and varies from $[-16, -16, -16]^{\top}$ in the deepest formations to $[-13, -13, -13]^{\top}$ in the shallowest formations. In all cases, the marginal standard deviations in each direction are set to $0.5\,\log_{10}(\metres^{2})$. We impose a strong correlation ($\rho = 0.8$) between the components of the permeability in the horizontal ($x_{1}$ and $x_{2}$) directions, and a moderate correlation ($\rho = 0.5$) between each horizontal component of the permeability and the vertical permeability. In all cases, we use truncation points of $\bMean_{-}^{b} = [-17, -17, -17]^{\top}$ and $\bMean_{+}^{b} = [-11, -11, -11]^{\top}$.

\begin{figure*}
    \centering
    \includegraphics[width=0.75\textwidth]{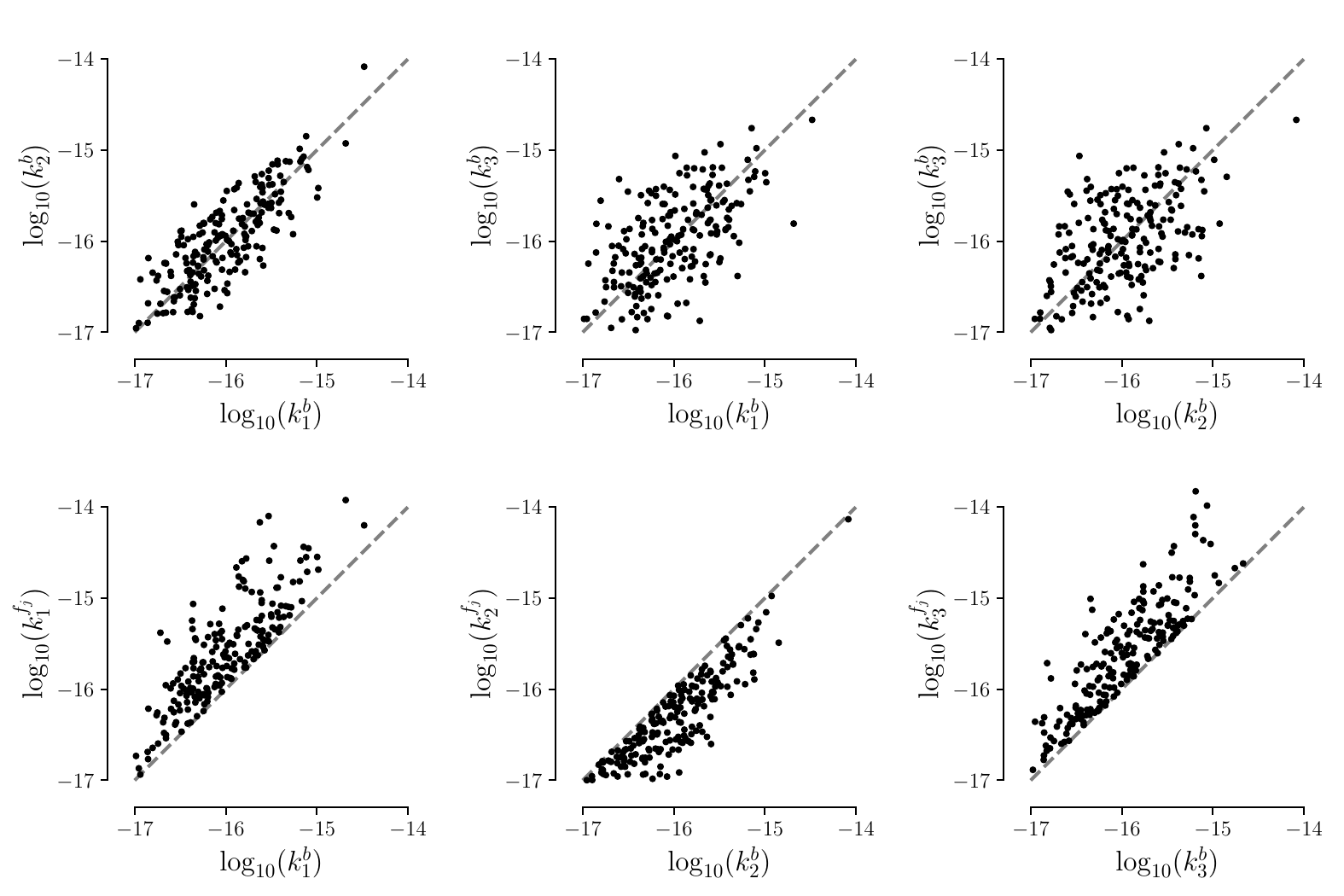}
    \caption{Prior samples of the log-permeability of the base rocktype (\emph{top row}), and the base rocktype and a fault rocktype (\emph{bottom row}) of one of the formations of the volcanic model. Note that in this case, the $x_{1}$ direction runs along the strike of the fault and the $x_{2}$ direction runs across the strike of the fault.}
    \label{fig:geo_rules_vol}
\end{figure*}

The log-permeability of each remaining rocktype is also modelled using a truncated Gaussian distribution. The parameters of each of these distributions are listed in Table \ref{tab:perm_rules}. Notably, the permeability of the clay cap is parametrised such that it is always less than or equal to that of the corresponding base rocktype. The component of the permeability in the direction across the strike of a fault is parametrised such that it is less than or equal to that of the surrounding rock, and the components of the permeability along the strike and up the dip of a fault are parametrised such that they are greater than or equal to those of the surrounding formation.\footnote{The direction along the strike of a given fault is considered to be the horizontal axis that the fault is closest to being parallel to. We note that the model uses a rotated coordinate system so that the horizontal axes are aligned with the key fault structures.} The horizontal components of the permeability at the intersection of multiple faults are parametrised such that they can be greater than or less than those of the surrounding rock. The vertical permeability, however, is parametrised such that it is always greater than or equal to that of the surrounding rock. Figure \ref{fig:geo_rules_vol} shows a set of prior samples of the log-permeability of the base rocktype, and one of the fault rocktypes, of one of the formations of the model. 

\begin{table*}
    \centering
    \caption{The parameters, $\Mean$, $\Std$, $\Mean_{-}$, and $\Mean_{+}$, of the truncated Gaussian distributions used to parametrise the log-permeability of the clay cap, fault and fault intersection rocktypes of each formation in each direction.}
    \scriptsize
    \begin{adjustbox}{center}
    \begin{tabular}{l c c c c c c}
        \toprule
        \textbf{Rocktype} & \textbf{Symbol} & \textbf{Direction} & $\Mean\,[\log_{10}(\metres^{2})]$ & $\Std\,[\log_{10}(\metres^{2})]$ & $\Mean_{-}\,[\log_{10}(\metres^{2})]$ & $\Mean_{+}\,[\log_{10}(\metres^{2})]$ \\
        \midrule
        Clay cap & $\Perm^{c}_{i}$ & All ($i = 1, 2, 3$) & $-16$ & $0.25$ & $-17$ & $\log_{10}(\Perm_{i}^{b})$ \\
        \midrule
        \multirow{3}{*}{Fault $j$} & \multirow{3}{*}{$\Perm^{f_{j}}_{i}$} & Along strike ($i\in\{1, 2\}$) & $\log_{10}(\Perm_{i}^{b})$ & $0.5$ & $\log_{10}(\Perm_{i}^{b})$ & $-11$ \\
        & & Across strike ($i\in\{1, 2\}$) & $\log_{10}(\Perm_{i}^{b})$ & $0.5$ & $-17$ & $\log_{10}(\Perm_{i}^{b})$ \\
        & & Up dip ($i=3$) & $\log_{10}(\Perm_{i}^{b})$ & $0.5$ & $\log_{10}(\Perm_{i}^{b})$ & $-11$ \\
        \midrule
        \multirow{3}{*}{Clay cap fault $j$} & \multirow{3}{*}{$\Perm^{cf_{j}}_{i}$} & Along strike ($i\in\{1, 2\}$) & $\log_{10}(\Perm_{i}^{c})$ & $0.5$ & $\log_{10}(\Perm_{i}^{c})$ & $-11$ \\
        & & Across strike ($i\in\{1, 2\}$) & $\log_{10}(\Perm_{i}^{c})$ & $0.5$ & $-17$ & $\log_{10}(\Perm_{i}^{c})$ \\
        & & Up dip ($i=3$) & $\log_{10}(\Perm_{i}^{c})$ & $0.5$ & $\log_{10}(\Perm_{i}^{c})$ & $-11$ \\
        \midrule
        \multirow{2}{*}{Intersection $j$} & \multirow{2}{*}{$\Perm^{n_{j}}_{i}$} & Horizontal ($i=1,2$) & $\log_{10}(\Perm_{i}^{b})$ & $0.5$ & $-17$ & $-11$ \\
        & & Vertical ($i=3$) & $\log_{10}(\Perm_{i}^{b})$ & $0.5$ & $\log_{10}(\Perm_{i}^{b})$ & $-11$ \\
        \midrule
        \multirow{2}{*}{Clay cap intersection $j$} & \multirow{2}{*}{$\Perm^{cn_{j}}_{i}$} & Horizontal ($i=1,2$) & $\log_{10}(\Perm_{i}^{c})$ & $0.5$ & $-17$ & $-11$ \\
        & & Vertical ($i=3$) & $\log_{10}(\Perm_{i}^{c})$ & $0.5$ & $\log_{10}(\Perm_{i}^{c})$ & $-11$ \\
        \bottomrule
    \end{tabular}
    \end{adjustbox}
    \label{tab:perm_rules}
\end{table*}

To generate a prior realisation of the permeability structure of a given formation, we first sample independent variates from the unit Gaussian distribution corresponding to each component of the diagonal of the permeability tensor for each rocktype. We then apply a transformation of the form outlined in Section \ref{sec:constraints} to each variate, to map it to a variate distributed according to the truncated Gaussian distribution with the desired mean, variance and truncation points. We note that the target distributions of the permeabilities of the fault rocktypes depend on the permeability of the clay cap and base rocktypes, and the target distributions of the permeability of the clay cap rocktype depends on the permeabilities of the base rocktype. For this reason, we first compute the permeability of the base rocktype, followed by the clay cap rocktype, then the fault rocktypes. To introduce the desired correlations between each component of the permeability tensor of the base rocktype, we apply an additional rotation prior to transforming the variates to the desired target distribution, to give these variates the required correlation structure. We note that the correlation structure of the variates is slightly modified when they are transformed. However, because the truncations are not severe, this effect is insignificant. 

During our initial tests of EKI using this model (discussed further in Section \ref{sec:results_vol}), it proved challenging to introduce uncertainty into the mass flux at the base of the model without increasing the simulation failure rate significantly. For this reason, we treat the magnitudes and locations of the fluxes plotted in Figure \ref{fig:upflow_vol} (obtained using manual calibration as part of the previous modelling study) as known. We do, however, introduce uncertainty into the enthalpy of the flux, using a Whittle-Mat{\'e}rn field with a mean of $1400\,\mathrm{kJ}\,\mathrm{kg}^{-1}$, a standard deviation of $50\,\mathrm{kJ}\,\mathrm{kg}^{-1}$, and an (isotropic) lengthscale of $750\,\mathrm{m}$.

\section{Results} \label{sec:results}

We now show the results of EKI applied to each model problem. In all cases, we discuss both the posterior and posterior predictive distributions. For additional results for the synthetic case studies that illustrate the effects of using localisation and inflation, and examples of the posterior distributions of the hyperparameters of the Whittle Mat{\'e}rn fields used as part of the prior parametrisations, the reader is referred to the Supplementary Material.

\subsection{Vertical Slice Model}

We run EKI on the vertical slice model using an ensemble of $\NumEns=100$ particles. The algorithm converges after $7$ iterations, and approximately $96$ percent of the simulations are successful.

\begin{figure*}
    \centering
    \includegraphics[width=0.75\textwidth]{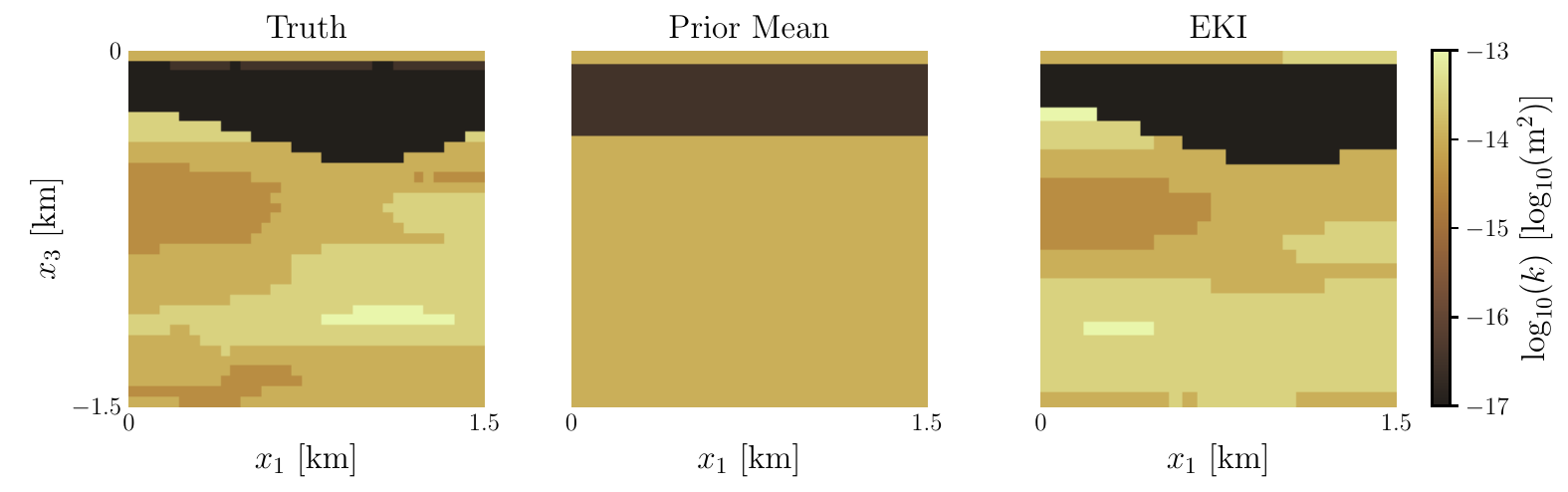}
    \caption{The true permeability of the vertical slice model (\emph{left}), the prior mean (\emph{centre}), and the mean of the approximation to the posterior generated using EKI (\emph{right}).}
    \label{fig:mean_eki_slice}
\end{figure*}

\begin{figure}
    \centering
    \includegraphics[width=0.5\linewidth]{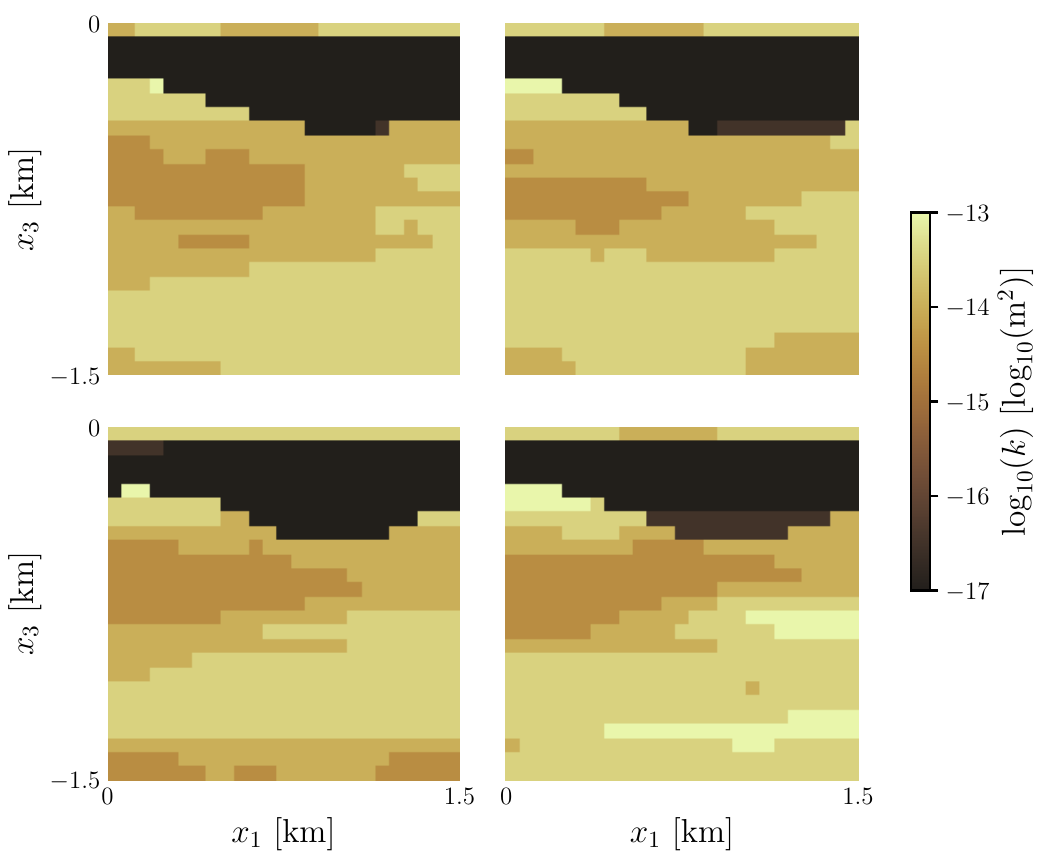}
    \caption{The permeability structures of particles drawn from the EKI approximation to the posterior of the slice model.}
    \label{fig:particles_post_slice}
\end{figure}

Figure \ref{fig:mean_eki_slice} shows the permeability structure of the true system, and the EKI estimate of the posterior mean (the prior mean is also shown for comparison). We note that our estimate of the posterior mean is obtained by first computing the ensemble mean of the axillary, untransformed parameters, $\bParamAux$, then applying the mapping $\Parametrisation$ (as discussed in Section \ref{sec:constraints}) to obtain the corresponding permeability structure. The EKI estimate of the posterior mean shows a high degree of similarity to the true permeability structure; the bottom boundary of the clay cap is well recovered, as is the permeability within each subdomain. There are, of course, differences between the truth and the reconstructions; notably, the region of low permeability at the base of the true system is almost completely absent in the conditional mean generated using EKI. It is reassuring, however, to see that this region is present in some of the particles of the final ensemble. A sample of these is shown in Figure \ref{fig:particles_post_slice}. As expected, these show a far greater degree of similarity to the truth, and to one another, than the draws from the prior (see Fig.\ \ref{fig:particles_pri_2d}).

\begin{figure}
    \centering
    \includegraphics[width=0.5\linewidth]{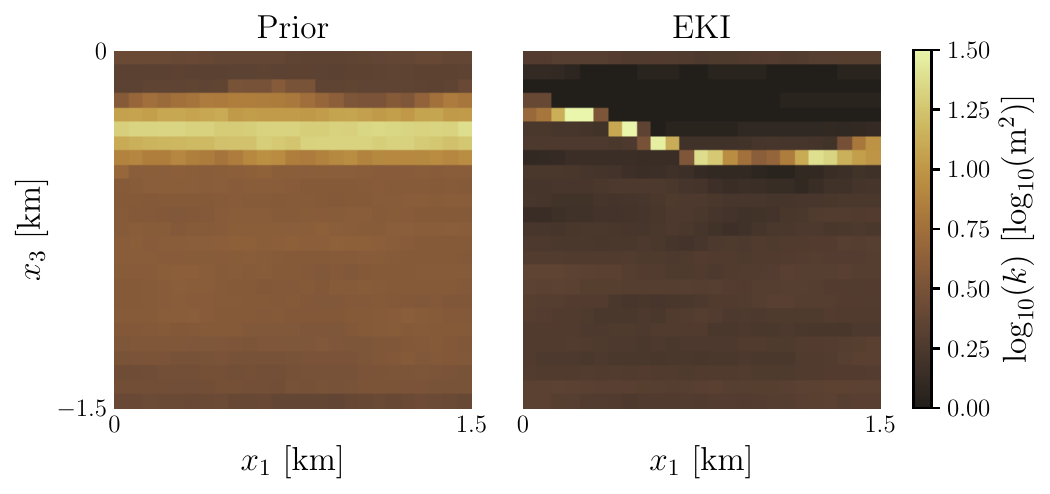}
    \caption{The marginal prior (\emph{left}) and approximate posterior (\emph{right}) standard deviations of the log-permeability of the slice model.}
    \label{fig:stds_slice}
\end{figure}

Figure \ref{fig:stds_slice} shows the standard deviations of the log-permeability of each cell in the model mesh, for the prior ensemble and the approximate posterior generated using EKI. The particles from the EKI posterior show significantly less variability than those sampled from the prior. The greatest uncertainty is in the permeability of the region surrounding the bottom boundary of the clay cap. This is to be expected; the permeability of the rock on either side of this boundary tends to be very different, so even a small degree of uncertainty in the location of the boundary will result in large uncertainties in the permeability structure of this region. 

\begin{figure}
    \centering
    \includegraphics[width=0.5\linewidth]{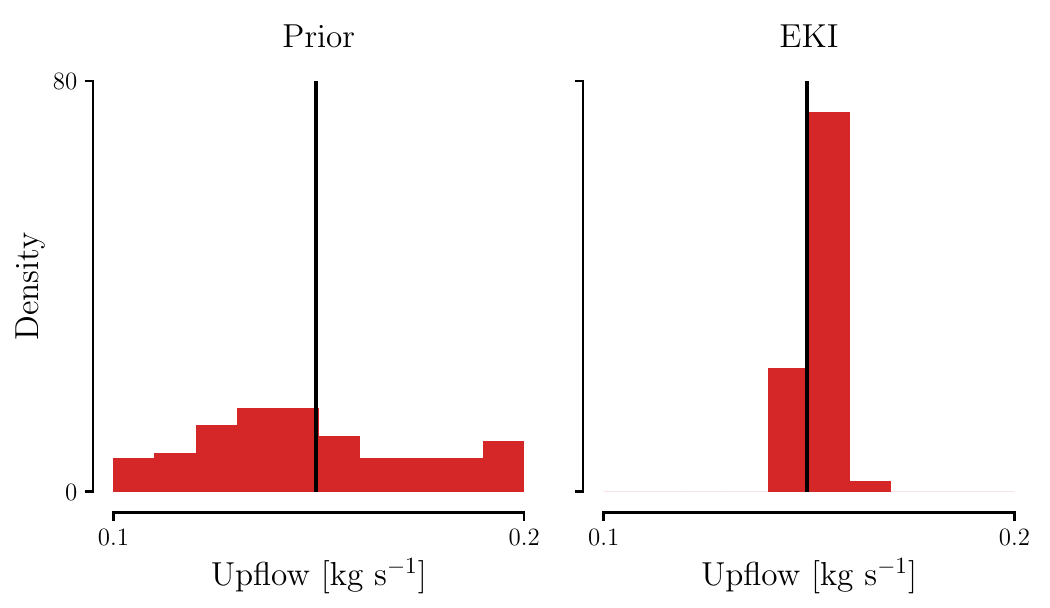}
    \caption{The marginal prior (\emph{left}) and posterior (\emph{right}) probability densities of the mass upflow in the cell at the centre of the bottom boundary of the slice model. The vertical line in each plot denotes the true mass rate.}
    \label{fig:upflows_slice}
\end{figure}

Figure \ref{fig:upflows_slice} shows the EKI estimate of the mass upflow in the cell in the centre of the bottom boundary of the model. Again, the posterior uncertainty is significantly reduced in comparison to the prior uncertainty and the true upflow rate is contained within the support of the posterior.

\begin{figure}
    \centering
    \includegraphics[width=0.5\linewidth]{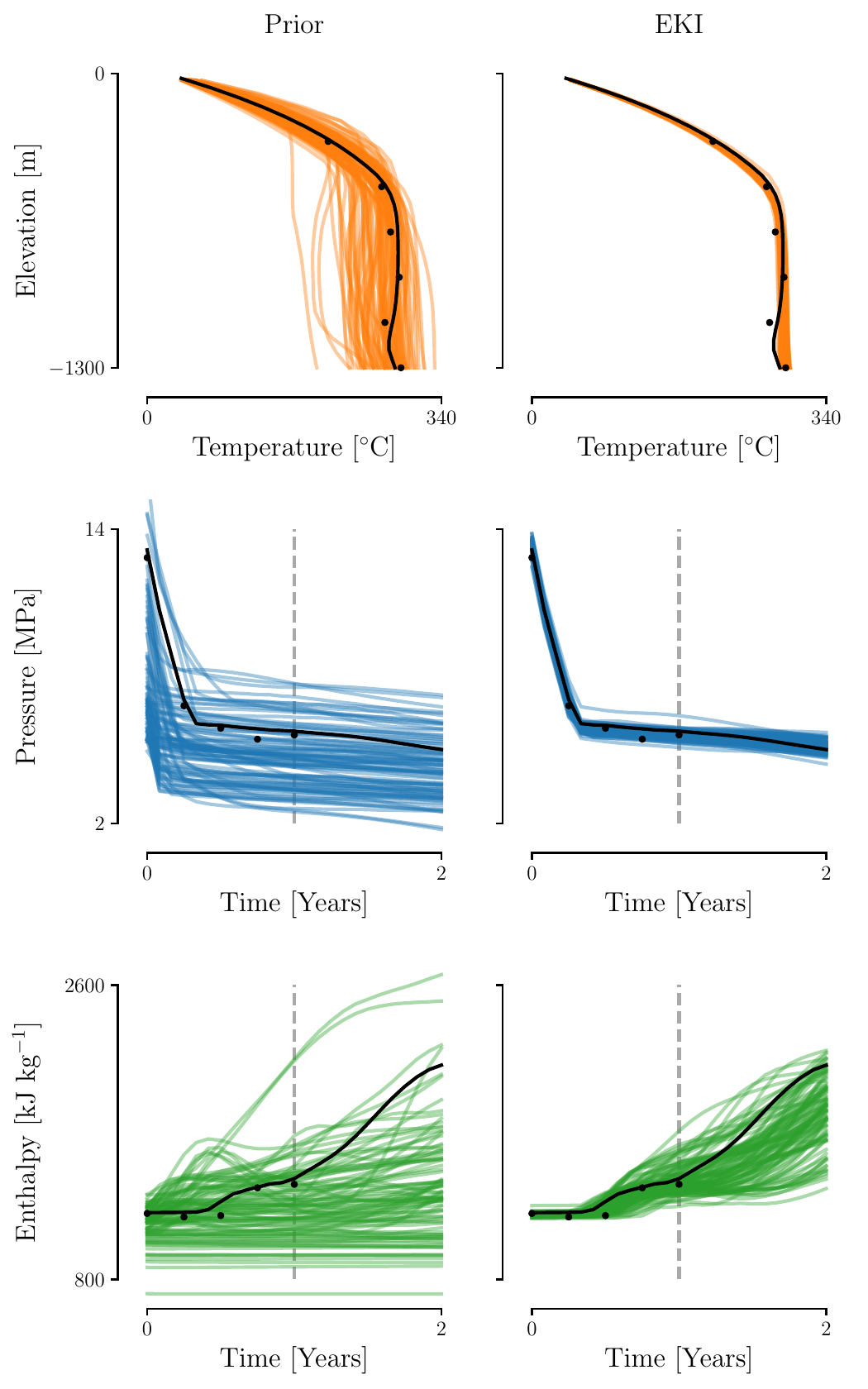}
    \caption{The modelled downhole temperature profiles (\emph{top row}) and production pressures (\emph{middle row}) and enthalpies (\emph{bottom row}) at well 3 associated with particles sampled from the prior (\emph{left}) and approximation to the posterior characterised using EKI (\emph{right}) for the vertical slice model. In the pressure and enthalpy plots, the dashed grey line denotes the end of the data collection period.}
    \label{fig:preds_slice}
\end{figure}

Figure \ref{fig:preds_slice} shows the posterior predictions of temperatures, pressures and enthalpies at one of the wells of the system generated using EKI. The corresponding prior predictions are also shown for comparison. In all cases, the posterior uncertainty is significantly reduced in comparison to the prior uncertainty, and the true state of the system is contained within the predictions. 

\begin{figure}
    \centering
    \includegraphics[width=0.5\linewidth]{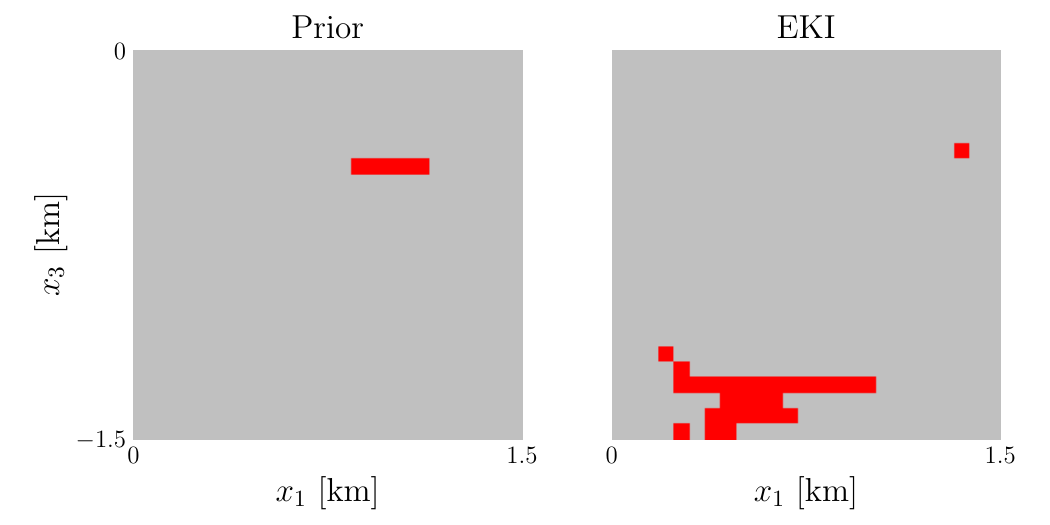}
    \caption{Heatmaps which indicate, for the prior and approximation to the posterior generated using EKI, whether the true value of the permeability in each cell of the model mesh is contained within the central 95 percent of permeabilities of the ensemble. Red cells are cells for which the permeability is not contained within the central 95 percent of the ensemble.}
    \label{fig:intervals_slice}
\end{figure}

Finally, Figure \ref{fig:intervals_slice} shows a heatmap which indicates, for each cell in the model mesh, whether the true permeability of the cell is contained within the central 95 percent of the permeabilities of the final EKI ensemble. The corresponding heatmap for the prior ensemble is also shown for comparison. We observe that the majority of the true permeabilities are contained within the final EKI ensemble. This suggests that, in this instance, EKI is able to significantly reduce our uncertainty in the permeability structure without discounting the true permeabilities. We emphasize, however, that while this is a desirable property, this does not imply that the final EKI ensemble provides an accurate approximation to the posterior. Investigating this would require us to characterise the posterior using a sampling method such as MCMC; the computational cost, however, of computing such a characterisation would be prohibitively high given the complexity of the model and the high dimensionality of the parameter space.

\subsection{Synthetic Three-Dimensional Model}

We run EKI on the synthetic three-dimensional reservoir model using an ensemble of $\NumEns = 100$ particles. The algorithm converges after $6$ iterations, and fewer than $1$ percent of the simulations fail.

\begin{figure*}
    \centering
    \includegraphics[width=0.75\textwidth]{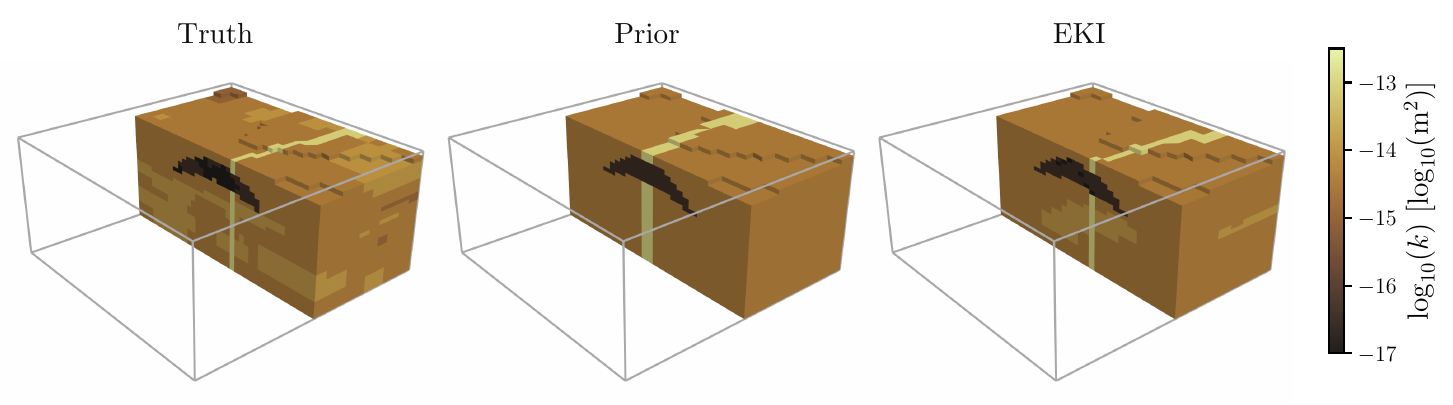}
    \caption{The true permeability structure of the synthetic three-dimensional model (\emph{left}), plotted alongside the prior mean (\emph{centre}) and the mean of the approximation to the posterior generated using EKI (\emph{right}).}
    \label{fig:means_fault}
\end{figure*}

\begin{figure}
    \centering
    \includegraphics[width=0.5\linewidth]{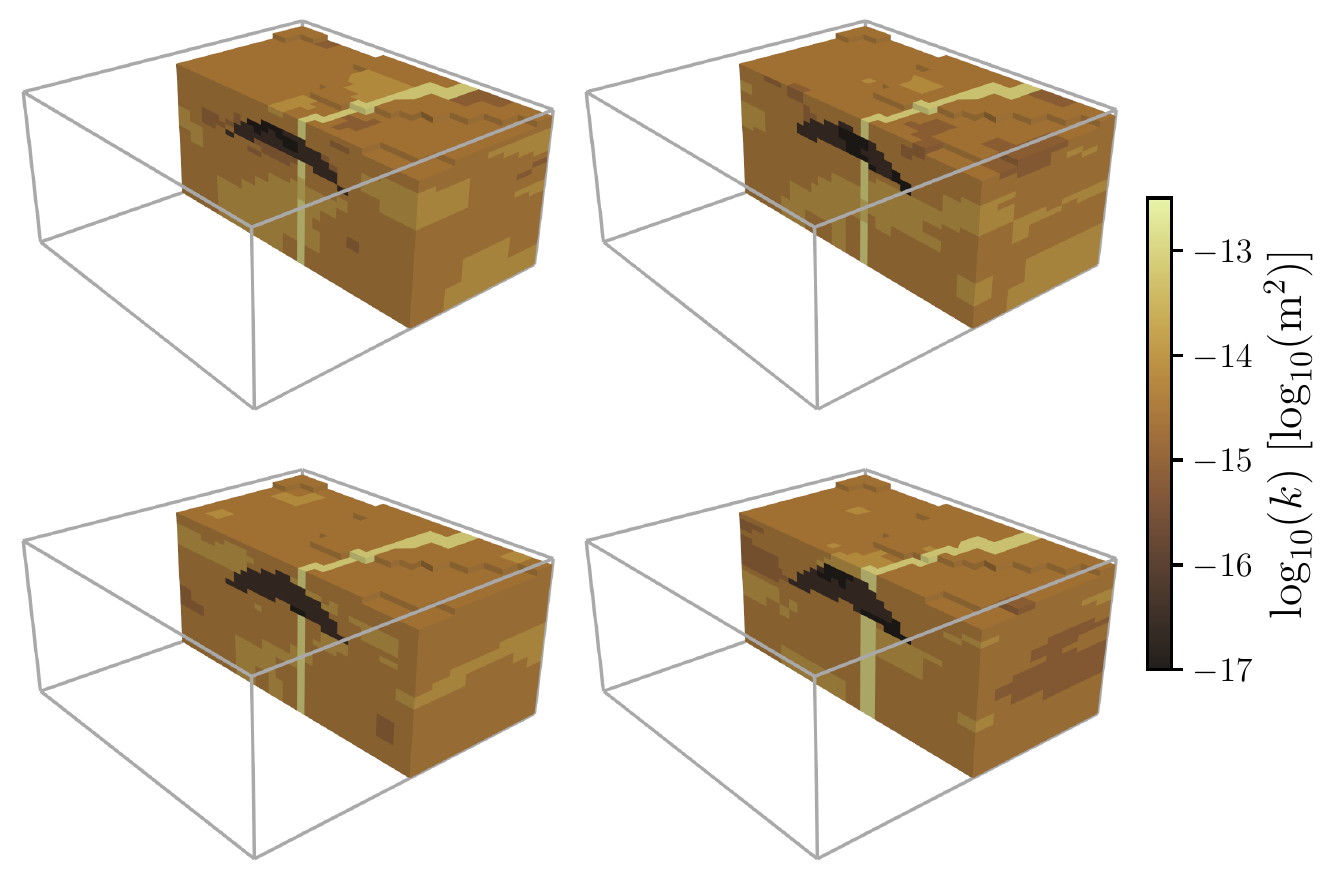}
    \caption{The permeability structures of particles drawn from the EKI approximation to the posterior of the synthetic three-dimensional model.}
    \label{fig:particles_post_fault}
\end{figure}

Figure \ref{fig:means_fault} shows the prior mean of the permeability in each cell of the model mesh, and the estimate of the posterior mean generated using EKI. The true permeability structure of the system is also shown. The permeability structures of the true system and the prior mean are quite similar; the most significant difference is in the location of the fault. The location of the fault in the estimate of the posterior mean is far closer to the true location than the prior mean. There are a number of high-permeability regions in the true system that are not reflected in the estimate of the posterior mean. Many of these, however, are present in the individual particles of the final ensemble, a sample of which are shown in Figure \ref{fig:particles_post_fault}. As expected, these show a far greater degree of similarity to the truth, and to one another, than the draws from the prior (see Fig.~\ref{fig:particles_pri_fault}).

\begin{figure}
    \centering
    \includegraphics[width=0.5\linewidth]{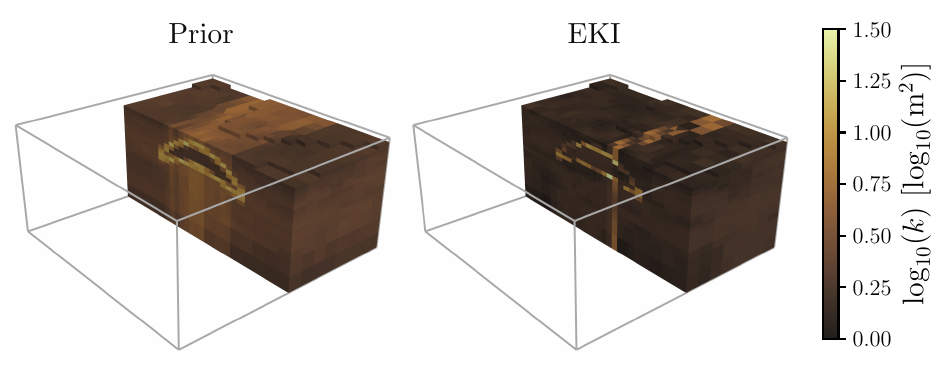}
    \caption{The prior (\emph{left}) and posterior (\emph{right}) standard deviations of the permeability structure for the synthetic three-dimensional model.}
    \label{fig:stds_fault}
\end{figure}

Figure \ref{fig:stds_fault} shows the standard deviations of the permeability in each cell of the model mesh, for the prior ensemble and the approximation to the posterior generated using EKI. The uncertainty in the final EKI ensemble is significantly reduced compared to the prior, both in terms of the permeability within each subdomain of the reservoir and the locations of the interfaces between each subdomain.

\begin{figure*}
    \centering
    \includegraphics[width=0.8\linewidth]{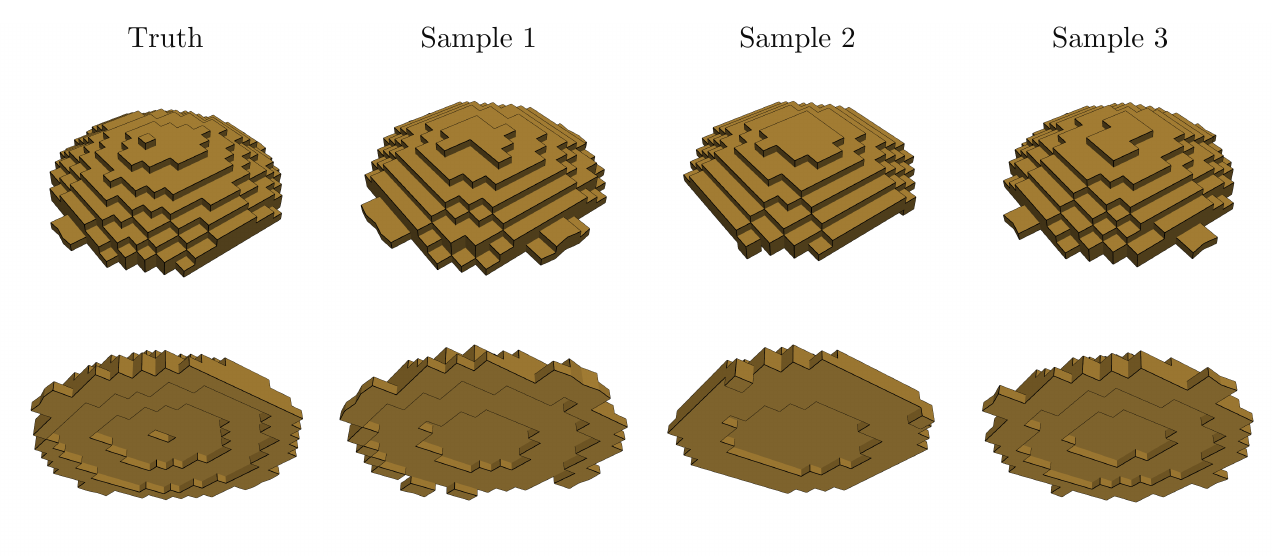}
    \caption{The top surfaces (\emph{top row}) and bottom surfaces (\emph{bottom row}) of the true clay cap (\emph{left}), and clay cap geometries corresponding to particles from the EKI approximation to the posterior of the synthetic three-dimensional model.}
    \label{fig:caps_fault_post}
\end{figure*}

\begin{figure}
    \centering
    \includegraphics[width=0.5\linewidth]{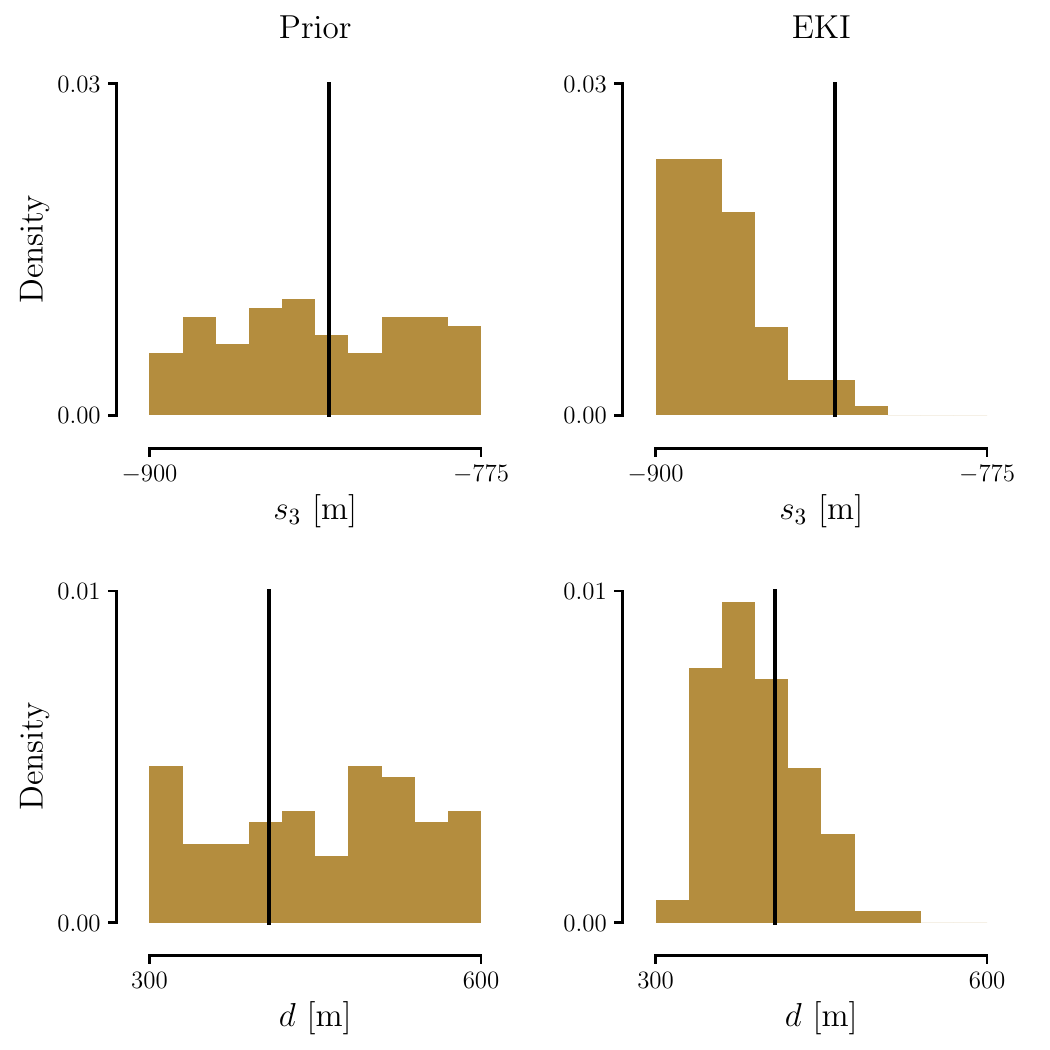}
    \caption{The marginal prior (\emph{left}) and posterior (\emph{right}) probability densities of parameter $s_{3}$ (\emph{top row}), which controls the elevation of the clay cap, and parameter $d$ (\emph{bottom row}), which controls the curvature of the clay cap. The vertical line in each plot denotes the true value of the parameter.}
    \label{fig:cap_params}
\end{figure}

Figure \ref{fig:caps_fault_post} shows a set of clay cap geometries corresponding to particles from the EKI approximation to the posterior. As expected, these show a far greater degree of similarity to one another, and the true clay cap geometry than the geometries drawn from the prior (see Fig.~\ref{fig:caps_fault_pri}). Additionally, Figure \ref{fig:cap_params} shows the EKI estimate of two of the key parameters that control the clay cap geometry; $s_{1}$ (which controls the elevation of the clay cap) and $d$ (which controls the curvature of the clay cap). Results for other parameters are similar. In both cases, we see that the posterior uncertainty in the parameter is significantly reduced in comparison to the prior uncertainty, and the true value of the parameter is contained within the posterior.

\begin{figure}
    \centering
    \includegraphics[width=0.5\linewidth]{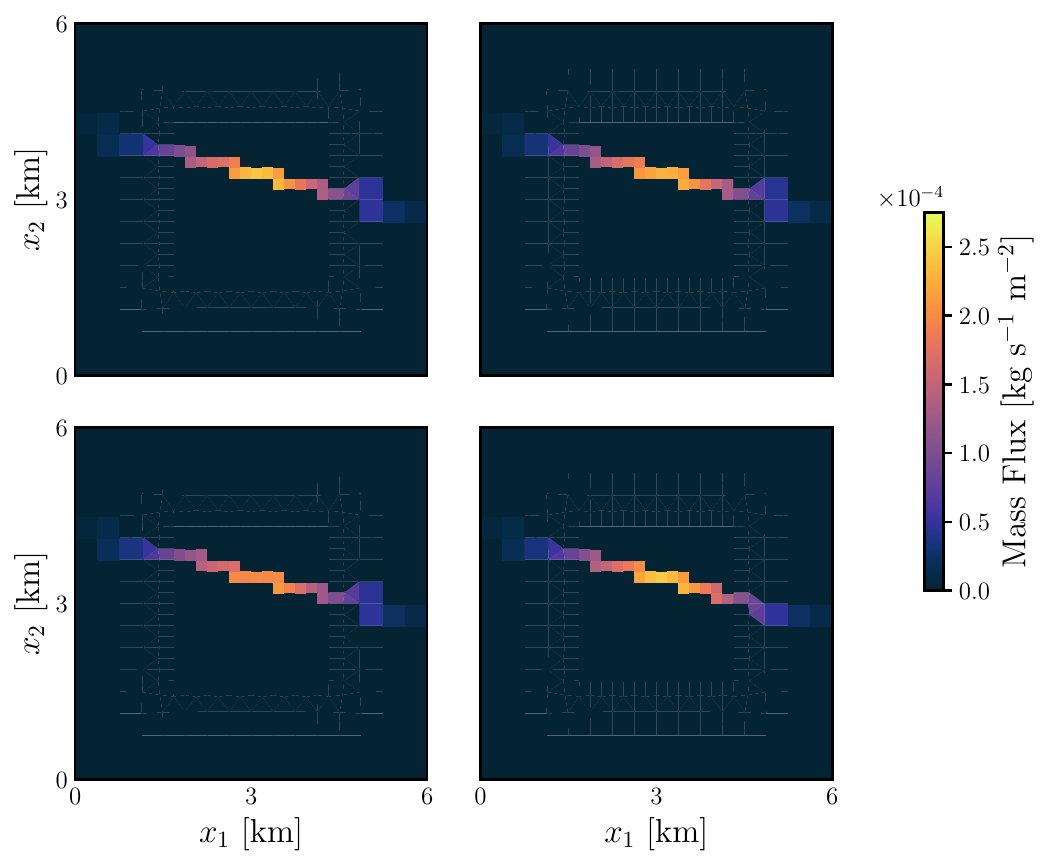}
    \caption{The location and magnitude of the mass flux associated with particles drawn from the EKI approximation to the posterior of the synthetic three-dimensional model.}
    \label{fig:upflows_post_fault}
\end{figure}

Figure \ref{fig:upflows_post_fault} shows the magnitude and location of the mass flux at the base of particles sampled from the approximation to the posterior characterised using EKI. These are significantly less variable than the samples from the prior (see Fig.~\ref{fig:upflows_pri_fault}) and show a high degree of similarity to the flux at the base of the true system (see Fig.~\ref{fig:truth_fault}).

\begin{figure}
    \centering
    \includegraphics[width=0.5\linewidth]{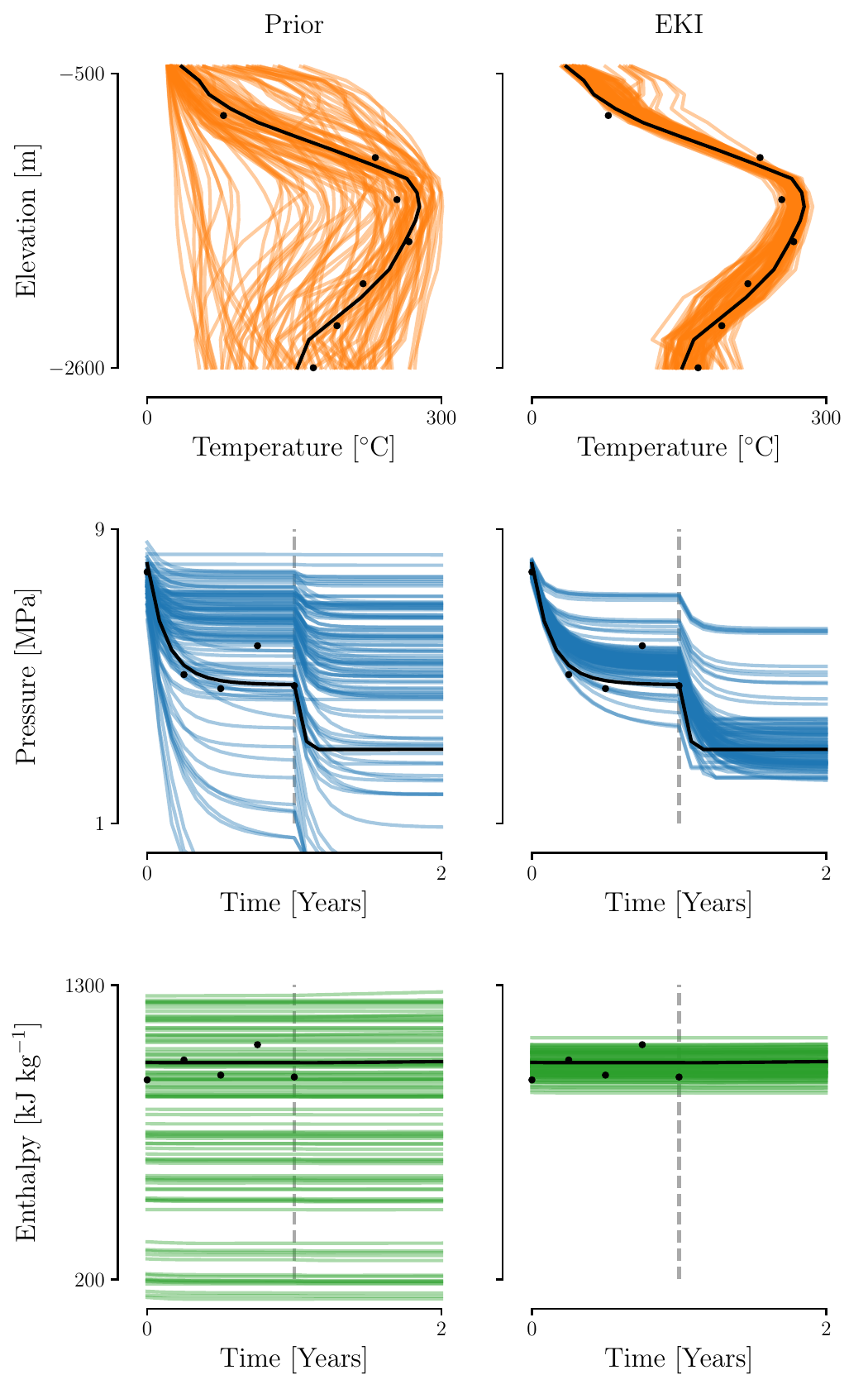}
    \caption{The modelled downhole temperature profiles (\emph{top row}) and production pressures (\emph{middle row}) and enthalpies (\emph{bottom row}) at well 3 associated with particles sampled from the prior (\emph{left}) and approximation to the posterior characterised using EKI (\emph{right}) for the synthetic three-dimensional model. In the pressure and enthalpy plots, the dashed grey line denotes the end of the data collection period.}
    \label{fig:preds_fault}
\end{figure}

Figure \ref{fig:preds_fault} shows the posterior predictions, generated using the final EKI ensemble, of the temperatures, pressures and enthalpies at well 3. The predictions of the prior ensemble are also shown for comparison. In all cases, the posterior uncertainty is significantly reduced compared to the prior uncertainty, and the true system state is well contained within the support of the ensemble predictions. We note, however, that the uncertainty in the modelled pressures remains slightly greater than we would expect given the magnitude of the observation error; there are a number of particles for which the modelled pressure is consistently far greater than both the true pressure and the observations.  

\begin{figure}
    \centering
    \includegraphics[width=0.5\linewidth]{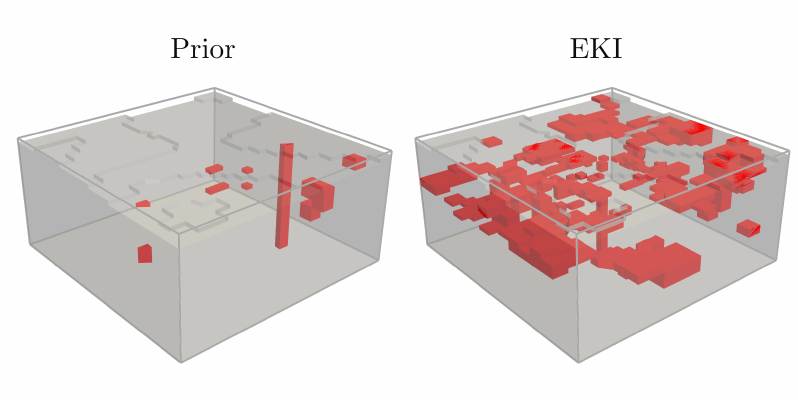}
    \caption{Heatmaps which indicate, for the prior and approximation to the posterior generated using EKI, whether the true value of the permeability in each cell of the model mesh is contained within the central 95 percent of permeabilities of the ensemble. Red cells are cells for which the permeability is not contained within the central 95 percent of the ensemble.}
    \label{fig:intervals_fault}
\end{figure}

Finally, Figure \ref{fig:intervals_fault} shows heatmaps which indicate, for each cell in the model mesh, whether the true permeability of the cell is contained within the central 95 percent of the posterior ensemble generated using EKI. The corresponding heatmap for the prior ensemble is also shown for comparison. The permeabilities of 99.6 percent of the cells are contained within the central 95 percent of the prior ensemble, while the permeabilities of 94.7 percent of the cells are contained within the support of the posterior ensemble; as in the vertical slice case, application of EKI reduces the uncertainty in the permeability structure of the system significantly without discounting the true permeabilities.

\subsection{Real Volcanic Geothermal System} 
\label{sec:results_vol}

Applying EKI to the volcanic geothermal system requires some additional considerations. Because the discretisation of the synthetic models in this study is significantly coarser than that of this model, they tend to converge to a steady state quickly, regardless of the parameters and initial condition used. For this reason, we simply initialise these simulations with the entire reservoir at atmospheric pressure and temperature. By contrast, the amount of computation for the volcanic model to converge to a steady-state solution is highly dependent on the choice of initial condition; if a poor choice is made, the model requires an excessively long time to converge, or even fails to converge entirely as a result of the simulation time-step reducing to increasingly small values to handle the complexity of the resulting reservoir dynamics.

To increase the convergence rate of simulations throughout the EKI algorithm, we first sample a set of particles from the prior and simulate the reservoir dynamics for each of these using an arbitrary initial condition. We then run each failed simulation again. This time, however, we use an initial condition sampled randomly from the final states of the converged simulations, in the hope that these will be similar to the steady states of the failed particles. For subsequent iterations of EKI, we initialise particles that were simulated successfully at the previous iteration with the steady state solution found during the previous iteration, while we initialise the resampled particles (used to replace the particles that failed at the previous iteration) with a randomly-chosen steady state solution associated with a successful particle. This method of selecting the initial conditions for each particle reduces the failure rate significantly in comparison to using an arbitrary initial condition. However, even when running each simulation with parallel processing, using $80$ CPUs for up to 45 minutes, we still encounter a large number of simulation failures. At the first iteration of EKI, approximately $72$ percent of simulations fail, and at subsequent iterations, between $20$ and $50$ percent of simulations fail. For this reason, we use an ensemble of $\NumEns = 200$ particles, rather than the $\NumEns = 100$ used in the previous case studies. Our implementation of EKI converges in $4$ iterations, requiring a total of $1000$ simulations.

\begin{figure}
    \centering
    \includegraphics[width=0.35\linewidth]{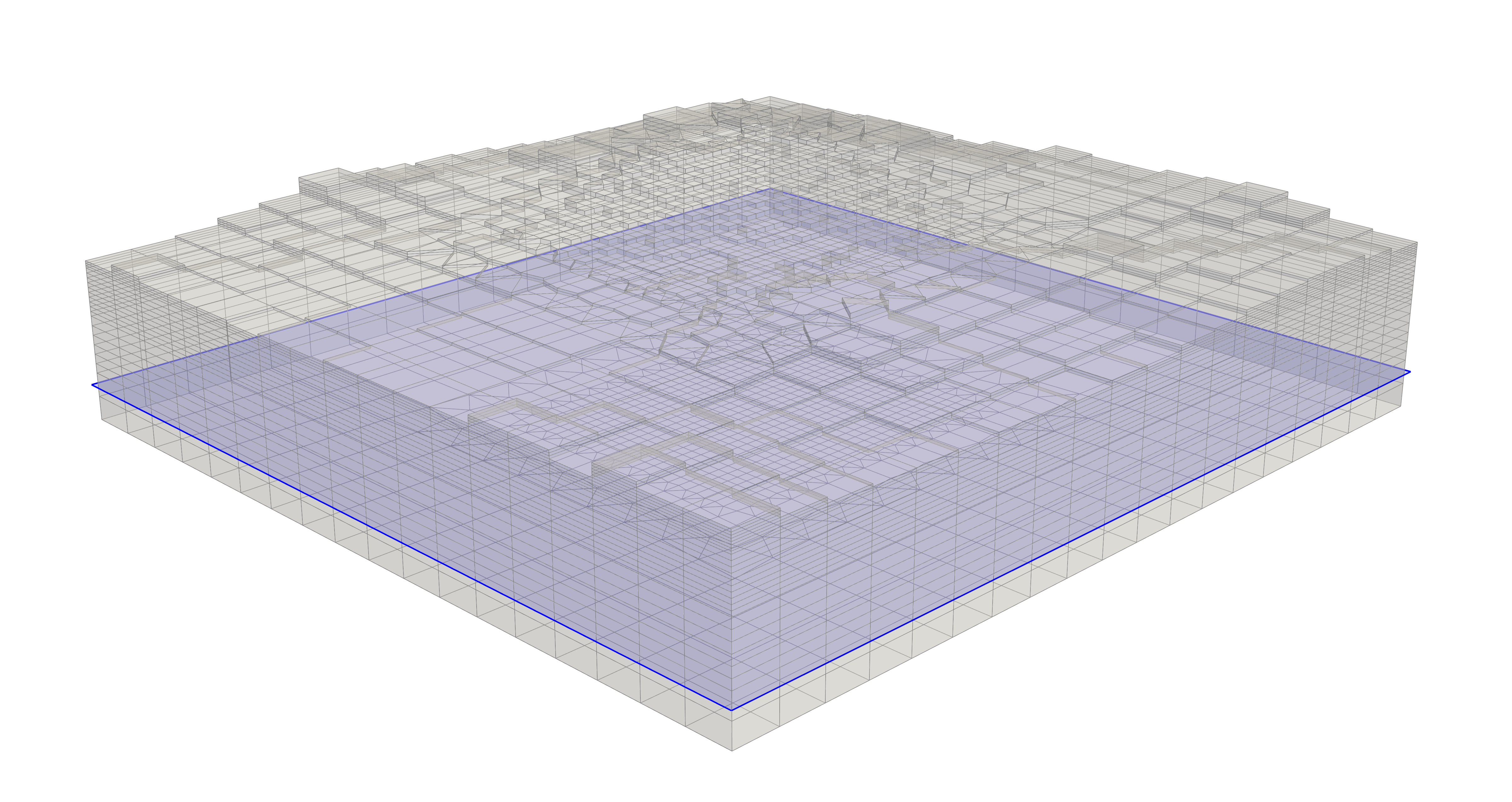} \\
    \includegraphics[width=0.5\linewidth]{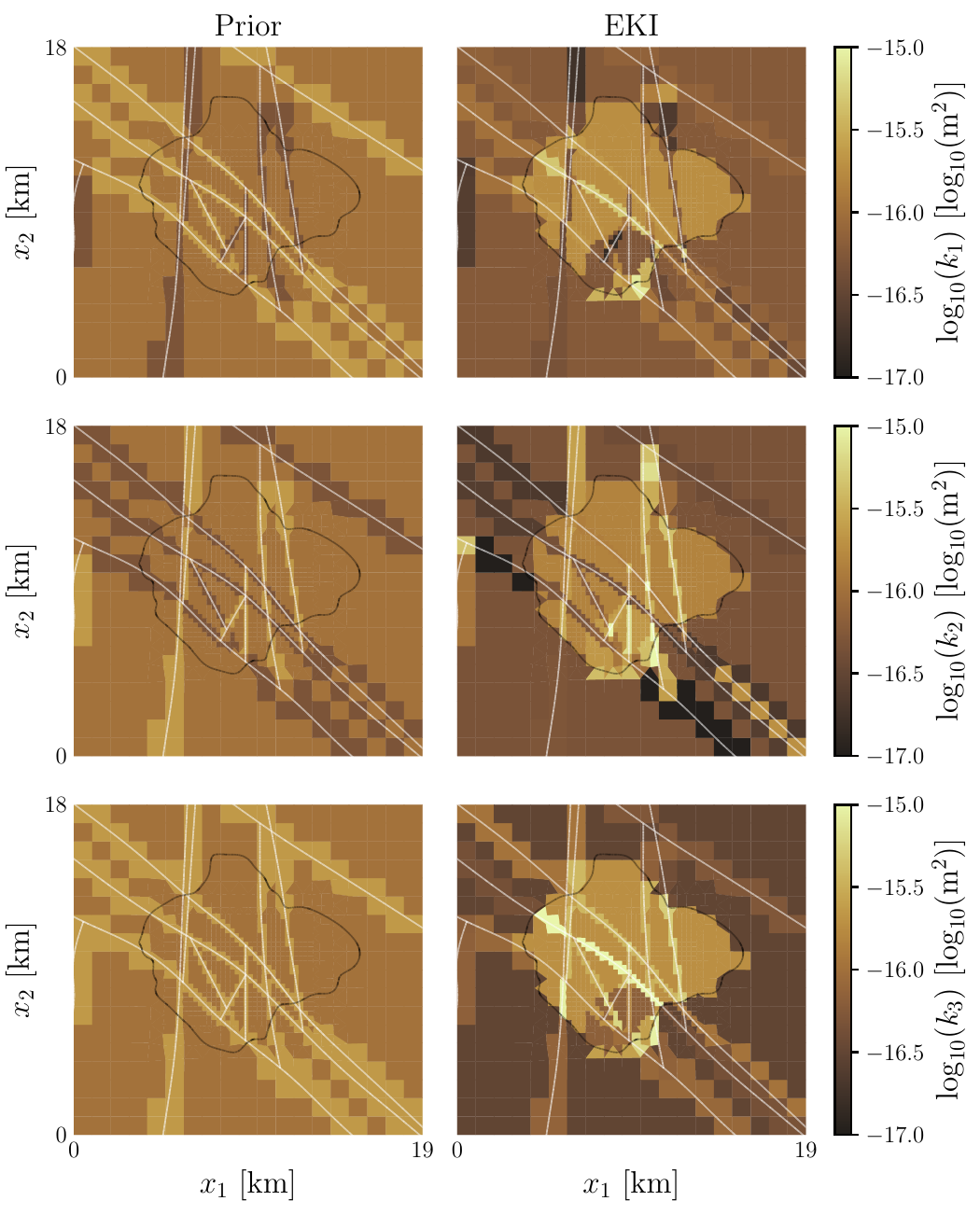}
    \caption{The prior mean (\emph{left}) and the mean of the approximation to the posterior characterised using EKI (\emph{right}) of the permeability in each direction of the horizontal cross-section indicated by the blue plane at the top of the figure. The white lines in each plot indicate faults, and the black line denotes the lateral extent of the clay cap.}
    \label{fig:perms_bottom_vol}
\end{figure}

Figure \ref{fig:perms_bottom_vol} shows the prior mean, and the mean of the approximation to the posterior generated using EKI, of the permeability of the system in a horizontal cross-section near the bottom of the model domain. As expected, we observe that the interior of the reservoir (i.e., the region beneath the lateral extent of the clay cap) appears to be, in general, more permeable than the exterior; while our choice of prior parametrisation models each region as similarly permeable, the posterior mean of each of the components of the permeability tensor in all three directions exceeds the prior mean in most regions inside the reservoir, but is less than the prior mean in most regions outside the reservoir. In addition, we observe that the posterior mean of the permeability of many of the fault structures of the system is significantly different to the prior mean.

\begin{figure}
    \centering
    \includegraphics[width=0.35\linewidth]{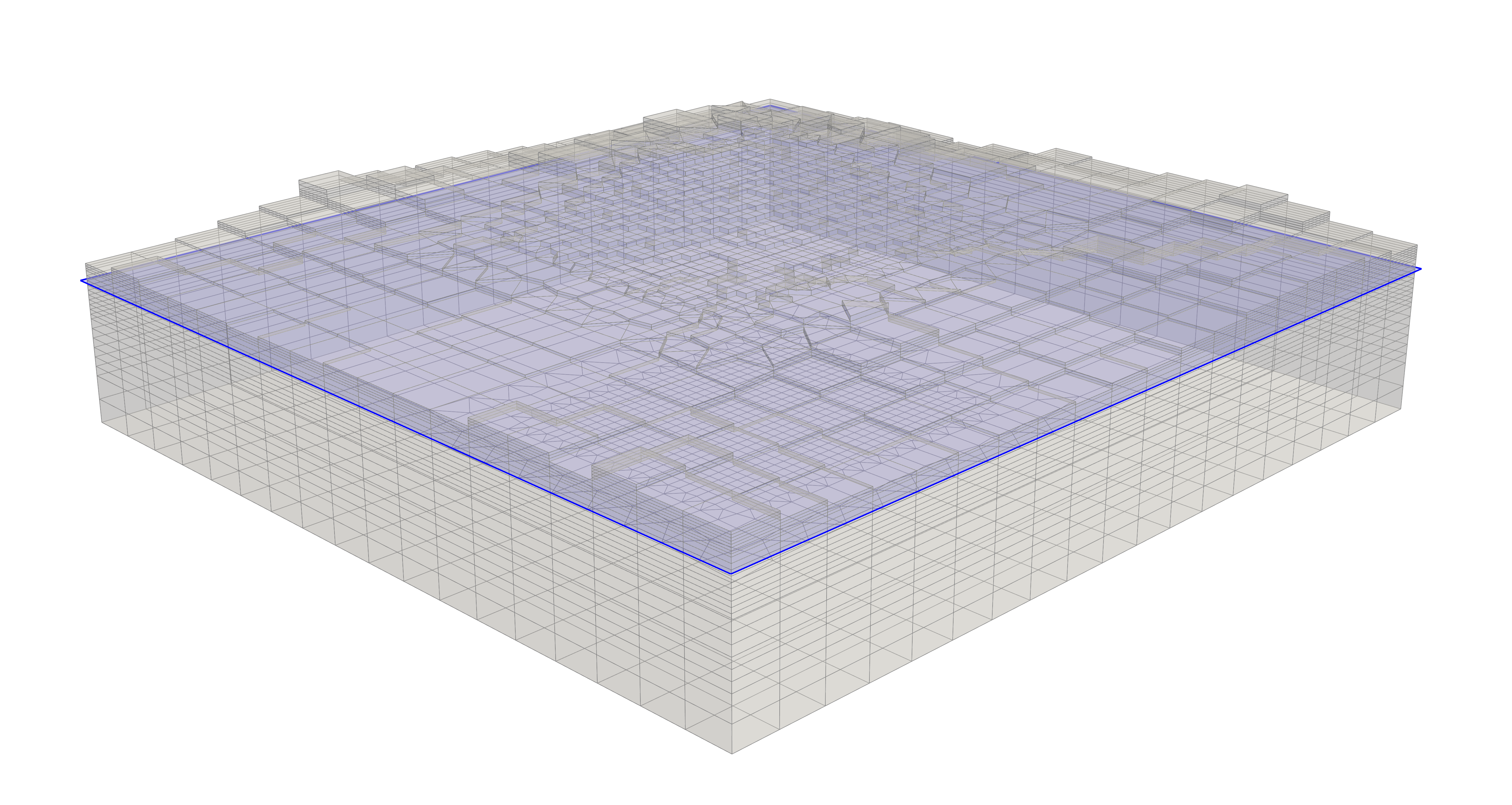} \\
    \includegraphics[width=0.5\linewidth]{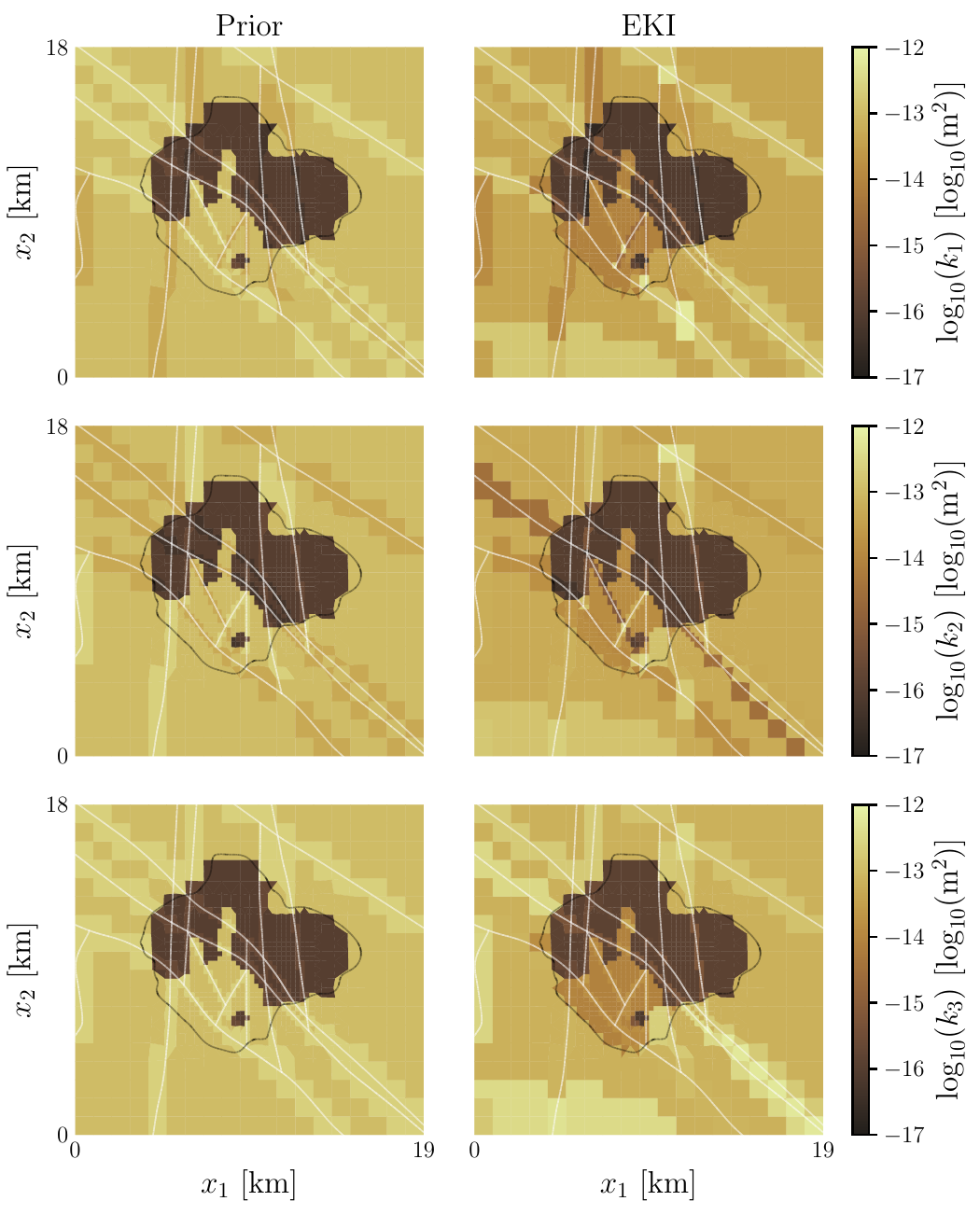}
    \caption{The prior mean (\emph{left}) and the mean of the approximation to the posterior characterised using EKI (\emph{right}) of the permeability in each direction of the horizontal cross-section indicated by the blue plane at the top of the figure. The white lines in each plot indicate faults, and the black line denotes the lateral extent of the clay cap.}
    \label{fig:perms_top_vol}
\end{figure}

Figure \ref{fig:perms_top_vol} presents the same quantities as Figure \ref{fig:perms_bottom_vol}, but for a horizontal cross-section close to the surface of the system. Note that part of this cross-section is intersected by the clay cap. The mean permeability in most regions of the section of the clay cap shown in the plots appear to change little from prior to posterior. The permeability outside the reservoir, however, tends to decrease slightly on average, while the permeability inside the reservoir reduces significantly in most places. This could be indicative that in this region, the clay cap extends deeper than was estimated during the development of the conceptual model of the system.

\begin{figure}
    \centering
    \includegraphics[width=0.35\linewidth]{figures/volcanic/mesh_plane_bottom.png} \\
    \includegraphics[width=0.5\linewidth]{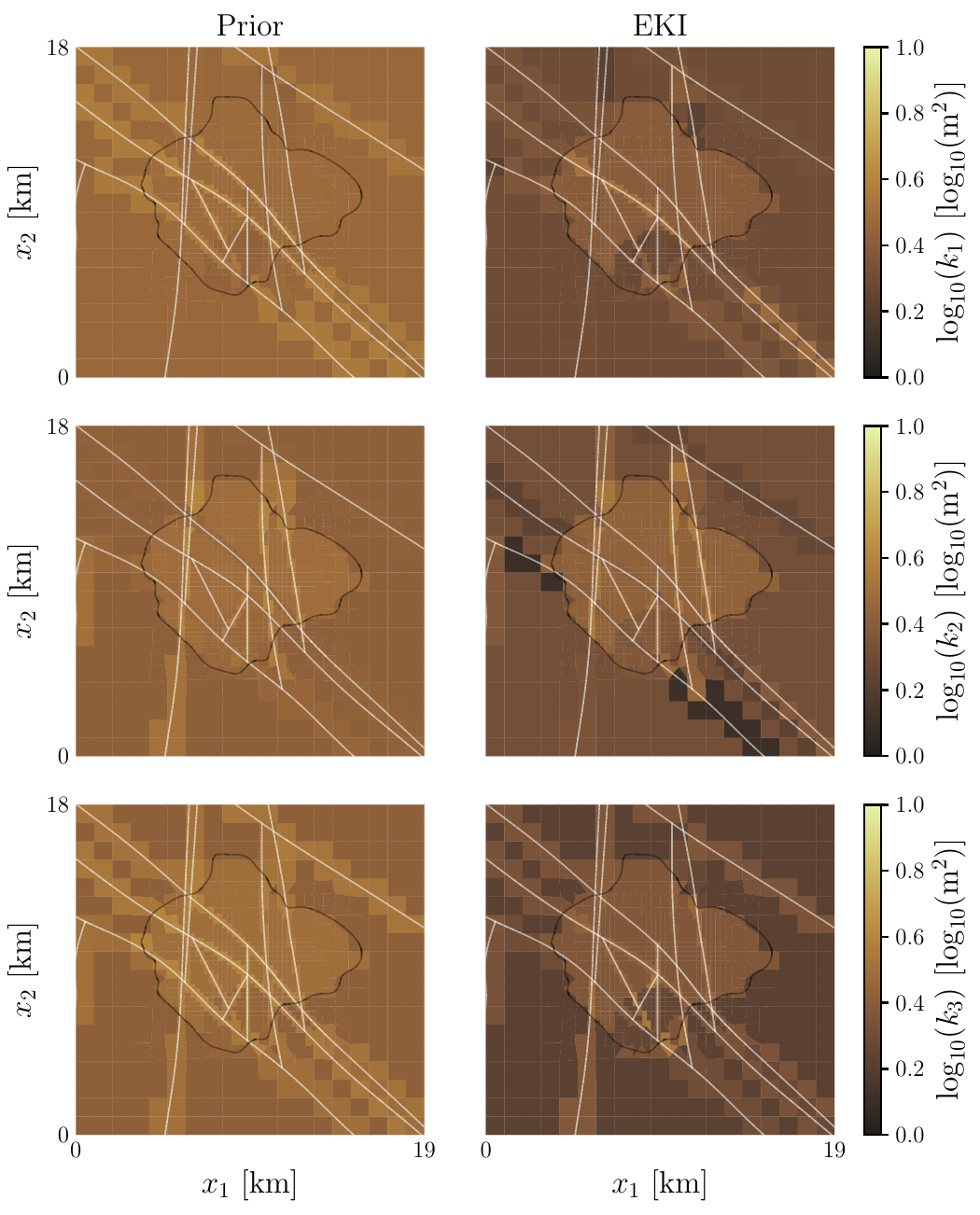}
    \caption{The prior standard deviations (\emph{left}) and the posterior standard deviations characterised using EKI (\emph{right}) of the permeability in each direction of the horizontal cross-section of the volcanic system indicated by the blue plane at the top of the figure. The white lines in each plot indicate faults, and the black line denotes the lateral extent of the clay cap.}
    \label{fig:stds_bottom_vol}
\end{figure}

\begin{figure}
    \centering
    \includegraphics[width=0.35\linewidth]{figures/volcanic/mesh_plane_top.png} \\
    \includegraphics[width=0.5\linewidth]{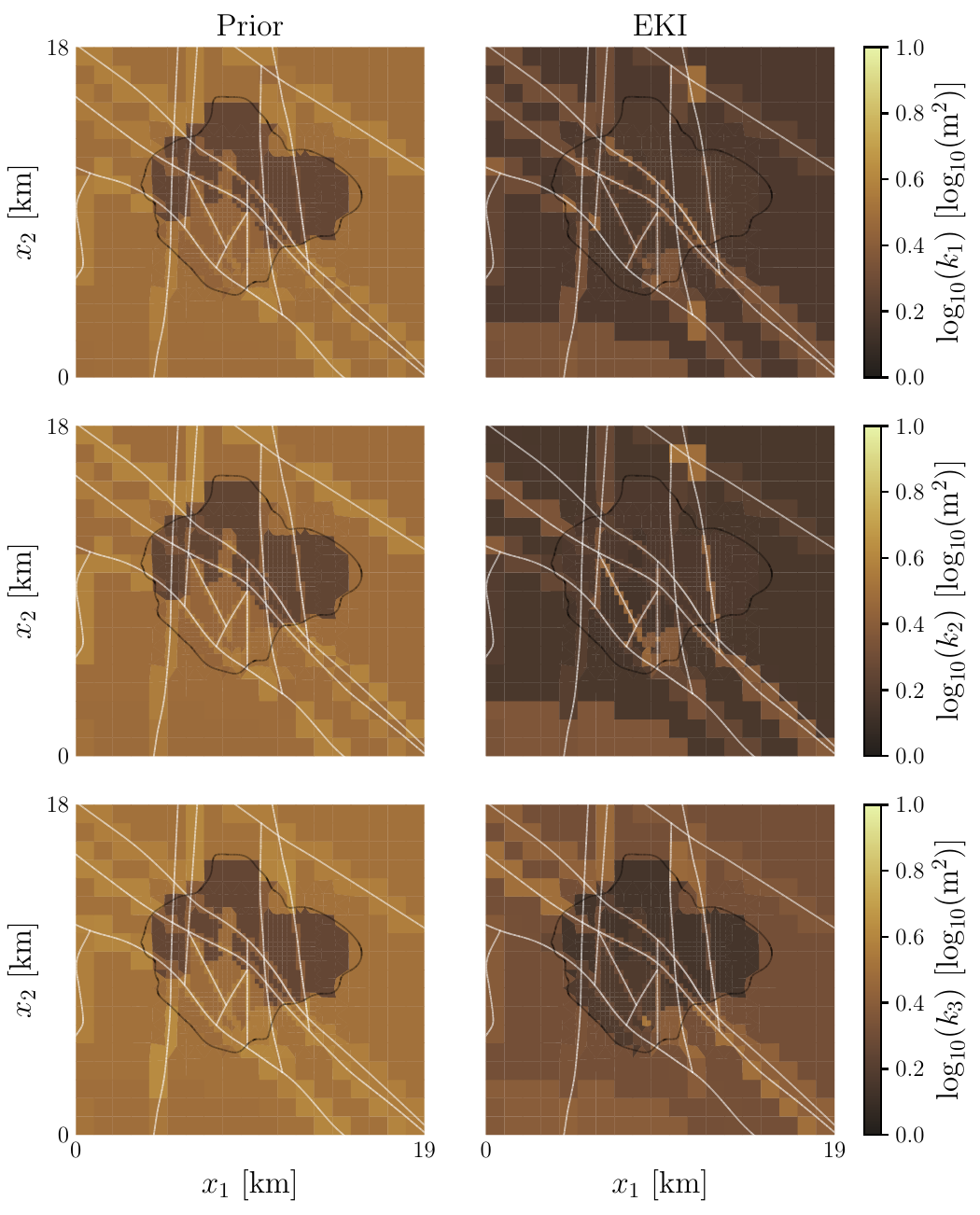}
    \caption{The prior standard deviations (\emph{left}) and the posterior standard deviations characterised using EKI (\emph{right}) of the permeability in each direction of the horizontal cross-section of the volcanic system indicated by the blue plane at the top of the figure. The white lines in each plot indicate faults, and the black line denotes the lateral extent of the clay cap.}
    \label{fig:stds_top_vol}
\end{figure}

Figure \ref{fig:stds_bottom_vol} shows the prior and posterior standard deviations of the permeability in each direction for the cross-section plotted in Figure \ref{fig:perms_bottom_vol}. We observe that there is a reasonable reduction in the uncertainty of the permeability in all directions across most parts of the system after applying EKI. This reduction is, in general, greater in the region outside the reservoir than in the region contained within reservoir. Figure \ref{fig:stds_top_vol} shows the same quantities for the cross-section plotted in Figure \ref{fig:perms_top_vol}. Again, there is a significant reduction in the uncertainty of the permeability in all directions across most parts of the system, with the exception of the clay cap. Interestingly, the reduction in uncertainty of the vertical component of the permeability outside the reservoir is, in most places, less than that of the horizontal components.

\begin{figure}
    \centering
    \includegraphics[width=0.5\linewidth]{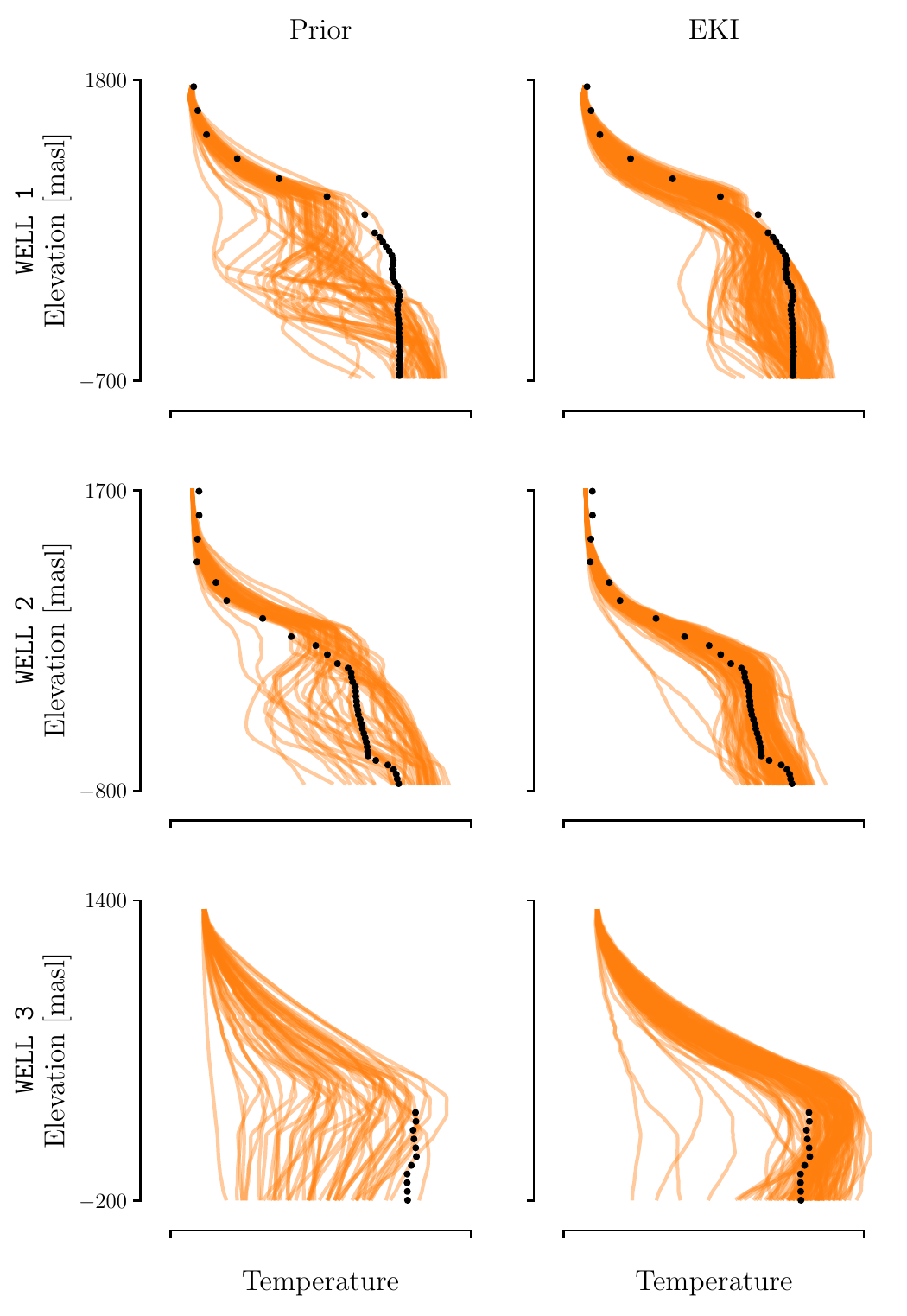}
    \caption{The modelled downhole temperature profiles at each well associated with particles sampled from the prior (\emph{left}) and the EKI approximation to the posterior (\emph{right}) for the volcanic system. In all plots, the black dots denote the observations. In all plots, the black dots denote the observations. Note that the units of the observations have been redacted for confidentiality reasons.}
    \label{fig:preds_temp_vol}
\end{figure}

\begin{figure}
    \centering
    \includegraphics[width=0.5\linewidth]{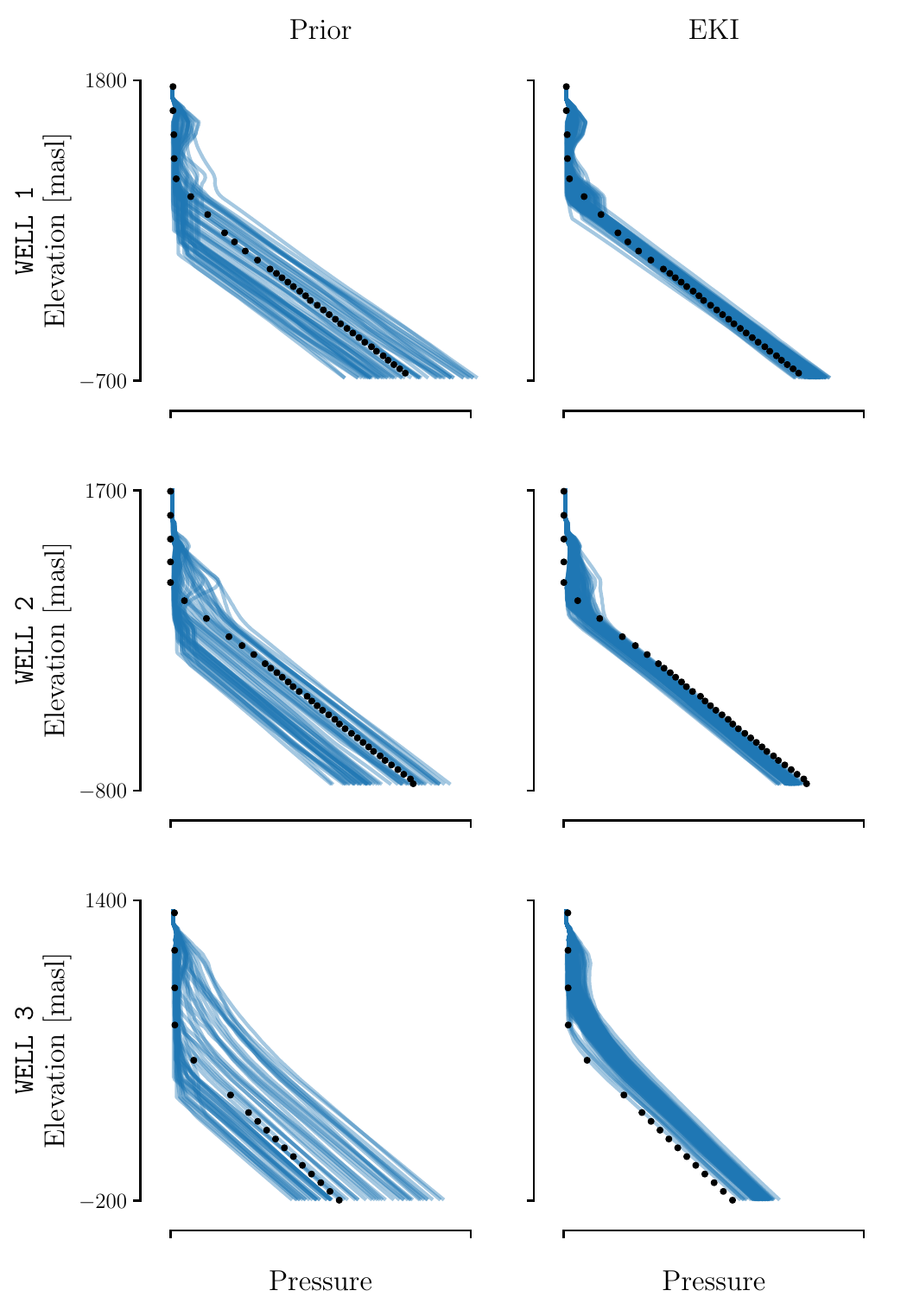}
    \caption{The modelled downhole pressure profiles at each well associated with particles sampled from the prior (\emph{left}) and the EKI approximation to the posterior (\emph{right}) for the volcanic system. In all plots, the black dots denote the observations. Note that the units of the observations have been redacted for confidentiality reasons.}
    \label{fig:preds_pres_vol}
\end{figure}

Finally, Figures \ref{fig:preds_temp_vol} and \ref{fig:preds_pres_vol} show the predicted downhole temperature and pressure profiles corresponding to particles sampled from the prior and the EKI approximation to the posterior. In all cases, the posterior predictions show a significant reduction in uncertainty and appear to fit the data well.

\section{Conclusions and Outlook} \label{sec:conclusions}

In this study, we have demonstrated that ensemble Kalman inversion is an efficient technique for obtaining estimates of the parameters of large-scale geothermal reservoir models, as well as measures of the uncertainty in these estimates. To do this, we have considered a variety of case studies, including a large-scale, real-world reservoir model with real data. In all cases, EKI required significantly less computation than conventional techniques for parameter estimation and uncertainty quantification; in each of the computational experiments reported on in this work, the algorithm required $\mathcal{O}(10^{3})$ simulations of the forward model. We have also illustrated how EKI can be used in combination with a variety of prior parametrisations that provide a realistic representation of geophysical characteristics, and the robustness of the method to simulation failures. It is worth noting that many of the elements of our framework, such as the use of variable transformations to enforce constraints and geological realism, and the idea of using resampling to replace the particles corresponding to failed simulations, can be used with ensemble methods other than EKI. These include those used for state estimation, such as the ensemble Kalman filter.

We reiterate that, in contrast to conventional sampling techniques such as Markov chain Monte Carlo and sequential Monte Carlo, EKI will not produce samples from the posterior when applied to general inverse problems (i.e., outside the linear-Gaussian setting), even in the limit of an infinite ensemble size. Nevertheless, the empirical results we have obtained here appear promising and suggest that EKI should be further studied as a tool for solving inverse problems arising in geothermal modelling.

\subsection{Future Work}

We have identified a variety of avenues for future work which could be investigated. 

It would be valuable to study the sensitivity of the EKI algorithm, using synthetic case studies such as those we have presented in this work, to factors including the number of observations, the size of the observation errors, and the ensemble size. It would also be valuable to study how the EKI estimate of the posterior changes if the set of samples from the prior used to form the initial ensemble is altered, and to investigate how to algorithm performs when the true values of the model parameters are located in a region of low prior probability. While the sensitivity of the EKI algorithm to some of these factors has been investigated in other areas of application \citep[see, e.g.,][]{Iglesias22, Tso24}, it would be of interest to study whether similar results are obtained in a geothermal setting.

An important component of our EKI-based framework for geothermal inverse problems is the resampling of particles corresponding to failed simulations. In particular, the choice of the parameter $\delta$ could, as discussed in Section \ref{sec:resampling}, have a significant effect on the approximation to the posterior generated using EKI in situations where simulations fail frequently. It would be valuable to investigate the effect of changing $\delta$ numerically and to develop an principled approach to the selection of $\delta$ for general inverse problems.

A key benefit of the numerical modelling of geothermal reservoirs is the ability of these models, once calibrated, to estimate the potential power output of a geothermal field, or to predict how the field will respond under potential future management scenarios \citep{Dekkers22, Dekkers23, OSullivan16, OSullivan23}. It would be of interest to use the particles from the EKI approximation to the posterior associated with a reservoir model as part of a resource assessment or future scenarios study, and to compare these results to those obtained using alternative techniques for parameter estimation and uncertainty quantification.

A significant challenge we encountered when applying EKI to the large-scale volcanic geothermal system was in reducing the number of natural state simulations which failed to converge. In particular, the failure rate was highest for the particles of the initial EKI ensemble, which were sampled from the prior; the failure rate was lower for subsequent iterations of EKI as the fit of the ensemble to the data began to improve. This phenomenon has also been observed in other studies \citep{Maclaren20}. Some ensemble methods, such as the ensemble Kalman sampler \citep{Garbuno20a, Garbuno20b}, do not require that the ensemble is initialised at the prior; any initial distribution can be used. It would be valuable to investigate whether the use of these alternative methods in combination with a well-chosen initial distribution could help to reduce the number of simulations that fail, potentially allowing us to incorporate additional uncertain parameters into the inversion.

While this work has illustrated a range of prior parametrisations for geothermal reservoir models, we have paid comparatively less attention to the specification of the likelihood. In each of our case studies, we have treated the errors associated with the data as independent and normally distributed, with known variances. In practice, however, there is generally significant uncertainty in the variances of the errors, and the errors associated with data collected at the same well are often correlated \citep{Maclaren22}. Variants of EKI that allow for characteristics of the errors to be treated as uncertain have been developed \citep{Botha23}; in future, it could be useful to apply these algorithms to the geothermal problems we have considered here, and to study the effect of treating the errors as uncertain on the resulting solution to the inverse problem. In particular, we anticipate that treating the errors associated with measurements made down a given well as correlated will result in a greater number of the particles of the final EKI ensemble with corresponding predictions that are offset from the observations but look similar otherwise, rather than predictions that lie close to the observations but look structurally quite different; the former kind of predictions are often more plausible in a geothermal setting.

Another possible extension to the modelling framework discussed in this paper is the pairing of ensemble methods with surrogate models. In particular, \citet{Cleary21} introduce a framework, referred to as calibrate, emulate, sample (CES), in which the outputs of the forward model from an initial run of an ensemble method are used to train a surrogate model, such as a Gaussian process or neural network, which can be applied in combination with a conventional sampling method, such as MCMC, to characterise an approximation to the posterior. In this setting, the ensemble method is viewed as an efficient technique to obtain sets of training points for the emulator distributed in regions of high posterior probability. It would be valuable to apply the CES methodology to the case studies we have used in this work, and to evaluate whether the use of a surrogate model provides improved results.

\section*{Acknowledgements}

The authors wish to acknowledge the use of New Zealand eScience Infrastructure (NeSI; \url{www.nesi.org.nz}) high performance computing facilities as part of this research. New Zealand's national facilities are provided by NeSI and funded jointly by NeSI's collaborator institutions and through the Ministry of Business, Innovation and Employment's Research Infrastructure programme.

\section*{Data Availability}

All models and code required to replicate the results of the synthetic case studies reported on in the paper are archived on Zenodo \citep{deBeer24b}, and are also available on GitHub (\url{https://github.com/alexgdebeer/GeothermalEnsembleMethods}) under the MIT license. The model and data for the volcanic geothermal system are commercially sensitive and so cannot be made available.

\appendix

\section{Level Set Mappings}

For completeness, this appendix summarises the level set mappings used to parametrise the permeability structure of the synthetic case studies reported on in the paper.

\subsection{Vertical Slice Model} \label{sec:level_set_slice}

The level set mapping that defines the permeability within the shallow region, $\Domain_{\Shal}$, of the vertical slice model, is given by
\begin{equation}
    \PermField_{\Shal}(\bm{x}) = \begin{cases}
        10^{-15.0}\,\metres^{2}, & \quad \phantom{+}\LevelFunc_{\Shal}(\bm{x}) < -1.5, \\
        10^{-14.5}\,\metres^{2}, & \quad -1.5 \leq \LevelFunc_{\Shal}(\bm{x}) < -0.5, \\
        10^{-14.0}\,\metres^{2}, & \quad -0.5 \leq \LevelFunc_{\Shal}(\bm{x}) < 0.5, \\
        10^{-13.5}\,\metres^{2}, & \quad \phantom{+}0.5 \leq \LevelFunc_{\Shal}(\bm{x}) < 1.5, \\
        10^{-13.0}\,\metres^{2}, & \quad \phantom{+}1.5 \leq \LevelFunc_{\Shal}(\bm{x}).
    \end{cases}
\end{equation}
The level set mapping for the deep region, $\Domain_{\Deep}$, is defined in an identical manner.

\subsection{Synthetic Three-Dimensional Model} \label{sec:level_set_fault}

The level set mapping that defines the permeability within the clay cap region, $\Domain_{\ClayCap}$, of the synthetic three-dimensional model, is given by

\begin{equation}
    \PermField_{\ClayCap}(\bm{x}) = \begin{cases}
        10^{-17.0}\,\metres^{2}, & \quad \phantom{+}\LevelFunc_{\ClayCap}(\bm{x}) < -0.5, \\
        10^{-16.5}\,\metres^{2}, & \quad -0.5 \leq \LevelFunc_{\ClayCap}(\bm{x}) < 0.5, \\
        10^{-16.0}\,\metres^{2}, & \quad \phantom{+}0.5 \leq \LevelFunc_{\ClayCap}(\bm{x}).
    \end{cases}
\end{equation}
The level set mapping that defines the permeability within the fault region, $\Domain_{\Fault}$, is given by
\begin{equation}
    \PermField_{\Fault}(\bm{x}) = \begin{cases}
        10^{-13.5}\,\metres^{2}, & \quad \phantom{+}\LevelFunc_{\Fault}(\bm{x}) < -0.5, \\
        10^{-13.0}\,\metres^{2}, & \quad -0.5 \leq \LevelFunc_{\Fault}(\bm{x}) < 0.5, \\
        10^{-12.5}\,\metres^{2}, & \quad \phantom{+}0.5 \leq \LevelFunc_{\Fault}(\bm{x}).
    \end{cases}
\end{equation}
Finally, the level set mapping that defines the permeability within the background region, $\Domain_{\Background}$, is given by
\begin{equation}
    \PermField_{\Background}(\bm{x}) = \begin{cases}
        10^{-15.5}\,\metres^{2}, & \quad \phantom{+}\LevelFunc_{\Background}(\bm{x}) < -1.5, \\
        10^{-15.0}\,\metres^{2}, & \quad -1.5 \leq \LevelFunc_{\Background}(\bm{x}) < -0.5, \\
        10^{-14.5}\,\metres^{2}, & \quad -0.5 \leq \LevelFunc_{\Background}(\bm{x}) < 0.5, \\
        10^{-14.0}\,\metres^{2}, & \quad \phantom{+}0.5 \leq \LevelFunc_{\Background}(\bm{x}) < 1.5, \\
        10^{-13.5}\,\metres^{2}, & \quad \phantom{+}1.5 \leq \LevelFunc_{\Background}(\bm{x}).
    \end{cases}
\end{equation}

\renewcommand*{\mkbibnamefamily}{\textsc}
\printbibliography

\end{document}


\maketitle

This supplement provides additional results for the synthetic reservoir models studied in the main body of the paper. Section \ref{sec:loc_inf} presents results for each model when EKI is used in combination with localisation and inflation techniques. Section \ref{sec:hyperparams} presents a selection of the posterior densities of the hyperparameters of the Whittle-Mat{\'e}rn fields used as part of the prior parametrisations of these models.

\section{Localisation and Inflation} \label{sec:loc_inf}

In this section, we demonstrate the effect of using localisation and inflation techniques on the performance of the EKI algorithm. We first outline the techniques we apply, before presenting the results of each technique applied to each synthetic reservoir model.

We note that the use of these techniques results in a slight increase in the amount of computation required to update the particles of the ensemble at each iteration of EKI. In our case studies, however, this is negligible in comparison to the cost of applying the forward model to the particles of the ensemble at each iteration. 

\subsection{Localisation}

The key idea behind localisation is to reduce the effect of spurious correlations when updating the particles of the ensemble at each iteration, by modifying the ensemble covariance matrices, $\bCov_{\Param\ForwardModel}$ and $\bCov_{\ForwardModel\ForwardModel}$, or the so-called \emph{Kalman gain}, $\bm{K} \in \mathbb{R}^{\ParamDim\times\DataDim}$, which, in the case of EKI, is defined by
\begin{equation}
    \bm{K}^{(i)} \defas \bCov_{\Param\ForwardModel}^{(i)}\left(\bCov_{\ForwardModel\ForwardModel}^{(i)} + \alpha^{(i)}\bCov_{\Error}\right)^{-1}. \label{eq:gain}
\end{equation}
The EKI update equation for particle $j$ can, using Equation \eqref{eq:gain}, be written as 
\begin{equation}
    \bParam^{(i+1)}_{j} = \bParam_{j}^{(i)} + \bm{K}^{(i)}\left(\bData + \bError^{(i)}_{j} - \ForwardModel(\bParam_{j}^{(i))}\right). \label{eq:eki_update_gain}
\end{equation}
From Equation \eqref{eq:eki_update_gain}, it is clear that element $\bm{K}_{rs}^{(i)}$ of the Kalman gain dictates how the difference between the $r$th observation and model prediction influences the update of parameter $s$. At each iteration, localisation replaces the ensemble estimates of the covariance matrices, or the Kalman gain, by their localised estimates, defined as
\begin{equation}
    \tilde{\bCov}^{(i)}_{\Param\ForwardModel} \defas \bLocMat^{(i)}_{\Param\ForwardModel} \odot \bCov^{(i)}_{\Param\ForwardModel}, \quad \tilde{\bCov}^{(i)}_{\ForwardModel\ForwardModel} \defas \bLocMat^{(i)}_{\ForwardModel\ForwardModel} \odot \bCov^{(i)}_{\ForwardModel\ForwardModel}, \quad \tilde{\bGainMat}^{(i)} \defas \bLocMat^{(i)}_{\GainMat} \odot \bGainMat^{(i)},
\end{equation}
where $\odot$ denotes the Schur (elementwise) product, and $\bLocMat^{(i)}_{\Param\ForwardModel} \in \mathbb{R}^{\ParamDim\times\DataDim}$, $\bLocMat^{(i)}_{\ForwardModel\ForwardModel} \in \mathbb{R}^{\DataDim\times\DataDim}$, and $\bLocMat^{(i)}_{\GainMat} \in \mathbb{R}^{\ParamDim\times\DataDim}$ are \emph{localisation matrices}; that is, matrices with entries that range between $0$ and $1$ depending on the degree of the expected correlations between the quantities they relate. 

Conventional localisation techniques enforce the prior belief that the magnitude of the correlations between quantities should reduce with distance \citep[see, e.g.,][]{Chen17, Houtekamer01, Schillings17, Tong23}. An alternative class of localisation methods, however, aim to estimate suitable localisation matrices without the use of information on the distances between model parameters and observations \citep[see, e.g.,][]{Anderson07, Flowerdew15, Lee21, Luo20, Morzfeld23, Zhang10}. Techniques from the latter class are more amenable to our EKI-based framework for geothermal reservoir modelling, for several reasons. The parameters we aim to estimate using EKI---for example, the white-noise representation of Whittle-Mat{\'ern} fields outlined in Section 2.4.1 in the main body of the paper, as well as the hyperparameters of these fields---are not associated with physical locations. Additionally, the fundamental assumption behind distance-based localisation (namely, that correlations between quantities reduce with distance) is unlikely to hold in a geothermal setting \citep{Bjarkason21a}; the deep mass upflow and the permeability structure of the system influence the shape of the natural state convective plume that forms \citep{OSullivan16}, meaning that the strength of the relationships between these parameters and the model predictions does not necessarily reduce with distance. 

In this supplement, we trial a commonly-used localisation method proposed by \citet{Zhang10}, which uses a resampling procedure to generate an estimate of a localisation matrix for the Kalman gain. In future, however, it would be valuable to conduct a systematic review and comparison of the many localisation techniques that are available. The method of \citet{Zhang10} proceeds by first generating a set of $\NumBoot$ bootstrapped realisations of the Kalman gain, $\{\hat{\GainMat}^{(i, b)}\}_{b=1}^{\NumBoot}$, by resampling with replacement from the current ensemble. In the present work, we use a value of $\NumBoot=50$. Then, element $\LocMat_{rs}^{(i)}$ (dropping the subscript $K$ for readability) of the localisation matrix is defined as the minimiser of the sum of the mean of the squared differences between the localised bootstrapped coefficients and the coefficient of the actual gain matrix, and a regularisation term that penalises large values of $\LocMat_{rs}^{(i)}$. That is,
\begin{equation}
    \LocMat_{rs}^{(i)} \defas \arg\,\min_{\psi}\left\{\frac{1}{\NumBoot} \sum_{b=1}^{\NumBoot}\left(\psi\hat{\GainMat}^{(i, b)}_{rs} - \GainMat^{(i)}_{rs}\right)^{2} + \left(\frac{\hat{\sigma}^{(i)}_{rs}}{\beta}\psi\right)^{2}\right\}. \label{eq:min_prob}
\end{equation}
Equation \eqref{eq:min_prob} has the approximate solution
\begin{equation}
    \LocMat_{rs}^{(i)} = \frac{1}{1 + (V_{rs}^{(i)})^{2}(1+1/\beta^{2})}, \quad \mathrm{where} \,\, V_{rs}^{(i)} \defas \frac{\hat{\sigma}^{(i)}_{rs}}{\GainMat^{(i)}_{rs}}. \label{eq:loc_mat_entry}
\end{equation}

The use of Equation \eqref{eq:loc_mat_entry} to form the entries of the localisation matrix requires a choice of the regularisation parameter, $\beta$. We use a value of $\beta=0.6$, which is suggested by \citet{Zhang10}, and is typically used in the literature \citep[see, e.g.,][]{Lacerda19, Oakley23}.

\subsection{Inflation}

The key idea behind inflation is to artificially increase the spread of the ensemble at each iteration of EKI, to correct for reduction in the variance of the ensemble that occurs as a result of sampling error \citep[see, e.g.,][]{Anderson99, Evensen09, Evensen22}. In its simplest form, the amount by which the spread of the ensemble is increased is determined through trial and error. This can, however, be computationally expensive; for this reason, a variety of methods for automatically selecting suitable inflation factors have been developed.

In this supplement, we use the adaptive inflation method introduced by \citet{Evensen09}, which uses a set of random variates to estimate the amount by which the variance of the ensemble is reduced at each iteration as a result of sampling error. The key assumption behind this procedure is that any reduction in the variance of the augmented variates, which are not correlated with the model outputs, provides an accurate estimate of the reduction in variance of the ensemble as a result of sampling error. When the ensemble is updated at the $i$th iteration of EKI, each particle is augmented with a vector of $\NumVariates$ random variates drawn from the unit normal distribution, shifted and scaled so that the ensemble mean and standard deviation of each additional variate are exactly $0$ and $1$ respectively. Then, the augmented ensemble is updated and the standard deviations of each of the updated variates, $\{\sigma^{(i+1)}_{k}\}_{k=1}^{\NumVariates}$, are computed. The inflation factor, $\InfFactor^{(i+1)}$, is then computed as
\begin{equation}
    \InfFactor^{(i+1)} = \left(\frac{1}{\NumVariates}\sum_{k=1}^{\NumVariates}\sigma^{(i+1)}_{k}\right)^{-1},
\end{equation}
and is used to adjust the updated ensemble; the inflated version of particle $j$, $\hat{\bParam}^{(i+1)}_{j}$, is defined as
\begin{equation}
    \hat{\bParam}^{(i+1)}_{j} \defas \bMean_{\Param}^{(i+1)} + \InfFactor^{(i+1)}\left(\bParam^{(i+1)}_{j} - \bMean_{\Param}^{(i+1)}\right). \label{eq:inflation}
\end{equation}
Generally, $\InfFactor^{(i+1)}$ is slightly greater than 1 (in our experience, values of $\InfFactor^{(i+1)}$ between $1.01$ and $1.05$ are typical). This means that Equation \eqref{eq:inflation} has the effect of increasing the variance of the ensemble without modifying the mean.

\subsection{Results: Vertical Slice Model}

When localisation is applied, our implementation of EKI converges in $8$ iterations, and when inflation is applied, it converges in $7$ (recall that our implementation without either technique applied converges in $7$ iterations).

\begin{figure}
    \centering
    \includegraphics[width=0.80\textwidth]{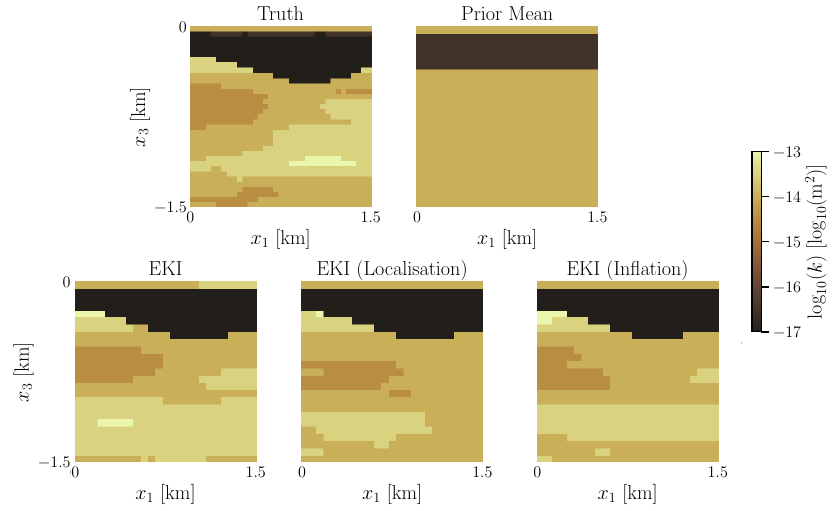}
    \caption{The true permeability structure of the vertical slice model (\emph{top left}), the prior mean (\emph{top right}), and the approximations to the posterior mean generated using EKI (\emph{bottom left}), EKI with localisation (\emph{bottom centre}), and EKI with inflation (\emph{bottom right}).}
    \label{fig:means_slice}
\end{figure}

Figure \ref{fig:means_slice} shows the approximations to the mean of the posterior permeability structure generated using EKI (without localisation or inflation applied), EKI with localisation applied, and EKI with inflation applied. All estimates look similar to one another, and show a far greater degree of similarity to the true permeability structure than the prior mean. In all cases, the permeability and location of the bottom surface of the clay cap are recovered well. The estimate of the permeability structure in the bottom third of the model domain appears to be most similar to the truth when EKI with inflation is used.

\begin{figure}
    \centering
    \includegraphics[width=1.0\textwidth]{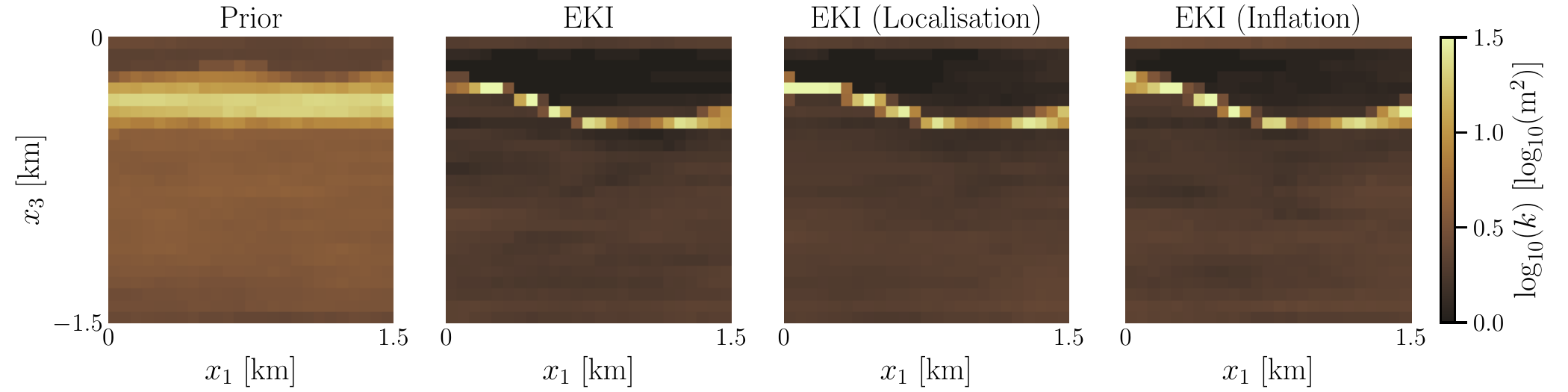}
    \caption{The prior standard deviations of the permeability structure of the vertical slice model (\emph{far left}), and the approximations to the posterior standard deviations generated using EKI (\emph{centre left}), EKI with localisation (\emph{centre right}), and EKI with inflation (\emph{far right}).}
    \label{fig:stds_slice}
\end{figure}

Figure \ref{fig:stds_slice} shows the approximations to the marginal posterior standard deviations of the permeability in each cell of the model mesh obtained using each variant of EKI. As expected, the marginal standard deviations appear to be slightly larger, on average, for EKI with localisation and EKI with inflation.

\begin{figure}
    \centering
    \includegraphics[width=1.0\textwidth]{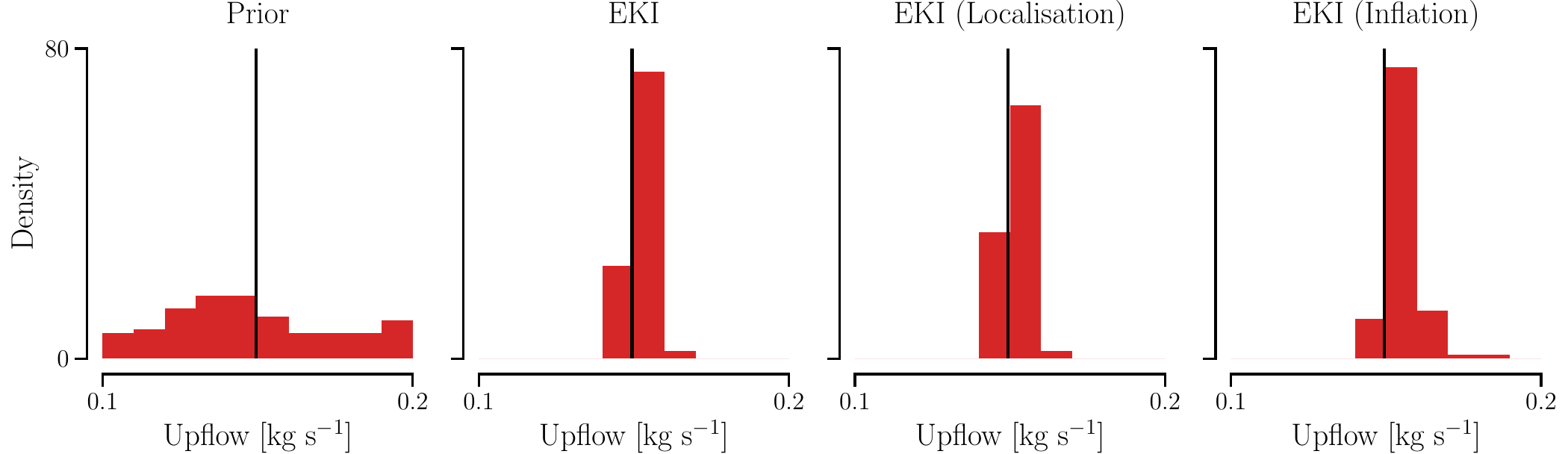}
    \caption{The prior density of the mass upflow of the vertical slice model (\emph{far left}), and the approximations to the posterior density generated using EKI (\emph{centre left}), EKI with localisation (\emph{centre right}), and EKI with inflation (\emph{far right}). The black vertical line in each plot denotes the true mass upflow.}
    \label{fig:upflows_slice}
\end{figure}

Figure \ref{fig:upflows_slice} shows the approximations to the posterior density of the mass upflow obtained using each variant of EKI. In all cases, the posterior uncertainty is significantly reduced in comparison to the prior uncertainty, and the true mass upflow rate is contained within the support of the approximation. The approximation generated using EKI with inflation applied has slightly greater variance than the approximations generated using the other forms of EKI.

\begin{figure}
    \centering
    \includegraphics[width=1.0\textwidth]{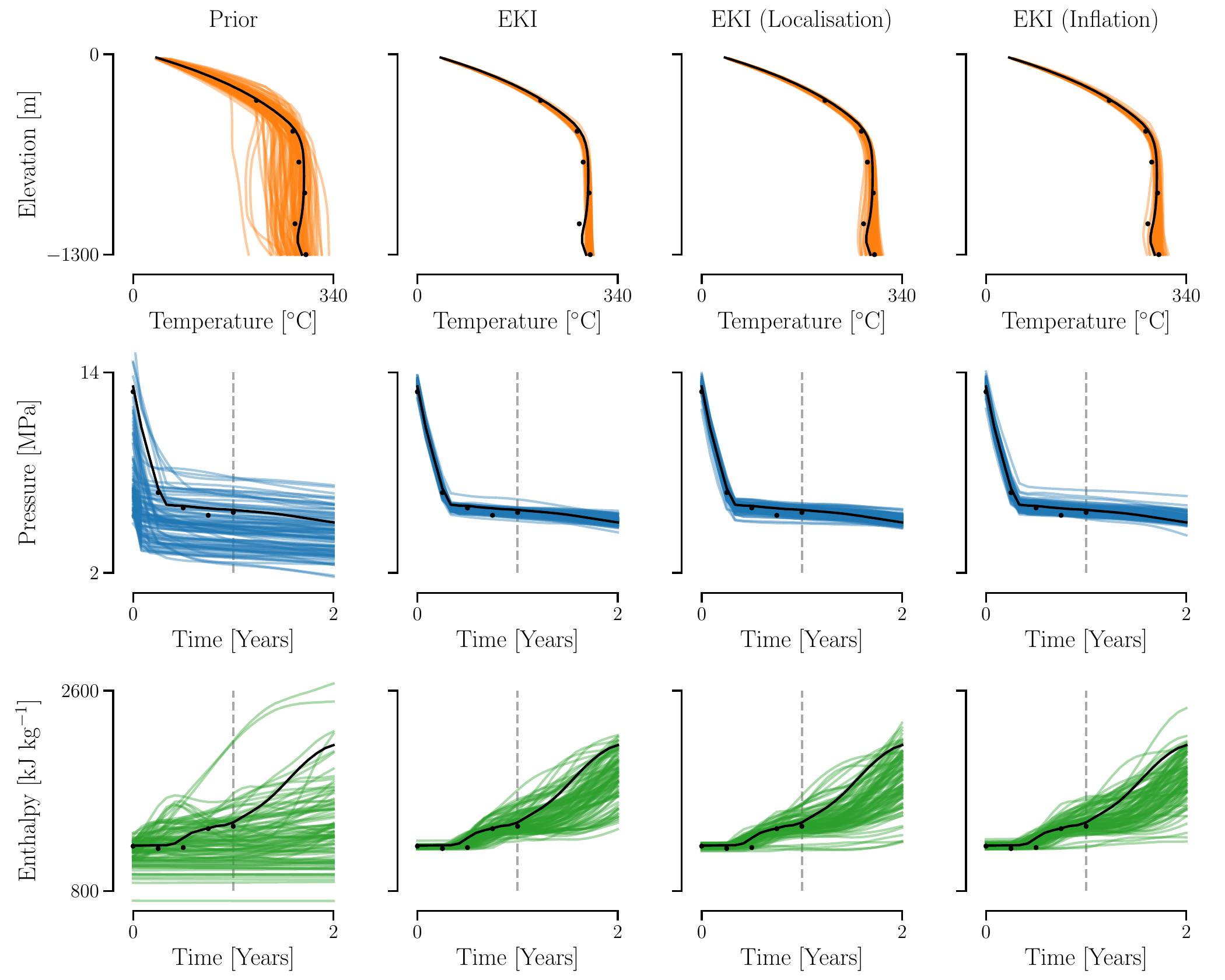}
    \caption{The predicted temperatures (\emph{top row}), pressures (\emph{middle row}), and enthalpies (\emph{bottom row}) for particles sampled from the prior (\emph{first column}) and the approximations to the posterior generated using EKI (\emph{second column}), EKI with localisation (\emph{third column}), and EKI with inflation (\emph{fourth column}), for well 3 of the vertical slice model.}
    \label{fig:preds_slice}
\end{figure}

Figure \ref{fig:preds_slice} shows the posterior predictions for the temperatures, pressures and enthalpies at well 3 obtained using each variant of EKI. In line with our previous findings, we observe that the predictions generated using EKI with localisation and EKI with inflation appear to exhibit, in general, slightly greater variance than those generated without the use of either technique. 

\begin{figure}
    \centering
    \includegraphics[width=1.0\textwidth]{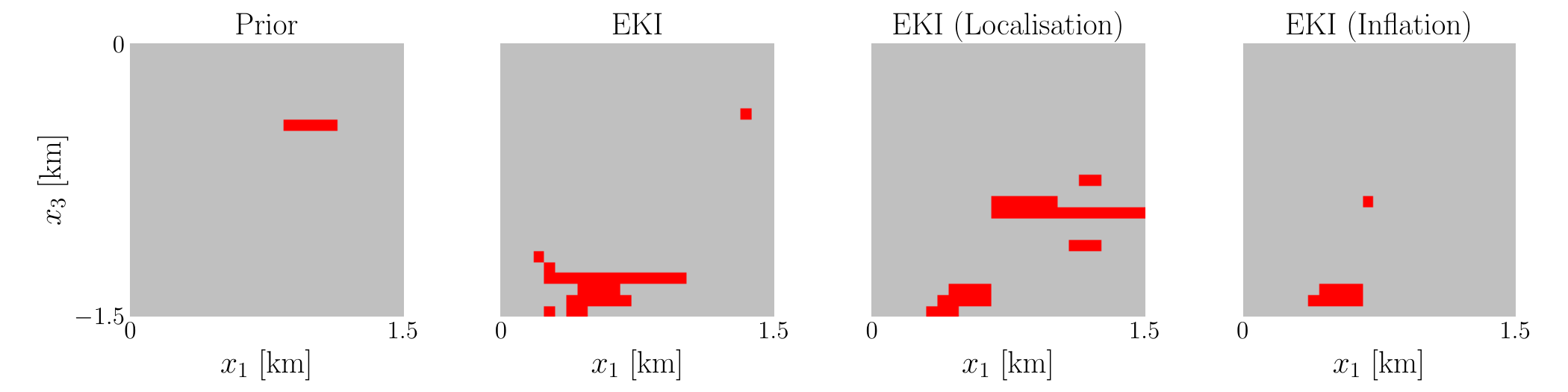}
    \caption{Heatmaps which indicate, for each cell in the mesh of the vertical slice model, whether the true permeability of the cell is contained within the central 95 percent of the ensemble, for the prior ensemble (\emph{far left}), and the approximations to the posterior density generated using EKI (\emph{centre left}), EKI with localisation (\emph{centre right}), and EKI with inflation (\emph{far right}). Red cells are cells for which the permeability is not contained within the central 95 percent of the ensemble.}
    \label{fig:intervals_slice}
\end{figure}

Finally, Figure \ref{fig:intervals_slice} shows heatmaps which indicate, for each cell in the model mesh, whether the true permeability of the cell is contained within the central 95 percent of the ensemble, for each variant of EKI. We observe that when EKI is used with inflation, a slightly greater proportion of the true permeability structure is contained within final ensemble.

\subsection{Results: Synthetic Three-Dimensional Model}

When localisation is applied, our implementation of EKI converges in 6 iterations, and when infla-
tion is applied, it converges in 7 (recall that our implementation without either technique applied
converges in 6 iterations).

\begin{figure}
    \centering
    \includegraphics[width=0.80\textwidth]{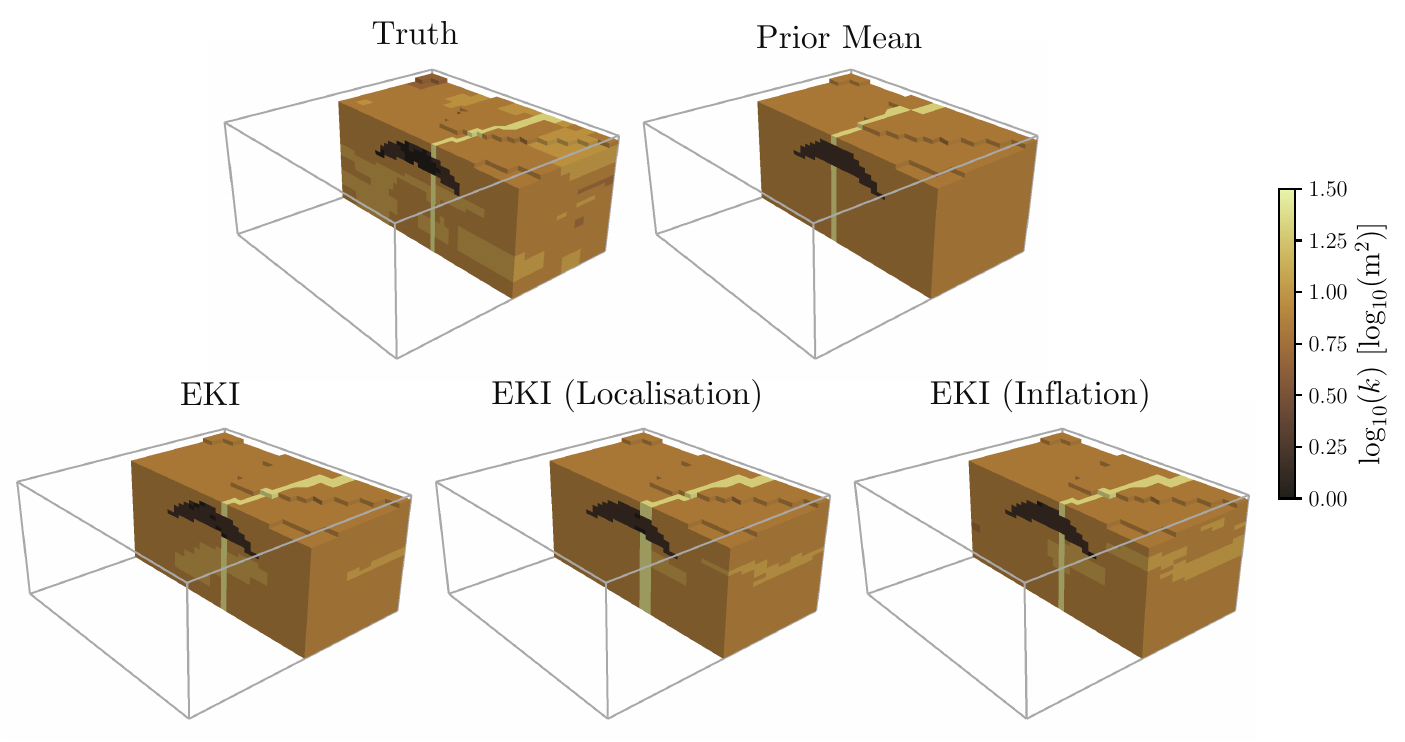}
    \caption{The true permeability structure of the three-dimensional model (\emph{top left}), the prior mean (\emph{top right}), and the approximations to the posterior mean generated using EKI (\emph{bottom left}), EKI with localisation (\emph{bottom centre}), and EKI with inflation (\emph{bottom right}).}
    \label{fig:means_fault}
\end{figure}

Figure \ref{fig:means_fault} shows the approximations to the posterior mean of the permeability structure of the model generated using each variant of EKI. We observe that each approximation appears fairly similar. In all cases, the location of the fault appears to be well-recovered. 

\begin{figure}
    \centering
    \includegraphics[width=1.0\textwidth]{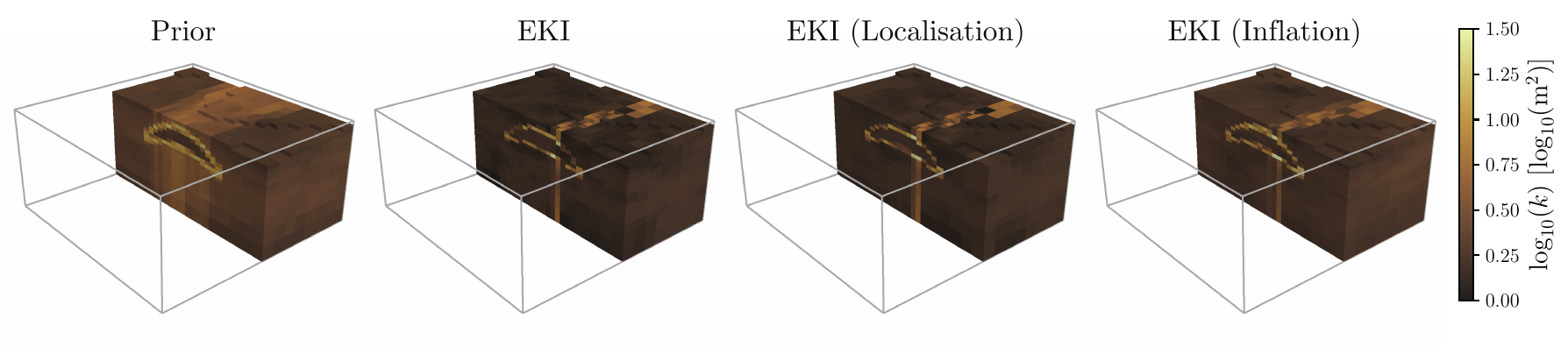}
    \caption{The prior standard deviations of the permeability structure of the three-dimensional model (\emph{far left}), and the approximations to the posterior standard deviations generated using EKI (\emph{centre left}), EKI with localisation (\emph{centre right}), and EKI with inflation (\emph{far right}).}
    \label{fig:stds_fault}
\end{figure}

Figure \ref{fig:stds_fault} shows the approximations to the marginal standard deviations of the permeability in each cell of the model mesh obtained using each variant of EKI. In all cases, the uncertainty is significantly reduced in comparison to the prior uncertainty. We do, however, observe that the estimates obtained using EKI with inflation appear, in general, to be slightly greater than those obtained using the other variants of EKI.

\begin{figure}
    \centering
    \includegraphics[width=1.0\textwidth]{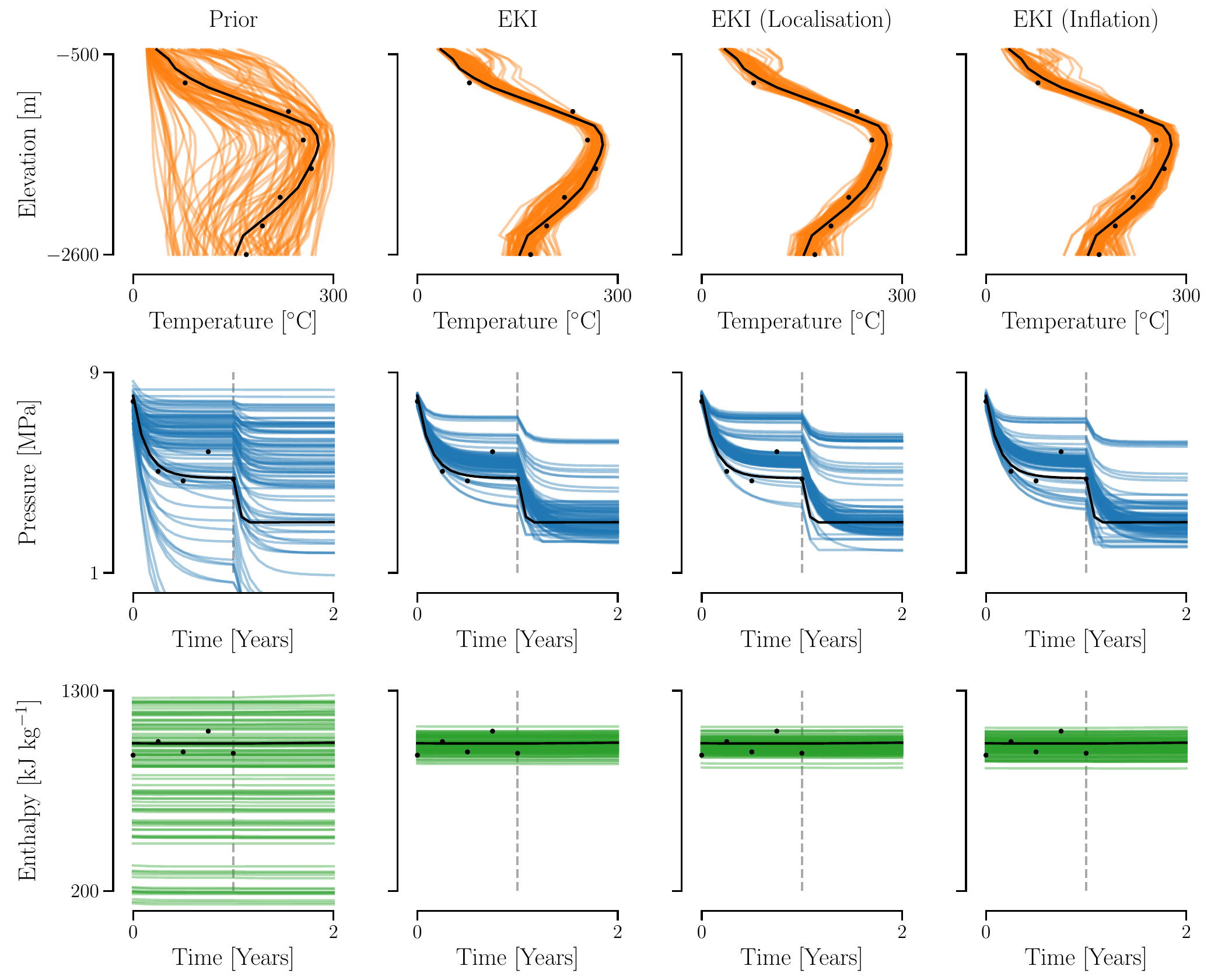}
    \caption{The predicted temperatures (\emph{top row}), pressures (\emph{middle row}), and enthalpies (\emph{bottom row}) for particles sampled from the prior (\emph{first column}) and the approximations to the posterior generated using EKI (\emph{second column}), EKI with localisation (\emph{third column}), and EKI with inflation (\emph{fourth column}), for well 3 of the three-dimensional model. The dashed grey line in the pressure and enthalpy plots denotes the end of the data collection period.}
    \label{fig:preds_fault}
\end{figure}

Figure \ref{fig:preds_fault} shows the posterior predictions for the temperatures, pressures and enthalpies at well 3 obtained using each variant of EKI. Again, in all cases, we observe a significant reduction in uncertainty from prior to posterior, and the true modelled states are contained within the predictions. It appears that the predictions generated using EKI with inflation are associated with a slightly greater variance than those generated using the other variants of EKI; this is perhaps most obvious when studying the temperature predictions.

\begin{figure}
    \centering
    \includegraphics[width=1.0\textwidth]{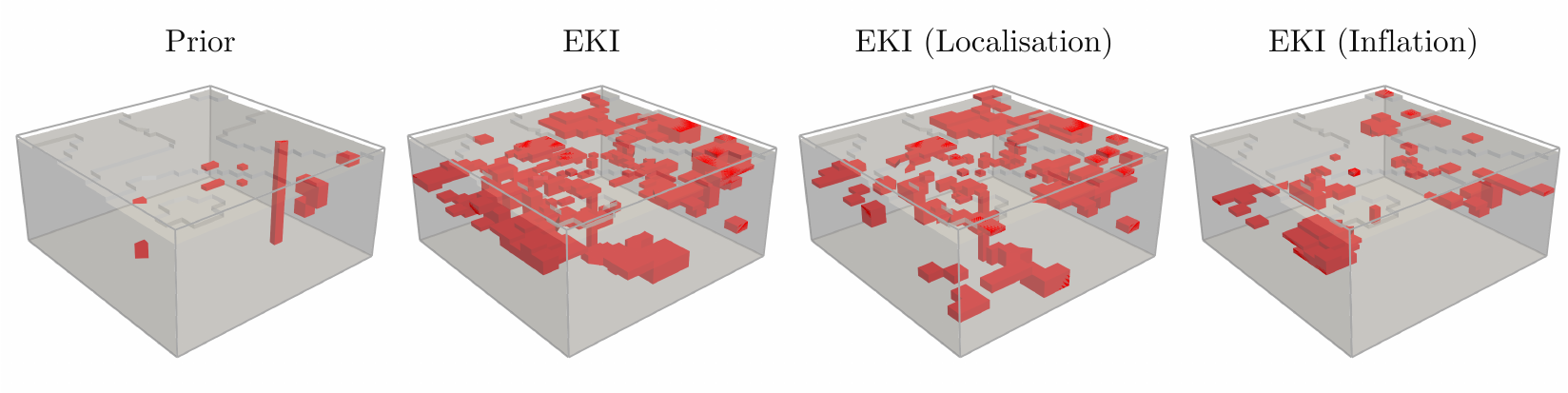}
    \caption{Heatmaps which indicate, for each cell in the mesh of the three-dimensional model, whether the true permeability of the cell is contained within the central 95 percent of the ensemble, for the prior ensemble (\emph{far left}), and the approximations to the posterior density generated using EKI (\emph{centre left}), EKI with localisation (\emph{centre right}), and EKI with inflation (\emph{far right}). Red cells are cells for which the permeability is not contained within the central 95 percent of the ensemble.}
    \label{fig:intervals_fault}
\end{figure}

Finally, Figure \ref{fig:intervals_fault} shows heatmaps which indicate, for each cell in the model mesh, whether the true permeability of the cell is contained within the central $95$ percent of the ensemble, for each variant of EKI. In all cases, the majority of the permeabilities associated with each cell are contained within the ensemble; as in the two-dimensional case, all variants of EKI are able to reduce the uncertainty in the permeability structure of the system significantly without discounting the true permeabilities. In line with our previous findings, the use of EKI with inflation results in a greater percentage of the permeabilities in each cell (98.7 percent) being contained within the final ensemble in comparison to EKI with localisation (96.3 percent) and EKI without either technique applied (94.6 percent).

These results demonstrate that, as expected, the use of localisation and inflation techniques tend to result in an increase in the variance of the posterior and posterior predictive distributions generated using EKI. If one's aim is to capture the true values of the model parameters and predictive quantities of interest within the final ensemble, it appears that the use of these techniques is advisable. We note, however, that these results tell us little in terms of whether the use of these techniques result in an improved approximation to the posterior; for this, we would need to characterise the posterior using a sampling technique such as MCMC \citep[see, e.g.,][]{deBeer24a, Emerick13b, Iglesias13b}.

\section{Hyperparameters} \label{sec:hyperparams}

In this section, we present the approximations to the posterior densities of selected hyperparameters of the Whittle-Mat{\'e}rn fields used as part of the prior parametrisations of each model. In all cases, we show the results obtained using EKI without localisation or inflation applied, EKI with localisation applied, and EKI with inflation applied.

\subsection{Results: Vertical Slice Model}

\begin{figure}
    \centering
    \includegraphics[width=1.0\textwidth]{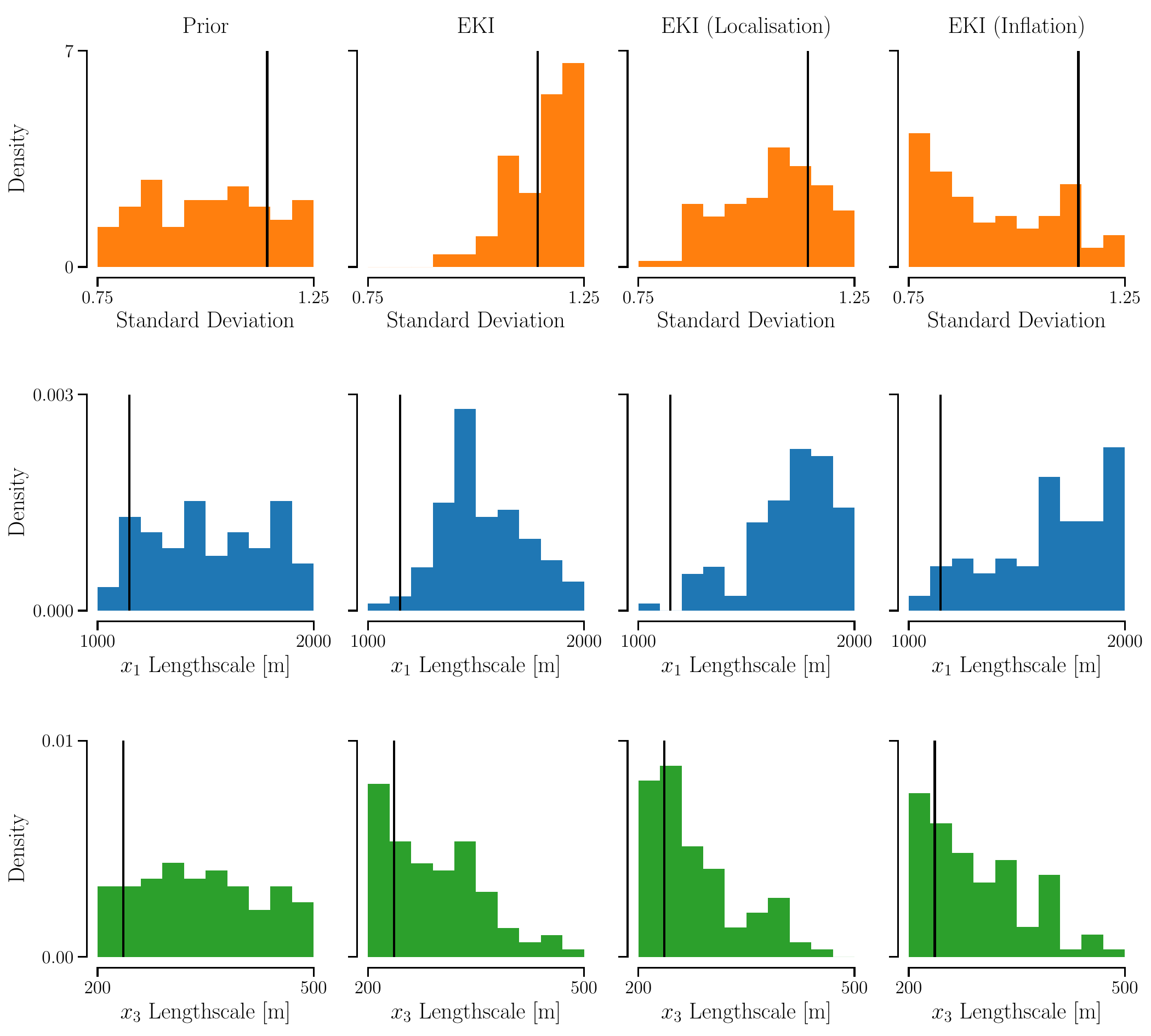}
    \caption{The standard deviation (\emph{top row}), $x_{1}$ (horizontal) lengthscale (\emph{middle row}) and $x_{3}$ (vertical) lengthscale (\emph{bottom row}) of the level set function in region $\Domain_{\Deep}$, for particles sampled from the prior (\emph{first column}) and the approximations to the posterior generated using EKI (\emph{second column}), EKI with localisation (\emph{third column}), and EKI with inflation (\emph{fourth column}), for the vertical slice model. The black vertical line in each plot denotes the true value of the hyperparameter.}
    \label{fig:hyperparams_slice}
\end{figure}

Figure \ref{fig:hyperparams_slice} shows the approximations to the posterior densities of the standard deviation, $x_{1}$ (horizontal) lengthscale, and $x_{3}$ (vertical) lengthscale of the Whittle-Mat{\'e}rn field used as the level set function in region $\Domain_{\Deep}$ obtained using each variant of EKI (results for the level set functions in the other regions of the domain are similar). The densities of the vertical ($x_{3}$) lengthscale produced using all variants of EKI appear similar and are concentrated around the true value. There are, however, significant differences in the estimates of the densities for the standard deviation and horizontal ($x_{1}$) lengthscale. Without knowledge of the true posterior, it is challenging to draw any further conclusions from these results.

\subsection{Results: Synthetic Three-Dimensional Model}

\begin{figure}
    \centering
    \includegraphics[width=1.0\textwidth]{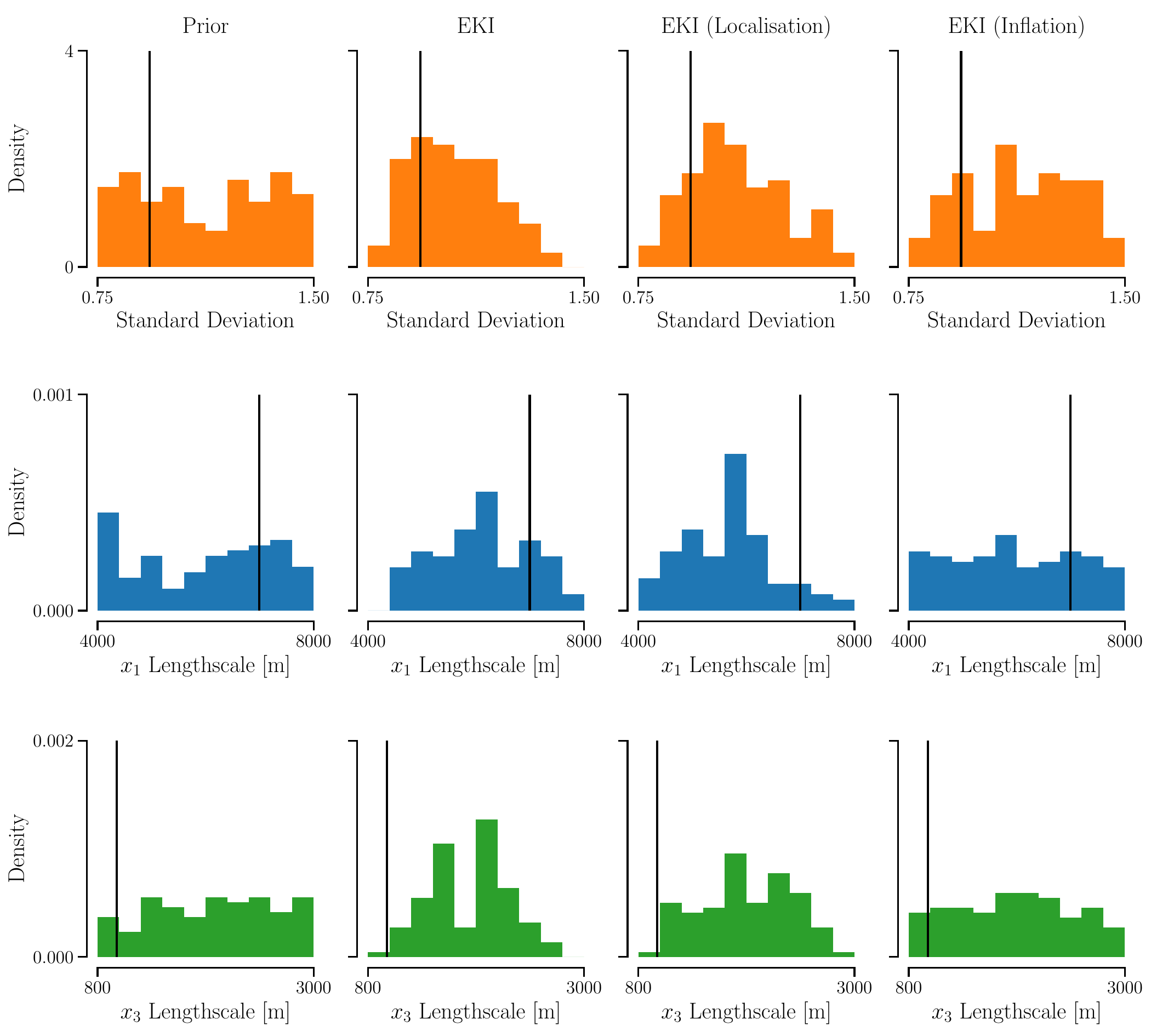}
    \caption{The standard deviation (\emph{top row}), $x_{1}$ (horizontal) lengthscale (\emph{middle row}) and $x_{3}$ (vertical) lengthscale (\emph{bottom row}) of the level set function in region $\Domain_{\Background}$, for particles sampled from the prior (\emph{first column}) and the approximations to the posterior generated using EKI (\emph{second column}), EKI with localisation (\emph{third column}), and EKI with inflation (\emph{fourth column}), for the three-dimensional model. The black vertical line in each plot denotes the true value of the hyperparameter.}
    \label{fig:hyperparams_fault}
\end{figure}

Figure \ref{fig:hyperparams_fault} shows the approximations to the posterior densities of the standard deviation, $x_{1}$ (horizontal) lengthscale, and $x_{3}$ (vertical) lengthscale of the Whittle-Mat{\'e}rn field used as the level set function in region $\Domain_{\Background}$ obtained using each variant of EKI (results for the level set functions in the other regions of the domain are similar). We observe that, in most cases, there appears to be little change from prior to posterior. 

\renewcommand*{\mkbibnamefamily}{\textsc}
\printbibliography